\documentclass[twocolumn,aps,prl,superscriptaddress,tightlines,nofootinbib]{revtex4-2}
\usepackage{hyperref}
\hypersetup{
colorlinks=true,
linkcolor=blue,         
citecolor=blue,         
filecolor=blue,      
urlcolor=blue
}
\usepackage{url}
\usepackage{float}
\usepackage{placeins}
\usepackage{color}
\usepackage{graphicx,epsfig,colordvi,epstopdf}
\usepackage{bm}
\usepackage{mathtools}
\usepackage{mathdots}
\usepackage{physics}

\usepackage{amsmath}
\usepackage{amsfonts}
\usepackage{amssymb}
\usepackage{cleveref}
\newcommand\numberthis{\addtocounter{equation}{1}\tag{\theequation}}

\begin{document}
\title{Mapping out phase diagrams with generative classifiers}
\author{Julian Arnold}
\affiliation{Department of Physics, University of Basel, Klingelbergstrasse 82, 4056 Basel, Switzerland}
\author{Frank Sch\"afer}
\affiliation{CSAIL, Massachusetts Institute of Technology, Cambridge, MA
	02139, USA}
\author{Alan Edelman}
\affiliation{CSAIL, Massachusetts Institute of Technology, Cambridge, MA
	02139, USA}
\affiliation{Department of Mathematics, Massachusetts Institute of Technology, Cambridge, MA
	02139, USA}
 \author{Christoph Bruder}
\affiliation{Department of Physics, University of Basel, Klingelbergstrasse 82, 4056 Basel, Switzerland}
\date{\today}

\begin{abstract}
One of the central tasks in many-body physics is the determination of phase diagrams. However, mapping out a phase diagram generally requires a great deal of human intuition and understanding. To automate this process, one can frame it as a classification task. Typically, classification problems are tackled using discriminative classifiers that explicitly model the probability of the labels for a given sample. Here we show that phase-classification problems are naturally suitable to be solved using generative classifiers based on probabilistic models of the measurement statistics underlying the physical system. Such a generative approach benefits from modeling concepts native to the realm of statistical and quantum physics, as well as recent advances in machine learning. This leads to a powerful framework for the autonomous determination of phase diagrams with little to no human supervision that we showcase in applications to classical equilibrium systems and quantum ground states.
\end{abstract}

\maketitle
\begin{figure}[bth!]
	\centering
		\includegraphics[width=0.75\linewidth]{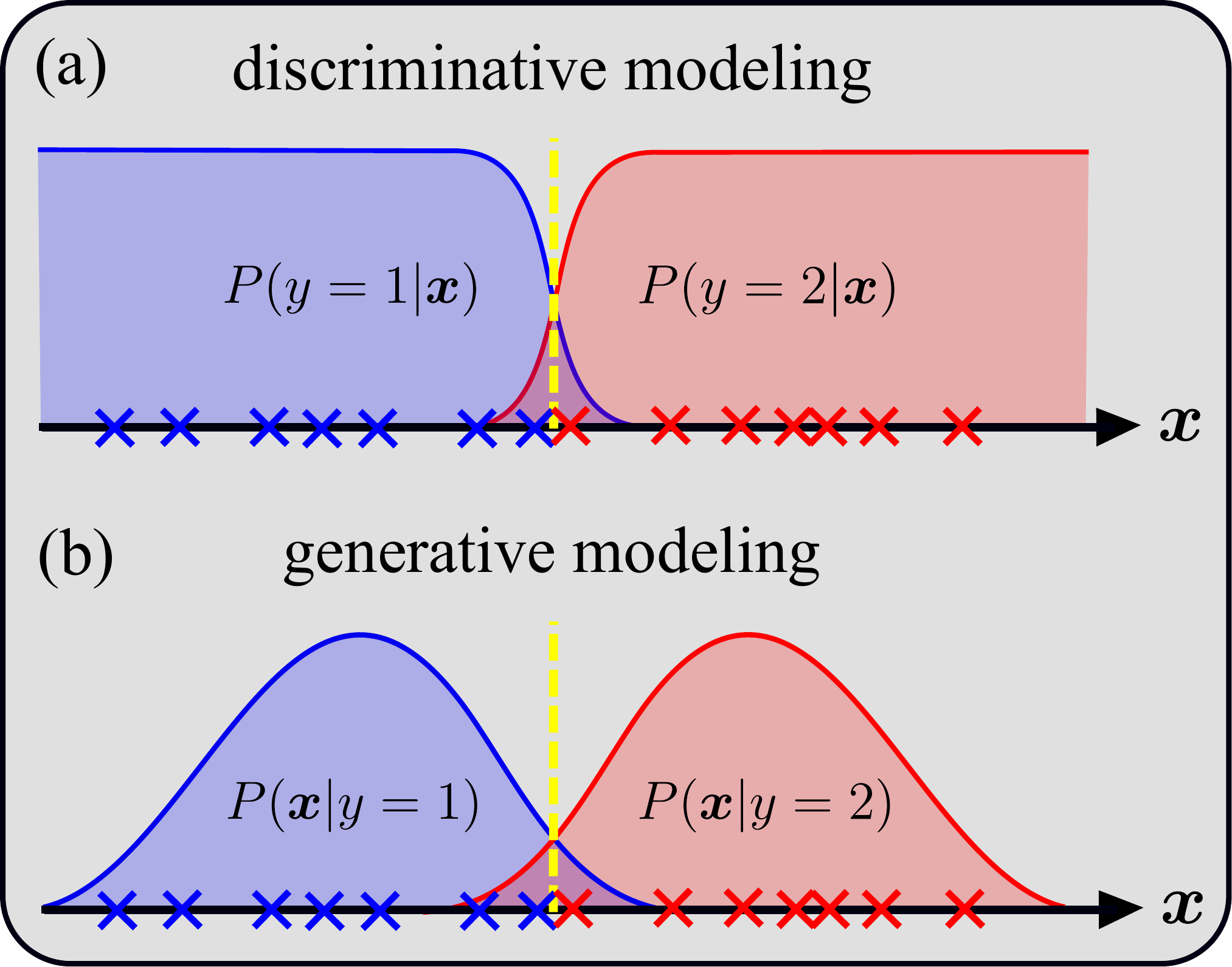}
		\caption{Schematic illustration of the probability distributions modeled in (a) the discriminative and (b) generative approach to a binary classification problem with labels $y \in \{1,2 \}$. In the discriminative approach, $P(y|\bm{x})$ is learned directly from data (crosses). In the generative approach, the class-conditional distributions $P(\bm{x}|y)$ are learned instead.}
		\label{fig:1}
\end{figure}

\emph{Introduction}.---A classification task can be approached in two fundamentally distinct ways~\cite{ng:2001}. Typically, a classifier is constructed by modeling the conditional probability $P(y|\bm{x})$ of label $y$ given a sample $\bm{x}$ directly [see Fig.~\ref{fig:1}(a)]. Models of this type are called discriminative and the corresponding classifier predicts the label $y$ with maximal $P(y|\bm{x})$. Due to the power of neural networks (NNs), particularly convolutional NNs~\cite{krizhevsky:2012}, discriminative classifiers have had tremendous success in fields such as image classification. The alternative is to model the conditional probability of a sample given a label $P(\bm{x}|y)$ [see Fig.~\ref{fig:1}(b)]. Such models are called generative models as they describe how to generate samples $\bm{x}$ conditioned on the class label $y$~\cite{gen_vs_disc}. Applying Bayes' rule,
\begin{equation}\label{eq:Bayes_1}
    P(y|\bm{x}) = \frac{P(\bm{x}|y)P(y)}{P(\bm{x})} = \frac{P(\bm{x}|y)P(y)}{\sum_{y'} P(\bm{x}|y')P(y')},
\end{equation}
one can use a generative model to construct a so-called generative classifier.

\indent Generative models have played a pivotal role in recent advances in machine learning, enabling applications such as the generation of images, composition of music, or translation of text~\cite{vaswani:2017}. Moreover, generative models have been used extensively to describe many-body systems. In statistical physics, Boltzmann distributions have been represented by generative models ranging from mean-field ansatzes to autoregressive networks~\cite{wu:2019}. Similarly, various approaches have been developed to express quantum states, including mean-field ansatzes and tensor networks~\cite{schollwock:2011}, as well as novel machine-learning-inspired generative models, such as restricted Boltzmann machines~\cite{carleo:2017}, or recurrent NNs~\cite{hibat:2020}, and transformers~\cite{cha:2021}.

\indent One of the most important tasks in many-body physics is the characterization of phase diagrams~\cite{goldenfeld:2018,sachdev:2011}. Traditionally, this is done by identifying a small set of low-dimensional physical quantities like response functions or order parameters. The large number of degrees of freedom makes finding such quantities difficult, and success generally requires a great deal of human intuition and understanding. In particular, it can be challenging to identify an order parameter without knowing the symmetry-breaking pattern or in the absence of local order. An exciting recent development has been the automation of this scientific process through machine learning methods~\cite{carleo:2019,terayama:2019,guan:2020,dawid:2022} by exploiting their ability to distill key information from large amounts of high-dimensional data. To this end, the problem of mapping out a phase diagram can be cast as a classification task~\cite{carrasquilla:2017,van:2017,wetzel2:2017,schaefer:2019}. By solving such classification tasks in a data-driven manner using discriminative classifiers, the phase diagrams of a large variety of physical systems have successfully been revealed in both simulation~\cite{carrasquilla:2017,van:2017,wetzel2:2017,schaefer:2019,liu:2018,beach:2018,suchsland:2018,lee:2019,kharkov:2020,guo:2020,greplova:2020,arnold:2021,gavreev:2022,zvyagintseva:2022,tibaldi:2023,guo:2023,schlomer:2023} and experiment~\cite{rem:2019,bohrdt:2021,miles:2023}. Because generic data, such as spin configurations or energy spectra, can be used as input, this constitutes a promising route toward the discovery of novel phases of matter and phase transitions with little to no human supervision.

\indent In this Letter, we show how generative models native to the realm of many-body physics can be used to construct classifiers that solve phase-classification tasks more efficiently compared to discriminative classifiers. The result is a powerful and general framework for the automated characterization of classical and quantum phase diagrams in arbitrary-dimensional parameter spaces featuring multiple phases. 

\indent \emph{Phase classification}.---We assume our physical system to be characterized by a vector $\bm{\gamma}\in \mathbb{R}^{d}$ of tuning parameters and the sampled set of tuning parameters is denoted by $\Gamma$, where $|\Gamma|$ is the number of sampled points. The probability of obtaining the results $\bm{x}\in \mathcal{X}$ when performing a measurement on the system described by $\bm{\gamma}$ is $P(\bm{x}|\bm{\gamma})$. Here, $\mathcal{X}$ is the relevant state space. In a phase-classification task, the label $y\in \mathcal{Y}$ specifies a set of points $\Gamma_{y}$ in the space of tuning parameters. Without loss of generality, we choose a uniform distribution over the set of parameters associated with each class, i.e., $P(\bm{\gamma}|y) = 1 / |\Gamma_{y}|$ for $\bm{\gamma}\in \Gamma_{y}$ and zero otherwise~\cite{suppl}. With this choice, the coarse-grained measurement probability is $P(\bm{x}|y) = \sum_{\bm{\gamma}} P(\bm{x}|\bm{\gamma}) P(\bm{\gamma}|y) = \frac{1}{|\Gamma_{y}|}\sum_{\bm{\gamma} \in \Gamma_{y}} P(\bm{x}|\bm{\gamma})$. Using Eq.~\eqref{eq:Bayes_1}, we have
\begin{equation}\label{eq:Bayes_2}
    P(y|\bm{x}) =  \frac{\frac{1}{|\Gamma_{y}|}\sum_{\bm{\gamma} \in \Gamma_{y}} P(\bm{x}|\bm{\gamma})}{\sum_{y' \in \mathcal{Y}}\frac{1}{|\Gamma_{y'}|}\sum_{\bm{\gamma}' \in \Gamma_{y'}} P(\bm{x}|\bm{\gamma}')},
\end{equation}
where, \emph{a priori}, we consider each label to be equally likely, i.e., $P(y)=1/|\mathcal{Y}|$. In previous work~\cite{van:2017}, the prior probability $P(y)$ was not uniform and instead chosen as $P(y)=\frac{|\Gamma_{y}|}{\sum_{y' \in \mathcal{Y}}|\Gamma_{y'}|}$. In this case, class imbalances are not accounted for and can lead to an unfavorable signal for detecting phase boundaries~\cite{suppl}. In the following, we discuss three distinct approaches for mapping out phase diagrams that differ in how the parameter space is labeled, i.e., in the choice of $\{ \Gamma_{y}\}_{y \in \mathcal{Y}}$.

First, let us assume that we have some prior knowledge of the phase diagram. In particular, assuming we know the number of distinct phases $K$ and their rough location in parameter space, we can choose $\mathcal{Y} = \{1,\dots,K \}$ and $\Gamma_{y}$ to be composed of a set of points characteristic of each phase $y$~\cite{carrasquilla:2017}. Next, we compute the posterior probability $P(y|\bm{\gamma}) = \mathbb{E}_{\bm{x} \sim P(\bm{x}|\bm{\gamma})}\left[ P(y|\bm{x}) \right]$ associated with each phase across a parameter region of interest [see Eq.~\eqref{eq:Bayes_2}]. Rapid changes in the posterior probability are characteristic of phase boundaries. We capture these by the following scalar indicator of phase transitions
\begin{align*}\label{eq:ISL}
    I_{1}(\bm{\gamma}) &= \frac{1}{K} \sum_{y \in \mathcal{Y}} \norm{\nabla_{\bm{\gamma}} P(y|\bm{\gamma}) }_{2},\\
    &= \frac{1}{K}\sum_{y \in \mathcal{Y}} \norm{\mathbb{E}_{\bm{x} \sim P(\bm{x}|\bm{\gamma})}\left[P(y|\bm{x}) \nabla_{\bm{\gamma}} \ln P(\bm{x}|\bm{\gamma})\right]}_{2} \numberthis,
\end{align*}
where $\| \cdot \|_{2}$ denotes the Euclidean norm and we used the log-derivative trick.

\indent The above approach requires partial knowledge of the phase diagram which may be unavailable. To get around this, the authors of Ref.~\cite{van:2017} have proposed a different labeling strategy for one-dimensional parameter spaces which is phase-agnostic. Let $\mathcal{Y} = \{1,2\}$ and for each point $\gamma$ partition the parameter space into two sets $\Gamma_{1} = \{ \gamma' \in \Gamma 
 |\gamma' \leq \gamma\}$ and $\Gamma_{2} = \{ \gamma' \in \Gamma | \gamma' > \gamma\}$. Each parameter $\gamma$ defines a bipartition and in turn a classification task. The associated (optimal) average error probability can be computed as
\begin{equation}\label{eq:LBC_err}
    p_{\rm err}(\gamma) = \frac{1}{2}\sum_{y \in \{1,2 \}} \frac{1}{|\Gamma_{y}|}\sum_{\gamma' \in \Gamma_{y}}\mathbb{E}_{\bm{x} \sim P(\bm{x}|\gamma')} \left[p_{\rm err}(\bm{x})\right],
\end{equation}
where $p_{\rm err}(\bm{x}) = {\rm min}\left\{ P(1|\bm{x}), P(2|\bm{x}) \right\}$ is the (optimal) average error probability when predicting the label of sample $\bm{x}$. Intuitively, one expects $p_{\rm err}(\gamma)$ to be lowest at a phase boundary where the data is partitioned according to the underlying phase structure. Thus, phase boundaries can be detected as local maxima in the indicator $I(\gamma) =  1- 2 p_{\rm err}(\gamma)$ which takes on values between 0 and 1 given that a classifier \emph{at worst} achieves $p_{\rm err}=1/2$.

\begin{figure*}[tbh!]
	\centering
		\includegraphics[width=0.99\linewidth]{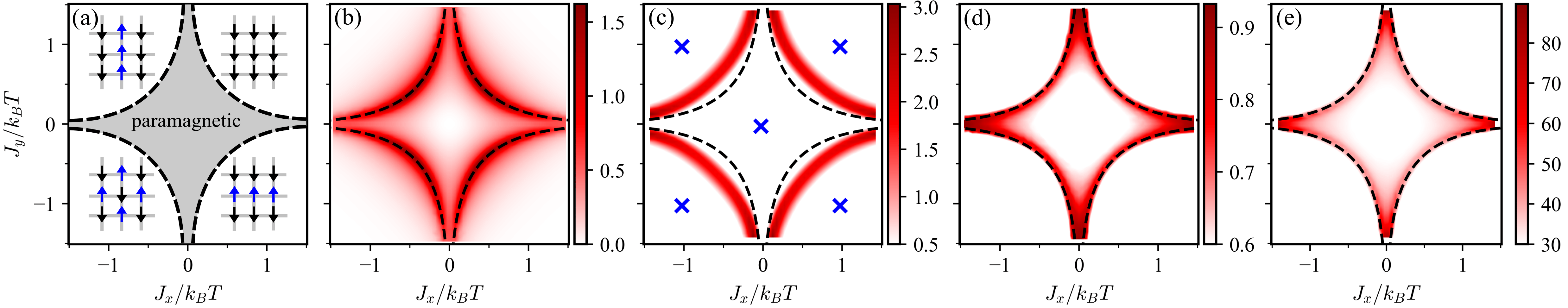}
		\caption{Results for the anisotropic Ising model on a square lattice [Eq.~\eqref{eq:ising_hamiltonian}, $L=20$]. (a) Schematic illustration of the phase diagram with the characteristic spin configurations of each of the four ordered phases. (b) Heat capacity per spin $C_{v}(\bm{\gamma})/Nk_{\rm B}= \left(\langle E^2 \rangle_{\bm{\gamma}} - \langle E \rangle_{\bm{\gamma}}^2\right)/Nk_{\rm B}^2T^2$, where $N=L^2$ is the number of spins. (c) $I_{1}(\bm{\gamma})$ [Eq.~\eqref{eq:ISL}] where the set of points $\{ \Gamma_{y}\}_{y \in \mathcal{Y}}$ representative of each phase is marked by blue crosses ($\mathcal{Y} = \{1,2,3,4,5\}$). (d) $I_{2}(\bm{\gamma})$ [Eq.~\eqref{eq:ILBC}] with $l=1$. (e) $I_{3}(\bm{\gamma})$ [Eq.~\eqref{eq:IPBM}] where $\hat{\bm{\gamma}}(\bm{x})$ is estimated element-wise from line scans. The set $\Gamma$ is composed of a uniform grid with 60 points for each axis and $|\mathcal{D}_{\bm{\gamma}}| = 10^5 \; \forall \bm{\gamma} \in \Gamma$. Onsager's analytical solution~\cite{onsager:1944} for the phase boundary is shown as a black dashed line ($J_y/k_{\rm B}T=-\ln[\tanh(J_x/k_{\rm B}T)]/2 $ for $J_x, J_y >0$; similarly for the other three sectors).}
		\label{fig:ising}
\end{figure*}

\indent In Ref.~\cite{liu:2018}, this procedure was extended to two-dimensional parameter spaces by considering the predicted phase boundary as a parametrized curve that partitions the parameter space locally and is driven via internal forces (e.g., preventing bending and stretching), as well as external forces aiming to minimize the overall classification error. This approach turns out to be unreliable and computationally costly without some partial prior knowledge of the phase diagram (details in Ref.~\cite{suppl}). Instead, we propose an alternative robust generalization that is applicable to parameter spaces of arbitrary dimension building upon the idea of a local bipartition introduced in Ref.~\cite{liu:2018}. At each sampled point $\bm{\gamma}= \left(\gamma_{1}, \gamma_{2},\dots,\gamma_{d}\right) \in \Gamma$, we bipartition the parameter space along each direction. For a given direction $1 \leq i \leq d$, this yields two sets, $\Gamma_{1}^{(i)}(\bm{\gamma})$ and $\Gamma_{2}^{(i)}(\bm{\gamma})$, each comprised of the $l$ points closest to $\bm{\gamma}$ in part 1 and 2 of the split parameter space, respectively (see~\cite{suppl} for an illustration). Based on these sets, we compute an indicator component $I_{2}^{(i)}(\bm{\gamma}) = 1 - 2 p_{\rm err}^{(i)}(\bm{\gamma})$ [see Eq.~\eqref{eq:LBC_err}]. The overall indicator is then given as 
\begin{equation}\label{eq:ILBC}
    I_{2}(\bm{\gamma}) = \sqrt{\sum_{i=1}^{d} \left(I_{2}^{(i)}(\bm{\gamma})\right)^2 }.
\end{equation}

\indent While this procedure does not require partial knowledge of the phase diagram, it requires solving a multitude of phase-classification problems (one for each bipartition). This has been addressed by the phase-agnostic labeling strategy proposed in Ref.~\cite{schaefer:2019} where each sampled value of the tuning parameter is considered its own class $\Gamma_{y} = \{ \bm{\gamma}_{y}\}, \; y \in \mathcal{Y} = \{1,\dots,|\Gamma| \}$. The mean predicted value of the tuning parameter
\begin{equation}
    \hat{\bm{\gamma}}(\bm{\gamma}) = \mathbb{E}_{\bm{x} \sim P(\bm{x}|\bm{\gamma})}\left[ \hat{\bm{\gamma}}(\bm{x})\right] = \mathbb{E}_{\bm{x} \sim P(\bm{x}|\bm{\gamma})}\left[ \sum_{y \in \mathcal{Y}} P(y|\bm{x})\bm{\gamma}_{y}\right]
\end{equation}
is expected to be most sensitive at phase boundaries. We capture this susceptibility by the following indicator
\begin{align*}\label{eq:IPBM}
    I_{3}(\bm{\gamma}) &= \sqrt{\sum_{i=1}^{d} \left(\frac{\partial \hat{\gamma}_{i}(\bm{\gamma}) / \partial \gamma_{i} }{\sigma_{i}(\bm{\gamma})}\right)^2},\\
    &=  \norm{ \mathbb{E}_{\bm{x} \sim P(\bm{x}|\bm{\gamma})}\left[\frac{\hat{\bm{\gamma}}(\bm{x})}{\bm{\sigma}(\bm{\gamma})} \nabla_{\bm{\gamma}} \ln P(\bm{x}|\bm{\gamma}) \right]}_{2} \numberthis,
\end{align*}
where $\bm{\sigma}(\bm{\gamma}) = \sqrt{\mathbb{E}_{\bm{x} \sim P(\bm{x}|\bm{\gamma})}\left[ \hat{\bm{\gamma}}(\bm{x})^2\right] - \left(\mathbb{E}_{\bm{x} \sim P(\bm{x}|\bm{\gamma})}\left[ \hat{\bm{\gamma}}(\bm{x})\right]\right)^2}$ is the associated standard deviation. Here, operations are carried out element-wise and we used the log-derivative trick. Dividing the signal $\partial \hat{\gamma}_{i}/\partial \gamma_{i}$ by its standard deviation $\sigma_{i}$ [cf.~Eq.~\eqref{eq:IPBM}] is found to yield more reliable phase diagrams and alleviate problems encountered in previous studies~\cite{schaefer:2019,arnold:2022} (see~\cite{suppl} for a comparison).

\indent \emph{Discriminative vs. generative modeling}.---By casting the determination of a phase diagram as a classification task, we have reduced the problem to the computation of a scalar indicator of phase transitions $I(\bm{\gamma})$ across the region of interest [see Eqs.~\eqref{eq:ISL}, \eqref{eq:ILBC}, and~\eqref{eq:IPBM}]. Up to now, this computation has typically been approached in a data-driven discriminative way. Given a set of samples $\mathcal{D}_{\bm{\gamma}}$ drawn from $P(\bm{x}|\bm{\gamma})$ for each $\bm{\gamma} \in \Gamma_{y}$, a (parametric) model $\tilde{P}(y|\bm{x})$ is constructed. Typically, $\tilde{P}(y|\bm{x})$ is represented as an NN whose parameters are optimized in a supervised fashion to solve the respective classification task~\cite{suppl}. An estimate of the indicator can be computed by substituting $P(y|\bm{x})$ with $\tilde{P}(y|\bm{x})$ and replacing expected values with a sample mean $\mathbb{E}_{\bm{x} \sim P(\bm{x}|\bm{\gamma})} \rightarrow \frac{1}{|\mathcal{D}_{\bm{\gamma}}|}\sum_{\bm{x} \in \mathcal{D}_{\bm{\gamma}}}$.

\indent The main idea of this Letter is to approach the problem of mapping out a phase diagram in a generative manner. Given a model of the probability distributions underlying the measurement statistics at various discrete points in parameter space $\{ \tilde{P}(\bm{x}|\bm{\gamma}) \}_{\bm{\gamma} \in \Gamma}$ from which one can efficiently sample, an estimate of an indicator [Eqs.~\eqref{eq:ISL}, \eqref{eq:ILBC}, and~\eqref{eq:IPBM}] can be computed by substituting $P(\bm{x}|\bm{\gamma})$ with $\tilde{P}(\bm{x}|\bm{\gamma})$ and replacing expected values with a sample mean $\mathbb{E}_{\bm{x} \sim P(\bm{x}|\bm{\gamma})} \rightarrow \frac{1}{|\tilde{\mathcal{D}}_{\bm{\gamma}}|}\sum_{\bm{x} \in \tilde{\mathcal{D}}_{\bm{\gamma}}}$, where $\tilde{\mathcal{D}}_{\bm{\gamma}}$ denotes the set of samples drawn from the model $\tilde{P}(\bm{x}|\bm{\gamma})$. In machine learning terms, one desires generative models with explicit, tractable densities~\cite{goodfellow2:2016}. Popular examples belonging to this class are autoregressive networks, such as fully visible belief networks or recurrent NNs, normalizing flows, or tensor networks. However, one can also consider non-parametric models, e.g., based on histogram binning, as well as numerically exact (or even analytical) probability distributions if available.

 \begin{figure*}[tbh!]
	\centering
		\includegraphics[width=\linewidth]{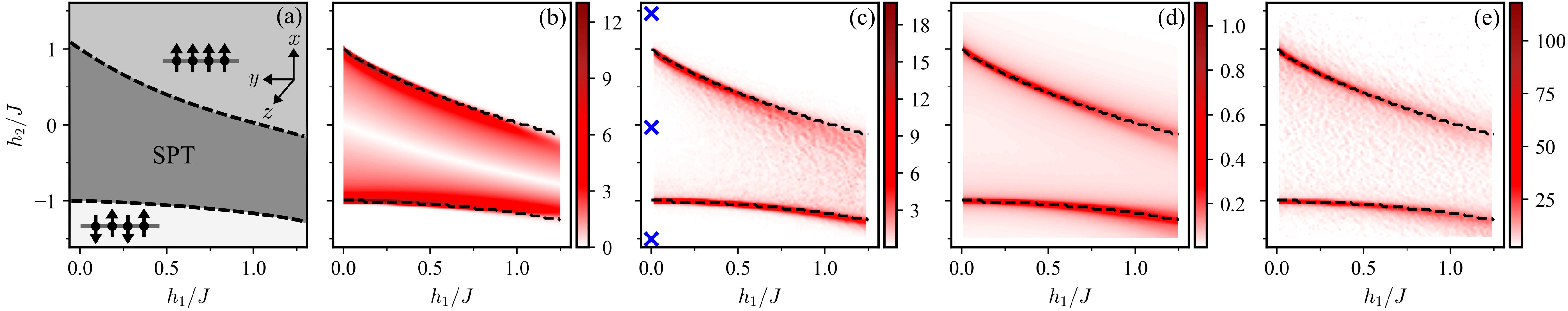}
		\caption{Results for the cluster-Ising model [Eq.~\eqref{eq:cluster_hamiltonian}, $L=71$]. (a) Schematic illustration of the phase diagram featuring three distinct phases: a Néel-type antiferromagnetic phase (bottom), an SPT phase (middle), and a paramagnetic phase (top). (b) Magnitude of the derivative of the string order parameter $|\partial \langle \mathcal{S}\rangle_{\bm{\gamma}}/\partial \gamma_{2}|$. (c) $I_{1}(\bm{\gamma})$ [Eq.~\eqref{eq:ISL}] where the sets of points $\{\Gamma_{y}\}_{y\in \mathcal{Y}}$ representative of each phase are marked by blue crosses ($\mathcal{Y} = \{1,2,3\}$). (d) $I_{2}(\bm{\gamma})$ [Eq.~\eqref{eq:ILBC}] with $l=1$. (e) $I_{3}(\bm{\gamma})$ [Eq.~\eqref{eq:IPBM}] where $\hat{\bm{\gamma}}(\bm{x})$ is estimated element-wise from line scans. The set $\Gamma$ is composed of a uniform grid with 101 points for each axis and $|\tilde{\mathcal{D}}_{\bm{\gamma}}| = 10^3 \;\forall \bm{\gamma} \in \Gamma$. Estimated phase boundaries (black dashed lines) are determined from maxima in $|\partial^2 \langle H \rangle_{\bm{\gamma}} /\partial \gamma_{2}^2|$.}
		\label{fig:SPT}
\end{figure*}

\indent Note that $\tilde{P}(\bm{x}|\bm{\gamma})$ models the measurement statistics underlying the physical system. Thus, in contrast to $\tilde{P}(y|\bm{x})$, $\tilde{P}(\bm{x}|\bm{\gamma})$ is of fundamental nature and can enable various downstream tasks, such as the computation of distinct indicators of phase transitions or physical observables. In numerical investigations, one often has direct access to a description of the system in terms of a generative model that acts as the source of data. The generative approach to classification can use this information to directly construct a classifier instead of learning it iteratively from data, which is necessary in the discriminative approach. In particular, if $\tilde{P}(\bm{x}|\bm{\gamma}) = P(\bm{x}|\bm{\gamma})$ the generative classifier is Bayes optimal~\cite{devroye2:1996,goodfellow:2016} \textit{by construction}, meaning no other classifier can perform better (on average) on the classification task at hand. In contrast, the discriminative approach yields a Bayes optimal classifier in the limit of infinite dataset size and model capacity~\cite{goodfellow:2016} (e.g., utilizing sufficiently large, well-trained NNs), see proof in~\cite{suppl}. In practice, this results in a large computational overhead compared to the generative approach~\cite{suppl}. In the following, we showcase how to map out phase diagrams of classical equilibrium systems and quantum ground states in a generative manner.

\indent \emph{Classical equilibrium systems}.---For a classical system at equilibrium with a large thermal reservoir, the probability to find the system in state $\bm{x} \in \mathcal{X}$ is given by $P(\bm{x}| \bm{\gamma}) = e^{-\mathcal{H}(\bm{x},\bm{\gamma})}/Z(\bm{\gamma})$,
where $Z(\bm{\gamma})$ is the partition function. Here, we consider dimensionless Hamiltonians of the form $\mathcal{H} = H/k_{\rm B}T = \sum_{i=1}^{d} \gamma_{i} X_{i}(\bm{x})$. The large state space $\mathcal{X}$ renders modeling such distributions a hard task. However, such Boltzmann distributions belong to the exponential family and $\bm{X} = \left( X_{1}(\bm{x}), \dots, X_{d}(\bm{x}) \right)$ is a minimal sufficient statistic for $\bm{\gamma}$, i.e., the map $\bm{x} \rightarrow \bm{X}$ corresponds to an optimal lossless compression with respect to $\bm{\gamma}$~\cite{non_optimal_compression}. Thus, to map out phase diagrams by computing an indicator it suffices to model $\{ P(\bm{X}|\bm{\gamma})\}_{\bm{\gamma} \in \Gamma}$ (proof in~\cite{suppl}). Crucially, the dimensionality of the minimal sufficient statistic $\bm{X}$, ${\rm dim}(\bm{X}) = d$, can be independent of the system size. This enables simple non-parametric modeling approaches that are asymptotically unbiased and fast to evaluate. 

\indent As a concrete example, we study the prototypical anisotropic Ising model on an $L \times L$ square lattice described by the Hamiltonian
\begin{equation}\label{eq:ising_hamiltonian}
H= -\sum_{j,k=1}^L  \left( J_x \sigma_{j,k} \sigma_{j, k+1} + J_y \sigma_{j+1,k} \sigma_{j, k} \right),
\end{equation}
where $\sigma_{j,k} \in \{ \pm 1\}$ denote Ising spins and $J_x$ and $J_y$ are the coupling strengths in the horizontal and vertical direction, respectively. At low temperatures, there exist four ordered phases (related by symmetry) each of which undergoes a second-order phase transition to a paramagnetic phase as the temperature is increased [Fig.~\ref{fig:ising}(a)]. The two tuning parameters are $\bm{\gamma} = \left( J_x/k_{\rm B}T, J_y/k_{\rm B}T \right)$ and for a given spin configuration $\bm{\sigma}$ the minimal sufficient statistic is
\begin{equation*}
    \bm{X}(\bm{\sigma}) = \left( -\sum_{j,k=1}^L \sigma_{j,k} \sigma_{j, k+1}\;,\; -\sum_{j,k=1}^L \sigma_{j+1,k} \sigma_{j, k} \right),
\end{equation*}
which corresponds to the nearest-neighbor correlation in $x$- and $y$-direction, respectively. We draw spin configurations $\bm{\sigma}$ from Boltzmann distributions $\{ P(\bm{\sigma}|\bm{\gamma})\}_{\bm{\gamma} \in \Gamma}$ via Markov chain Monte Carlo~\cite{suppl} from which we compute the sufficient statistic and construct empirical distributions  $\{ \tilde{P}(\bm{X}|\bm{\gamma})\}_{\bm{\gamma} \in \Gamma}$ using histogram binning $\tilde{P}(\bm{X}'|\bm{\gamma}) = 1/|\mathcal{D}_{\bm{\gamma}}|\sum_{\bm{\sigma} \in \mathcal{D}_{\bm{\gamma}}} \delta_{\bm{X}(\bm{\sigma}),\bm{X}'}$. Based on these models, we compute the three indicators of phase transitions [Eqs.~\eqref{eq:ISL}, \eqref{eq:ILBC}, and~\eqref{eq:IPBM}], see Figs.~\ref{fig:ising}(c)-(e). The newly proposed indicators $I_{2}$ and $I_{3}$ reproduce the known phase diagram quantitatively and are consistent with physics-informed quantities, such as the heat capacity [cf. Fig.~\ref{fig:ising}(b)], which is remarkable given that these indicators are generic in nature and do not require prior knowledge of the phase diagram. The simple indicator $I_{1}$ only reproduces the phase diagram qualitatively and the result depends on the choice of $\{\Gamma_{y}\}_{y\in \mathcal{Y}}$, i.e., on the amount of prior knowledge of the phase diagram being utilized.

\indent \emph{Quantum ground states}.---For a quantum system subjected to a measurement described by a positive operator-valued measure (POVM), the probability to obtain the measurement outcome $\bm{x} \in \mathcal{X}$ associated with the POVM element $\Pi_{\bm{x}}$ is given by $P(\bm{x}|\bm{\gamma}) = {\rm tr}\left(\Pi_{\bm{x}} \rho(\bm{\gamma})\right)$. Modeling such distributions is a hard task due to the exponential growth of the underlying Hilbert space. However, numerous ansatzes, ranging from mean-field to tensor networks~\cite{schollwock:2011}, as well as machine-learning inspired architectures based on autoregressive NNs~\cite{sharir:2020,hibat:2020} have been developed to approximate ground states on the basis of the variational principle. 

\indent As an example, we consider a spin-$1/2$ chain of length $L$ (odd) governed by a cluster-Ising Hamiltonian~\cite{smacchia:2011,verresen:2017}
\begin{equation}\label{eq:cluster_hamiltonian}
    H = -\sum_{i=1}^{L} \left(JZ_{i-1}X_{i}Z_{i+1} + h_{1}X_{i} + h_{2} X_{i}X_{i+1} \right),
\end{equation}
where $\{X_{i}, Y_{i}, Z_{i}\}$ are the Pauli operators acting on the spin at site $i$ and we consider open boundary conditions defined by $Z_{0} = Z_{L+1}=X_{L+1} = \mathbb{I}$. Here, the tuning parameters $\bm{\gamma} = \left( h_{1}/J,h_{2}/J \right)$ correspond to the external field strength and nearest-neighbor Ising-type coupling, respectively. The ground state phase diagram of this model features three distinct phases [Fig.~\ref{fig:SPT}(a)]: At finite $h_{1}/J$, for  $h_{2}/J \rightarrow \infty$ the system is in a paramagnetic phase with all spins pointing in the $x$-direction, whereas for $h_{2}/J \rightarrow - \infty$  it is in a Néel-type antiferromagnetic phase. At $h_{1}/J = h_{2}/J = 0$, the ground state is a cluster state~\cite{briegel:2001} giving rise to a symmetry-protected topological (SPT) quantum phase ($\mathbb{Z}_{2} \times \mathbb{Z}_{2}$ symmetry) characterized by a non-zero expectation value of the string order parameter $\mathcal{S} = Z_{1}X_{2}X_{4}\cdots X_{L-3}X_{L-1}Z_{L}$. The cluster-Ising model is a prime candidate for studying topological order with quantum computers~\cite{azses:2020,herrmann:2022,smith:2022,zapletal:2023} and has recently been investigated using both classical~\cite{sadoune:2023} and quantum discriminative classifiers~\cite{herrmann:2022,zapletal:2023}.

\indent To avoid any information loss and for the sake of generality, we consider informationally complete POVMs, i.e., measurements whose statistics completely specify the quantum state at hand. For a single qubit, the Pauli-$6$ POVM is a simple and common choice. It consists of the six POVM elements $\{ \frac{1}{3}| 0 \rangle \langle 0 |,\frac{1}{3}| 1 \rangle \langle 1 |,\frac{1}{3}| + \rangle \langle + |,\frac{1}{3}| - \rangle \langle - |,\frac{1}{3}| +i \rangle \langle +i |,\frac{1}{3}| -i \rangle \langle -i | \}$, where $\{ | 0 \rangle, | 1 \rangle\}$, $\{ | + \rangle, | - \rangle\}$, and $\{ | +i \rangle, | -i \rangle\}$ denote the eigenstates of the Pauli operators $Z$, $X$, and $Y$, respectively. A POVM for the entire many-qubit Hilbert space can be constructed using tensor products $\Pi_{\bm{x}} = \Pi_{x_{1}}^{(1)} \otimes \Pi_{x_{2}}^{(2)} \otimes \dots \otimes \Pi_{x_{L}}^{(L)}$, leading to $|\mathcal{X}| = 6^L$ possible measurement outcomes. We construct models for the measurement statistic $\{ \tilde{P}(\bm{x}|\bm{\gamma})\}_{\bm{\gamma} \in \Gamma}$ using a matrix product state ansatz that is optimized via the density matrix renormalization group algorithm~\cite{suppl}. Based on these generative models, we compute indicators of phase transitions [Eqs.~\eqref{eq:ISL}, \eqref{eq:ILBC}, and~\eqref{eq:IPBM}], see Figs.~\ref{fig:SPT}(c)-(e). The signals correctly reproduce the phase diagram obtained from physics-informed quantities, such as the \textit{nonlocal} string order parameter [cf. Fig.~\ref{fig:SPT}(b)], demonstrating the applicability of our framework to systems featuring topological order.

\emph{Conclusions}.---In this Letter, we have introduced generative classifiers as alternatives to discriminative classifiers to solve the classification tasks required for mapping out phase diagrams in an autonomous fashion. The generative approach naturally allows for the incorporation of system knowledge, such as the Hamiltonian or the functional form of the relevant family of probability distributions. This makes the approach favorable for numerical investigations where such information is readily available, as we have explicitly demonstrated for classical systems in equilibrium as well as quantum ground states.

Extensions to classical nonequilibrium systems, as well as other quantum states, are possible by adapting the generative model. For the dynamics of open quantum systems, for example, compatible machine-learning-inspired ansatzes~\cite{luo:2022} and time-dependent variational principles~\cite{reh:2021} have recently been proposed. If system knowledge is scarce, such as when characterizing quantum states prepared in an experiment~\cite{carrasquilla:2019,gomez:2022}, generative models can be constructed in a data-driven manner, e.g., via maximum likelihood estimation. This makes generative classifiers highly versatile and applicable in various contexts. In particular, because generative models play a fundamental role in many-body physics, it is expected that other tasks in this domain that can be cast as classification problems, such as testing physical theories~\cite{bohrdt:2019,zhang2:2019,munoz:2021}, detecting entanglement~\cite{lu:2018}, or investigating thermodynamic principles~\cite{seif:2021}, will benefit from a generative approach. 

\indent A \texttt{Julia}~\cite{bezanson:2012} implementation of our work is available at Ref.~\cite{github}.\\

\indent We would like to thank Niels Lörch and Flemming Holtorf for their helpful suggestions on the manuscript. We thank Alexander Gresch, Lennart Bittel, Martin Kliesch, Niels Lörch, Difei Zhang, Petr Zapletal, and Flemming Holtorf for stimulating discussions. J.A. and C.B. acknowledge financial support from the Swiss National Science Foundation individual grant (grant no. 200020 200481). Computation time at sciCORE (scicore.unibas.ch) scientific computing center at the University of Basel is gratefully acknowledged. We acknowledge financial support from the MIT-Switzerland Lockheed Martin Seed Fund and MIT International Science and Technology Initiatives (MISTI).

\bibliography{refs.bib}

\begin{thebibliography}{70}%
\makeatletter
\providecommand \@ifxundefined [1]{%
 \@ifx{#1\undefined}
}%
\providecommand \@ifnum [1]{%
 \ifnum #1\expandafter \@firstoftwo
 \else \expandafter \@secondoftwo
 \fi
}%
\providecommand \@ifx [1]{%
 \ifx #1\expandafter \@firstoftwo
 \else \expandafter \@secondoftwo
 \fi
}%
\providecommand \natexlab [1]{#1}%
\providecommand \enquote  [1]{``#1''}%
\providecommand \bibnamefont  [1]{#1}%
\providecommand \bibfnamefont [1]{#1}%
\providecommand \citenamefont [1]{#1}%
\providecommand \href@noop [0]{\@secondoftwo}%
\providecommand \href [0]{\begingroup \@sanitize@url \@href}%
\providecommand \@href[1]{\@@startlink{#1}\@@href}%
\providecommand \@@href[1]{\endgroup#1\@@endlink}%
\providecommand \@sanitize@url [0]{\catcode `\\12\catcode `\$12\catcode
  `\&12\catcode `\#12\catcode `\^12\catcode `\_12\catcode `\%12\relax}%
\providecommand \@@startlink[1]{}%
\providecommand \@@endlink[0]{}%
\providecommand \url  [0]{\begingroup\@sanitize@url \@url }%
\providecommand \@url [1]{\endgroup\@href {#1}{\urlprefix }}%
\providecommand \urlprefix  [0]{URL }%
\providecommand \Eprint [0]{\href }%
\providecommand \doibase [0]{https://doi.org/}%
\providecommand \selectlanguage [0]{\@gobble}%
\providecommand \bibinfo  [0]{\@secondoftwo}%
\providecommand \bibfield  [0]{\@secondoftwo}%
\providecommand \translation [1]{[#1]}%
\providecommand \BibitemOpen [0]{}%
\providecommand \bibitemStop [0]{}%
\providecommand \bibitemNoStop [0]{.\EOS\space}%
\providecommand \EOS [0]{\spacefactor3000\relax}%
\providecommand \BibitemShut  [1]{\csname bibitem#1\endcsname}%
\let\auto@bib@innerbib\@empty
\bibitem [{\citenamefont {Ng}\ and\ \citenamefont {Jordan}(2001)}]{ng:2001}%
  \BibitemOpen
  \bibfield  {author} {\bibinfo {author} {\bibfnamefont {A.}~\bibnamefont
  {Ng}}\ and\ \bibinfo {author} {\bibfnamefont {M.}~\bibnamefont {Jordan}},\
  }\bibfield  {title} {\bibinfo {title} {{On Discriminative vs. Generative
  Classifiers: A comparison of logistic regression and naive Bayes}},\ }in\
  \href
  {https://proceedings.neurips.cc/paper/2001/hash/7b7a53e239400a13bd6be6c91c4f6c4e-Abstract.html}
  {\emph {\bibinfo {booktitle} {Adv. Neural Inf. Process. Syst.}}},\
  Vol.~\bibinfo {volume} {14},\ \bibinfo {editor} {edited by\ \bibinfo {editor}
  {\bibfnamefont {T.}~\bibnamefont {Dietterich}}, \bibinfo {editor}
  {\bibfnamefont {S.}~\bibnamefont {Becker}},\ and\ \bibinfo {editor}
  {\bibfnamefont {Z.}~\bibnamefont {Ghahramani}}}\ (\bibinfo  {publisher} {MIT
  Press},\ \bibinfo {year} {2001})\BibitemShut {NoStop}%
\bibitem [{\citenamefont {Krizhevsky}\ \emph {et~al.}(2012)\citenamefont
  {Krizhevsky}, \citenamefont {Sutskever},\ and\ \citenamefont
  {Hinton}}]{krizhevsky:2012}%
  \BibitemOpen
  \bibfield  {author} {\bibinfo {author} {\bibfnamefont {A.}~\bibnamefont
  {Krizhevsky}}, \bibinfo {author} {\bibfnamefont {I.}~\bibnamefont
  {Sutskever}},\ and\ \bibinfo {author} {\bibfnamefont {G.~E.}\ \bibnamefont
  {Hinton}},\ }\bibfield  {title} {\bibinfo {title} {Image{N}et
  {C}lassification with {D}eep {C}onvolutional {N}eural {N}etworks},\ }in\
  \href
  {https://proceedings.neurips.cc/paper/2012/hash/c399862d3b9d6b76c8436e924a68c45b-Abstract.html}
  {\emph {\bibinfo {booktitle} {Adv. Neural Inf. Process. Syst.}}},\
  Vol.~\bibinfo {volume} {25},\ \bibinfo {editor} {edited by\ \bibinfo {editor}
  {\bibfnamefont {F.}~\bibnamefont {Pereira}}, \bibinfo {editor} {\bibfnamefont
  {C.~J.~C.}\ \bibnamefont {Burges}}, \bibinfo {editor} {\bibfnamefont
  {L.}~\bibnamefont {Bottou}},\ and\ \bibinfo {editor} {\bibfnamefont {K.~Q.}\
  \bibnamefont {Weinberger}}}\ (\bibinfo  {publisher} {Curran Associates,
  Inc.},\ \bibinfo {year} {2012})\BibitemShut {NoStop}%
\bibitem [{gen()}]{gen_vs_disc}%
  \BibitemOpen
  \href@noop {} {}\bibinfo {note} {Sometimes a generative model is also defined
  as a statistical model of the joint probability distribution $P(\bm{x},y)$.
  Here, we stick with the former definition given that $P(y)$ is chosen
  beforehand in phase-classification tasks.}\BibitemShut {Stop}%
\bibitem [{\citenamefont {Vaswani}\ \emph {et~al.}(2017)\citenamefont
  {Vaswani}, \citenamefont {Shazeer}, \citenamefont {Parmar}, \citenamefont
  {Uszkoreit}, \citenamefont {Jones}, \citenamefont {Gomez}, \citenamefont
  {Kaiser},\ and\ \citenamefont {Polosukhin}}]{vaswani:2017}%
  \BibitemOpen
  \bibfield  {author} {\bibinfo {author} {\bibfnamefont {A.}~\bibnamefont
  {Vaswani}}, \bibinfo {author} {\bibfnamefont {N.}~\bibnamefont {Shazeer}},
  \bibinfo {author} {\bibfnamefont {N.}~\bibnamefont {Parmar}}, \bibinfo
  {author} {\bibfnamefont {J.}~\bibnamefont {Uszkoreit}}, \bibinfo {author}
  {\bibfnamefont {L.}~\bibnamefont {Jones}}, \bibinfo {author} {\bibfnamefont
  {A.~N.}\ \bibnamefont {Gomez}}, \bibinfo {author} {\bibfnamefont
  {{\L}.}~\bibnamefont {Kaiser}},\ and\ \bibinfo {author} {\bibfnamefont
  {I.}~\bibnamefont {Polosukhin}},\ }\bibfield  {title} {\bibinfo {title}
  {{Attention is All you Need}},\ }in\ \href
  {https://papers.nips.cc/paper_files/paper/2017/hash/3f5ee243547dee91fbd053c1c4a845aa-Abstract.html}
  {\emph {\bibinfo {booktitle} {Adv. Neural Inf. Process. Syst.}}},\
  Vol.~\bibinfo {volume} {30},\ \bibinfo {editor} {edited by\ \bibinfo {editor}
  {\bibfnamefont {I.}~\bibnamefont {Guyon}}, \bibinfo {editor} {\bibfnamefont
  {U.~V.}\ \bibnamefont {Luxburg}}, \bibinfo {editor} {\bibfnamefont
  {S.}~\bibnamefont {Bengio}}, \bibinfo {editor} {\bibfnamefont
  {H.}~\bibnamefont {Wallach}}, \bibinfo {editor} {\bibfnamefont
  {R.}~\bibnamefont {Fergus}}, \bibinfo {editor} {\bibfnamefont
  {S.}~\bibnamefont {Vishwanathan}},\ and\ \bibinfo {editor} {\bibfnamefont
  {R.}~\bibnamefont {Garnett}}}\ (\bibinfo  {publisher} {Curran Associates,
  Inc.},\ \bibinfo {year} {2017})\BibitemShut {NoStop}%
\bibitem [{\citenamefont {Wu}\ \emph {et~al.}(2019)\citenamefont {Wu},
  \citenamefont {Wang},\ and\ \citenamefont {Zhang}}]{wu:2019}%
  \BibitemOpen
  \bibfield  {author} {\bibinfo {author} {\bibfnamefont {D.}~\bibnamefont
  {Wu}}, \bibinfo {author} {\bibfnamefont {L.}~\bibnamefont {Wang}},\ and\
  \bibinfo {author} {\bibfnamefont {P.}~\bibnamefont {Zhang}},\ }\bibfield
  {title} {\bibinfo {title} {{S}olving {S}tatistical {M}echanics {U}sing
  {V}ariational {A}utoregressive {N}etworks},\ }\href
  {https://doi.org/10.1103/PhysRevLett.122.080602} {\bibfield  {journal}
  {\bibinfo  {journal} {Phys. Rev. Lett.}\ }\textbf {\bibinfo {volume} {122}},\
  \bibinfo {pages} {080602} (\bibinfo {year} {2019})}\BibitemShut {NoStop}%
\bibitem [{\citenamefont {Schollwöck}(2011)}]{schollwock:2011}%
  \BibitemOpen
  \bibfield  {author} {\bibinfo {author} {\bibfnamefont {U.}~\bibnamefont
  {Schollwöck}},\ }\bibfield  {title} {\bibinfo {title} {The density-matrix
  renormalization group in the age of matrix product states},\ }\href
  {https://doi.org/https://doi.org/10.1016/j.aop.2010.09.012} {\bibfield
  {journal} {\bibinfo  {journal} {Ann. Phys.}\ }\textbf {\bibinfo {volume}
  {326}},\ \bibinfo {pages} {96} (\bibinfo {year} {2011})}\BibitemShut
  {NoStop}%
\bibitem [{\citenamefont {Carleo}\ and\ \citenamefont
  {Troyer}(2017)}]{carleo:2017}%
  \BibitemOpen
  \bibfield  {author} {\bibinfo {author} {\bibfnamefont {G.}~\bibnamefont
  {Carleo}}\ and\ \bibinfo {author} {\bibfnamefont {M.}~\bibnamefont
  {Troyer}},\ }\bibfield  {title} {\bibinfo {title} {Solving the quantum
  many-body problem with artificial neural networks},\ }\href
  {https://doi.org/10.1126/science.aag2302} {\bibfield  {journal} {\bibinfo
  {journal} {Science}\ }\textbf {\bibinfo {volume} {355}},\ \bibinfo {pages}
  {602} (\bibinfo {year} {2017})}\BibitemShut {NoStop}%
\bibitem [{\citenamefont {Hibat-Allah}\ \emph {et~al.}(2020)\citenamefont
  {Hibat-Allah}, \citenamefont {Ganahl}, \citenamefont {Hayward}, \citenamefont
  {Melko},\ and\ \citenamefont {Carrasquilla}}]{hibat:2020}%
  \BibitemOpen
  \bibfield  {author} {\bibinfo {author} {\bibfnamefont {M.}~\bibnamefont
  {Hibat-Allah}}, \bibinfo {author} {\bibfnamefont {M.}~\bibnamefont {Ganahl}},
  \bibinfo {author} {\bibfnamefont {L.~E.}\ \bibnamefont {Hayward}}, \bibinfo
  {author} {\bibfnamefont {R.~G.}\ \bibnamefont {Melko}},\ and\ \bibinfo
  {author} {\bibfnamefont {J.}~\bibnamefont {Carrasquilla}},\ }\bibfield
  {title} {\bibinfo {title} {Recurrent neural network wave functions},\ }\href
  {https://doi.org/10.1103/PhysRevResearch.2.023358} {\bibfield  {journal}
  {\bibinfo  {journal} {Phys. Rev. Res.}\ }\textbf {\bibinfo {volume} {2}},\
  \bibinfo {pages} {023358} (\bibinfo {year} {2020})}\BibitemShut {NoStop}%
\bibitem [{\citenamefont {Cha}\ \emph {et~al.}(2021)\citenamefont {Cha},
  \citenamefont {Ginsparg}, \citenamefont {Wu}, \citenamefont {Carrasquilla},
  \citenamefont {McMahon},\ and\ \citenamefont {Kim}}]{cha:2021}%
  \BibitemOpen
  \bibfield  {author} {\bibinfo {author} {\bibfnamefont {P.}~\bibnamefont
  {Cha}}, \bibinfo {author} {\bibfnamefont {P.}~\bibnamefont {Ginsparg}},
  \bibinfo {author} {\bibfnamefont {F.}~\bibnamefont {Wu}}, \bibinfo {author}
  {\bibfnamefont {J.}~\bibnamefont {Carrasquilla}}, \bibinfo {author}
  {\bibfnamefont {P.~L.}\ \bibnamefont {McMahon}},\ and\ \bibinfo {author}
  {\bibfnamefont {E.-A.}\ \bibnamefont {Kim}},\ }\bibfield  {title} {\bibinfo
  {title} {Attention-based quantum tomography},\ }\href
  {https://doi.org/10.1088/2632-2153/ac362b} {\bibfield  {journal} {\bibinfo
  {journal} {Mach. Learn.: Sci. Technol.}\ }\textbf {\bibinfo {volume} {3}},\
  \bibinfo {pages} {01LT01} (\bibinfo {year} {2021})}\BibitemShut {NoStop}%
\bibitem [{\citenamefont {Goldenfeld}(2018)}]{goldenfeld:2018}%
  \BibitemOpen
  \bibfield  {author} {\bibinfo {author} {\bibfnamefont {N.}~\bibnamefont
  {Goldenfeld}},\ }\href {https://doi.org/10.1201/9780429493492} {\emph
  {\bibinfo {title} {{Lectures On Phase Transitions And The Renormalization
  Group}}}}\ (\bibinfo  {publisher} {CRC Press},\ \bibinfo {year}
  {2018})\BibitemShut {NoStop}%
\bibitem [{\citenamefont {Sachdev}(2011)}]{sachdev:2011}%
  \BibitemOpen
  \bibfield  {author} {\bibinfo {author} {\bibfnamefont {S.}~\bibnamefont
  {Sachdev}},\ }\href {https://doi.org/10.1017/CBO9780511973765} {\emph
  {\bibinfo {title} {{Quantum Phase Transitions}}}}\ (\bibinfo  {publisher}
  {Cambridge University Press},\ \bibinfo {year} {2011})\BibitemShut {NoStop}%
\bibitem [{\citenamefont {Carleo}\ \emph {et~al.}(2019)\citenamefont {Carleo},
  \citenamefont {Cirac}, \citenamefont {Cranmer}, \citenamefont {Daudet},
  \citenamefont {Schuld}, \citenamefont {Tishby}, \citenamefont
  {Vogt-Maranto},\ and\ \citenamefont {Zdeborov\'a}}]{carleo:2019}%
  \BibitemOpen
  \bibfield  {author} {\bibinfo {author} {\bibfnamefont {G.}~\bibnamefont
  {Carleo}}, \bibinfo {author} {\bibfnamefont {I.}~\bibnamefont {Cirac}},
  \bibinfo {author} {\bibfnamefont {K.}~\bibnamefont {Cranmer}}, \bibinfo
  {author} {\bibfnamefont {L.}~\bibnamefont {Daudet}}, \bibinfo {author}
  {\bibfnamefont {M.}~\bibnamefont {Schuld}}, \bibinfo {author} {\bibfnamefont
  {N.}~\bibnamefont {Tishby}}, \bibinfo {author} {\bibfnamefont
  {L.}~\bibnamefont {Vogt-Maranto}},\ and\ \bibinfo {author} {\bibfnamefont
  {L.}~\bibnamefont {Zdeborov\'a}},\ }\bibfield  {title} {\bibinfo {title}
  {Machine learning and the physical sciences},\ }\href
  {https://doi.org/10.1103/RevModPhys.91.045002} {\bibfield  {journal}
  {\bibinfo  {journal} {Rev. Mod. Phys.}\ }\textbf {\bibinfo {volume} {91}},\
  \bibinfo {pages} {045002} (\bibinfo {year} {2019})}\BibitemShut {NoStop}%
\bibitem [{\citenamefont {Terayama}\ \emph {et~al.}(2019)\citenamefont
  {Terayama}, \citenamefont {Tamura}, \citenamefont {Nose}, \citenamefont
  {Hiramatsu}, \citenamefont {Hosono}, \citenamefont {Okuno},\ and\
  \citenamefont {Tsuda}}]{terayama:2019}%
  \BibitemOpen
  \bibfield  {author} {\bibinfo {author} {\bibfnamefont {K.}~\bibnamefont
  {Terayama}}, \bibinfo {author} {\bibfnamefont {R.}~\bibnamefont {Tamura}},
  \bibinfo {author} {\bibfnamefont {Y.}~\bibnamefont {Nose}}, \bibinfo {author}
  {\bibfnamefont {H.}~\bibnamefont {Hiramatsu}}, \bibinfo {author}
  {\bibfnamefont {H.}~\bibnamefont {Hosono}}, \bibinfo {author} {\bibfnamefont
  {Y.}~\bibnamefont {Okuno}},\ and\ \bibinfo {author} {\bibfnamefont
  {K.}~\bibnamefont {Tsuda}},\ }\bibfield  {title} {\bibinfo {title} {Efficient
  construction method for phase diagrams using uncertainty sampling},\ }\href
  {https://doi.org/10.1103/PhysRevMaterials.3.033802} {\bibfield  {journal}
  {\bibinfo  {journal} {Phys. Rev. Mater.}\ }\textbf {\bibinfo {volume} {3}},\
  \bibinfo {pages} {033802} (\bibinfo {year} {2019})}\BibitemShut {NoStop}%
\bibitem [{\citenamefont {Guan}\ and\ \citenamefont
  {Viswanathan}(2020)}]{guan:2020}%
  \BibitemOpen
  \bibfield  {author} {\bibinfo {author} {\bibfnamefont {P.-W.}\ \bibnamefont
  {Guan}}\ and\ \bibinfo {author} {\bibfnamefont {V.}~\bibnamefont
  {Viswanathan}},\ }\bibfield  {title} {\bibinfo {title} {{MeltNet}: Predicting
  alloy melting temperature by machine learning},\ }\href
  {https://arxiv.org/abs/2010.14048} {\bibfield  {journal} {\bibinfo  {journal}
  {arXiv:2010.14048}\ } (\bibinfo {year} {2020})}\BibitemShut {NoStop}%
\bibitem [{\citenamefont {Dawid}\ \emph {et~al.}(2022)\citenamefont {Dawid},
  \citenamefont {Arnold}, \citenamefont {Requena}, \citenamefont {Gresch},
  \citenamefont {P{\l}odzie{\'n}}, \citenamefont {Donatella}, \citenamefont
  {Nicoli}, \citenamefont {Stornati}, \citenamefont {Koch}, \citenamefont
  {B{\"u}ttner} \emph {et~al.}}]{dawid:2022}%
  \BibitemOpen
  \bibfield  {author} {\bibinfo {author} {\bibfnamefont {A.}~\bibnamefont
  {Dawid}}, \bibinfo {author} {\bibfnamefont {J.}~\bibnamefont {Arnold}},
  \bibinfo {author} {\bibfnamefont {B.}~\bibnamefont {Requena}}, \bibinfo
  {author} {\bibfnamefont {A.}~\bibnamefont {Gresch}}, \bibinfo {author}
  {\bibfnamefont {M.}~\bibnamefont {P{\l}odzie{\'n}}}, \bibinfo {author}
  {\bibfnamefont {K.}~\bibnamefont {Donatella}}, \bibinfo {author}
  {\bibfnamefont {K.~A.}\ \bibnamefont {Nicoli}}, \bibinfo {author}
  {\bibfnamefont {P.}~\bibnamefont {Stornati}}, \bibinfo {author}
  {\bibfnamefont {R.}~\bibnamefont {Koch}}, \bibinfo {author} {\bibfnamefont
  {M.}~\bibnamefont {B{\"u}ttner}}, \emph {et~al.},\ }\bibfield  {title}
  {\bibinfo {title} {Modern applications of machine learning in quantum
  sciences},\ }\href {https://arxiv.org/abs/2204.04198} {\bibfield  {journal}
  {\bibinfo  {journal} {arXiv:2204.04198}\ } (\bibinfo {year}
  {2022})}\BibitemShut {NoStop}%
\bibitem [{\citenamefont {Carrasquilla}\ and\ \citenamefont
  {Melko}(2017)}]{carrasquilla:2017}%
  \BibitemOpen
  \bibfield  {author} {\bibinfo {author} {\bibfnamefont {J.}~\bibnamefont
  {Carrasquilla}}\ and\ \bibinfo {author} {\bibfnamefont {R.~G.}\ \bibnamefont
  {Melko}},\ }\bibfield  {title} {\bibinfo {title} {Machine learning phases of
  matter},\ }\href {https://doi.org/10.1038/nphys4035} {\bibfield  {journal}
  {\bibinfo  {journal} {Nat. Phys.}\ }\textbf {\bibinfo {volume} {13}},\
  \bibinfo {pages} {431} (\bibinfo {year} {2017})}\BibitemShut {NoStop}%
\bibitem [{\citenamefont {van Nieuwenburg}\ \emph {et~al.}(2017)\citenamefont
  {van Nieuwenburg}, \citenamefont {Liu},\ and\ \citenamefont
  {Huber}}]{van:2017}%
  \BibitemOpen
  \bibfield  {author} {\bibinfo {author} {\bibfnamefont {E.~P.~L.}\
  \bibnamefont {van Nieuwenburg}}, \bibinfo {author} {\bibfnamefont {Y.-H.}\
  \bibnamefont {Liu}},\ and\ \bibinfo {author} {\bibfnamefont {S.~D.}\
  \bibnamefont {Huber}},\ }\bibfield  {title} {\bibinfo {title} {Learning phase
  transitions by confusion},\ }\href {https://doi.org/10.1038/nphys4037}
  {\bibfield  {journal} {\bibinfo  {journal} {Nat. Phys.}\ }\textbf {\bibinfo
  {volume} {13}},\ \bibinfo {pages} {435} (\bibinfo {year} {2017})}\BibitemShut
  {NoStop}%
\bibitem [{\citenamefont {Wetzel}\ and\ \citenamefont
  {Scherzer}(2017)}]{wetzel2:2017}%
  \BibitemOpen
  \bibfield  {author} {\bibinfo {author} {\bibfnamefont {S.~J.}\ \bibnamefont
  {Wetzel}}\ and\ \bibinfo {author} {\bibfnamefont {M.}~\bibnamefont
  {Scherzer}},\ }\bibfield  {title} {\bibinfo {title} {Machine learning of
  explicit order parameters: {F}rom the {I}sing model to {SU}(2) lattice gauge
  theory},\ }\href {https://doi.org/10.1103/PhysRevB.96.184410} {\bibfield
  {journal} {\bibinfo  {journal} {Phys. Rev. B}\ }\textbf {\bibinfo {volume}
  {96}},\ \bibinfo {pages} {184410} (\bibinfo {year} {2017})}\BibitemShut
  {NoStop}%
\bibitem [{\citenamefont {Sch\"afer}\ and\ \citenamefont
  {L\"orch}(2019)}]{schaefer:2019}%
  \BibitemOpen
  \bibfield  {author} {\bibinfo {author} {\bibfnamefont {F.}~\bibnamefont
  {Sch\"afer}}\ and\ \bibinfo {author} {\bibfnamefont {N.}~\bibnamefont
  {L\"orch}},\ }\bibfield  {title} {\bibinfo {title} {Vector field divergence
  of predictive model output as indication of phase transitions},\ }\href
  {https://doi.org/10.1103/PhysRevE.99.062107} {\bibfield  {journal} {\bibinfo
  {journal} {Phys. Rev. E}\ }\textbf {\bibinfo {volume} {99}},\ \bibinfo
  {pages} {062107} (\bibinfo {year} {2019})}\BibitemShut {NoStop}%
\bibitem [{\citenamefont {Liu}\ and\ \citenamefont {van
  Nieuwenburg}(2018)}]{liu:2018}%
  \BibitemOpen
  \bibfield  {author} {\bibinfo {author} {\bibfnamefont {Y.-H.}\ \bibnamefont
  {Liu}}\ and\ \bibinfo {author} {\bibfnamefont {E.~P.~L.}\ \bibnamefont {van
  Nieuwenburg}},\ }\bibfield  {title} {\bibinfo {title} {{D}iscriminative
  {C}ooperative {N}etworks for {D}etecting {P}hase {T}ransitions},\ }\href
  {https://doi.org/10.1103/PhysRevLett.120.176401} {\bibfield  {journal}
  {\bibinfo  {journal} {Phys. Rev. Lett.}\ }\textbf {\bibinfo {volume} {120}},\
  \bibinfo {pages} {176401} (\bibinfo {year} {2018})}\BibitemShut {NoStop}%
\bibitem [{\citenamefont {Beach}\ \emph {et~al.}(2018)\citenamefont {Beach},
  \citenamefont {Golubeva},\ and\ \citenamefont {Melko}}]{beach:2018}%
  \BibitemOpen
  \bibfield  {author} {\bibinfo {author} {\bibfnamefont {M.~J.~S.}\
  \bibnamefont {Beach}}, \bibinfo {author} {\bibfnamefont {A.}~\bibnamefont
  {Golubeva}},\ and\ \bibinfo {author} {\bibfnamefont {R.~G.}\ \bibnamefont
  {Melko}},\ }\bibfield  {title} {\bibinfo {title} {Machine learning vortices
  at the {K}osterlitz-{T}houless transition},\ }\href
  {https://doi.org/10.1103/PhysRevB.97.045207} {\bibfield  {journal} {\bibinfo
  {journal} {Phys. Rev. B}\ }\textbf {\bibinfo {volume} {97}},\ \bibinfo
  {pages} {045207} (\bibinfo {year} {2018})}\BibitemShut {NoStop}%
\bibitem [{\citenamefont {Suchsland}\ and\ \citenamefont
  {Wessel}(2018)}]{suchsland:2018}%
  \BibitemOpen
  \bibfield  {author} {\bibinfo {author} {\bibfnamefont {P.}~\bibnamefont
  {Suchsland}}\ and\ \bibinfo {author} {\bibfnamefont {S.}~\bibnamefont
  {Wessel}},\ }\bibfield  {title} {\bibinfo {title} {Parameter diagnostics of
  phases and phase transition learning by neural networks},\ }\href
  {https://doi.org/10.1103/PhysRevB.97.174435} {\bibfield  {journal} {\bibinfo
  {journal} {Phys. Rev. B}\ }\textbf {\bibinfo {volume} {97}},\ \bibinfo
  {pages} {174435} (\bibinfo {year} {2018})}\BibitemShut {NoStop}%
\bibitem [{\citenamefont {Lee}\ and\ \citenamefont {Kim}(2019)}]{lee:2019}%
  \BibitemOpen
  \bibfield  {author} {\bibinfo {author} {\bibfnamefont {S.~S.}\ \bibnamefont
  {Lee}}\ and\ \bibinfo {author} {\bibfnamefont {B.~J.}\ \bibnamefont {Kim}},\
  }\bibfield  {title} {\bibinfo {title} {Confusion scheme in machine learning
  detects double phase transitions and quasi-long-range order},\ }\href
  {https://doi.org/10.1103/PhysRevE.99.043308} {\bibfield  {journal} {\bibinfo
  {journal} {Phys. Rev. E}\ }\textbf {\bibinfo {volume} {99}},\ \bibinfo
  {pages} {043308} (\bibinfo {year} {2019})}\BibitemShut {NoStop}%
\bibitem [{\citenamefont {Kharkov}\ \emph {et~al.}(2020)\citenamefont
  {Kharkov}, \citenamefont {Sotskov}, \citenamefont {Karazeev}, \citenamefont
  {Kiktenko},\ and\ \citenamefont {Fedorov}}]{kharkov:2020}%
  \BibitemOpen
  \bibfield  {author} {\bibinfo {author} {\bibfnamefont {Y.~A.}\ \bibnamefont
  {Kharkov}}, \bibinfo {author} {\bibfnamefont {V.~E.}\ \bibnamefont
  {Sotskov}}, \bibinfo {author} {\bibfnamefont {A.~A.}\ \bibnamefont
  {Karazeev}}, \bibinfo {author} {\bibfnamefont {E.~O.}\ \bibnamefont
  {Kiktenko}},\ and\ \bibinfo {author} {\bibfnamefont {A.~K.}\ \bibnamefont
  {Fedorov}},\ }\bibfield  {title} {\bibinfo {title} {Revealing quantum chaos
  with machine learning},\ }\href {https://doi.org/10.1103/PhysRevB.101.064406}
  {\bibfield  {journal} {\bibinfo  {journal} {Phys. Rev. B}\ }\textbf {\bibinfo
  {volume} {101}},\ \bibinfo {pages} {064406} (\bibinfo {year}
  {2020})}\BibitemShut {NoStop}%
\bibitem [{\citenamefont {Guo}\ \emph {et~al.}(2020)\citenamefont {Guo},
  \citenamefont {Ai},\ and\ \citenamefont {He}}]{guo:2020}%
  \BibitemOpen
  \bibfield  {author} {\bibinfo {author} {\bibfnamefont {W.}~\bibnamefont
  {Guo}}, \bibinfo {author} {\bibfnamefont {B.}~\bibnamefont {Ai}},\ and\
  \bibinfo {author} {\bibfnamefont {L.}~\bibnamefont {He}},\ }\bibfield
  {title} {\bibinfo {title} {Reveal flocking of birds flying in fog by machine
  learning},\ }\href {https://arxiv.org/abs/2005.10505} {\bibfield  {journal}
  {\bibinfo  {journal} {arXiv:2005.10505}\ } (\bibinfo {year}
  {2020})}\BibitemShut {NoStop}%
\bibitem [{\citenamefont {Greplova}\ \emph {et~al.}(2020)\citenamefont
  {Greplova}, \citenamefont {Valenti}, \citenamefont {Boschung}, \citenamefont
  {Schäfer}, \citenamefont {Lörch},\ and\ \citenamefont
  {Huber}}]{greplova:2020}%
  \BibitemOpen
  \bibfield  {author} {\bibinfo {author} {\bibfnamefont {E.}~\bibnamefont
  {Greplova}}, \bibinfo {author} {\bibfnamefont {A.}~\bibnamefont {Valenti}},
  \bibinfo {author} {\bibfnamefont {G.}~\bibnamefont {Boschung}}, \bibinfo
  {author} {\bibfnamefont {F.}~\bibnamefont {Schäfer}}, \bibinfo {author}
  {\bibfnamefont {N.}~\bibnamefont {Lörch}},\ and\ \bibinfo {author}
  {\bibfnamefont {S.~D.}\ \bibnamefont {Huber}},\ }\bibfield  {title} {\bibinfo
  {title} {Unsupervised identification of topological phase transitions using
  predictive models},\ }\href {https://doi.org/10.1088/1367-2630/ab7771}
  {\bibfield  {journal} {\bibinfo  {journal} {New J. Phys.}\ }\textbf {\bibinfo
  {volume} {22}},\ \bibinfo {pages} {045003} (\bibinfo {year}
  {2020})}\BibitemShut {NoStop}%
\bibitem [{\citenamefont {Arnold}\ \emph {et~al.}(2021)\citenamefont {Arnold},
  \citenamefont {Sch\"afer}, \citenamefont {\ifmmode~\check{Z}\else
  \v{Z}\fi{}onda},\ and\ \citenamefont {Lode}}]{arnold:2021}%
  \BibitemOpen
  \bibfield  {author} {\bibinfo {author} {\bibfnamefont {J.}~\bibnamefont
  {Arnold}}, \bibinfo {author} {\bibfnamefont {F.}~\bibnamefont {Sch\"afer}},
  \bibinfo {author} {\bibfnamefont {M.}~\bibnamefont {\ifmmode~\check{Z}\else
  \v{Z}\fi{}onda}},\ and\ \bibinfo {author} {\bibfnamefont {A.~U.~J.}\
  \bibnamefont {Lode}},\ }\bibfield  {title} {\bibinfo {title} {Interpretable
  and unsupervised phase classification},\ }\href
  {https://doi.org/10.1103/PhysRevResearch.3.033052} {\bibfield  {journal}
  {\bibinfo  {journal} {Phys. Rev. Res.}\ }\textbf {\bibinfo {volume} {3}},\
  \bibinfo {pages} {033052} (\bibinfo {year} {2021})}\BibitemShut {NoStop}%
\bibitem [{\citenamefont {Gavreev}\ \emph {et~al.}(2022)\citenamefont
  {Gavreev}, \citenamefont {Mastiukova}, \citenamefont {Kiktenko},\ and\
  \citenamefont {Fedorov}}]{gavreev:2022}%
  \BibitemOpen
  \bibfield  {author} {\bibinfo {author} {\bibfnamefont {M.~A.}\ \bibnamefont
  {Gavreev}}, \bibinfo {author} {\bibfnamefont {A.~S.}\ \bibnamefont
  {Mastiukova}}, \bibinfo {author} {\bibfnamefont {E.~O.}\ \bibnamefont
  {Kiktenko}},\ and\ \bibinfo {author} {\bibfnamefont {A.~K.}\ \bibnamefont
  {Fedorov}},\ }\bibfield  {title} {\bibinfo {title} {Learning entanglement
  breakdown as a phase transition by confusion},\ }\href
  {https://doi.org/10.1088/1367-2630/ac7fb2} {\bibfield  {journal} {\bibinfo
  {journal} {New J. Phys.}\ }\textbf {\bibinfo {volume} {24}},\ \bibinfo
  {pages} {073045} (\bibinfo {year} {2022})}\BibitemShut {NoStop}%
\bibitem [{\citenamefont {Zvyagintseva}\ \emph {et~al.}(2022)\citenamefont
  {Zvyagintseva}, \citenamefont {Sigurdsson}, \citenamefont {Kozin},
  \citenamefont {Iorsh}, \citenamefont {Shelykh}, \citenamefont {Ulyantsev},\
  and\ \citenamefont {Kyriienko}}]{zvyagintseva:2022}%
  \BibitemOpen
  \bibfield  {author} {\bibinfo {author} {\bibfnamefont {D.}~\bibnamefont
  {Zvyagintseva}}, \bibinfo {author} {\bibfnamefont {H.}~\bibnamefont
  {Sigurdsson}}, \bibinfo {author} {\bibfnamefont {V.~K.}\ \bibnamefont
  {Kozin}}, \bibinfo {author} {\bibfnamefont {I.}~\bibnamefont {Iorsh}},
  \bibinfo {author} {\bibfnamefont {I.~A.}\ \bibnamefont {Shelykh}}, \bibinfo
  {author} {\bibfnamefont {V.}~\bibnamefont {Ulyantsev}},\ and\ \bibinfo
  {author} {\bibfnamefont {O.}~\bibnamefont {Kyriienko}},\ }\bibfield  {title}
  {\bibinfo {title} {Machine learning of phase transitions in nonlinear
  polariton lattices},\ }\href {https://doi.org/10.1038/s42005-021-00755-5}
  {\bibfield  {journal} {\bibinfo  {journal} {Commun. Phys.}\ }\textbf
  {\bibinfo {volume} {5}},\ \bibinfo {pages} {8} (\bibinfo {year}
  {2022})}\BibitemShut {NoStop}%
\bibitem [{\citenamefont {Tibaldi}\ \emph {et~al.}(2023)\citenamefont
  {Tibaldi}, \citenamefont {Magnifico}, \citenamefont {Vodola},\ and\
  \citenamefont {Ercolessi}}]{tibaldi:2023}%
  \BibitemOpen
  \bibfield  {author} {\bibinfo {author} {\bibfnamefont {S.}~\bibnamefont
  {Tibaldi}}, \bibinfo {author} {\bibfnamefont {G.}~\bibnamefont {Magnifico}},
  \bibinfo {author} {\bibfnamefont {D.}~\bibnamefont {Vodola}},\ and\ \bibinfo
  {author} {\bibfnamefont {E.}~\bibnamefont {Ercolessi}},\ }\bibfield  {title}
  {\bibinfo {title} {{Unsupervised and supervised learning of interacting
  topological phases from single-particle correlation functions}},\ }\href
  {https://doi.org/10.21468/SciPostPhys.14.1.005} {\bibfield  {journal}
  {\bibinfo  {journal} {SciPost Phys.}\ }\textbf {\bibinfo {volume} {14}},\
  \bibinfo {pages} {005} (\bibinfo {year} {2023})}\BibitemShut {NoStop}%
\bibitem [{\citenamefont {Guo}\ and\ \citenamefont {He}(2023)}]{guo:2023}%
  \BibitemOpen
  \bibfield  {author} {\bibinfo {author} {\bibfnamefont {W.}~\bibnamefont
  {Guo}}\ and\ \bibinfo {author} {\bibfnamefont {L.}~\bibnamefont {He}},\
  }\bibfield  {title} {\bibinfo {title} {Learning phase transitions from
  regression uncertainty: a new regression-based machine learning approach for
  automated detection of phases of matter},\ }\href
  {https://doi.org/10.1088/1367-2630/acef4e} {\bibfield  {journal} {\bibinfo
  {journal} {New J. Phys.}\ }\textbf {\bibinfo {volume} {25}},\ \bibinfo
  {pages} {083037} (\bibinfo {year} {2023})}\BibitemShut {NoStop}%
\bibitem [{\citenamefont {Schlömer}\ and\ \citenamefont
  {Bohrdt}(2023)}]{schlomer:2023}%
  \BibitemOpen
  \bibfield  {author} {\bibinfo {author} {\bibfnamefont {H.}~\bibnamefont
  {Schlömer}}\ and\ \bibinfo {author} {\bibfnamefont {A.}~\bibnamefont
  {Bohrdt}},\ }\bibfield  {title} {\bibinfo {title} {{Fluctuation based
  interpretable analysis scheme for quantum many-body snapshots}},\ }\href
  {https://doi.org/10.21468/SciPostPhys.15.3.099} {\bibfield  {journal}
  {\bibinfo  {journal} {SciPost Phys.}\ }\textbf {\bibinfo {volume} {15}},\
  \bibinfo {pages} {099} (\bibinfo {year} {2023})}\BibitemShut {NoStop}%
\bibitem [{\citenamefont {Rem}\ \emph {et~al.}(2019)\citenamefont {Rem},
  \citenamefont {K{\"a}ming}, \citenamefont {Tarnowski}, \citenamefont
  {Asteria}, \citenamefont {Fl{\"a}schner}, \citenamefont {Becker},
  \citenamefont {Sengstock},\ and\ \citenamefont {Weitenberg}}]{rem:2019}%
  \BibitemOpen
  \bibfield  {author} {\bibinfo {author} {\bibfnamefont {B.~S.}\ \bibnamefont
  {Rem}}, \bibinfo {author} {\bibfnamefont {N.}~\bibnamefont {K{\"a}ming}},
  \bibinfo {author} {\bibfnamefont {M.}~\bibnamefont {Tarnowski}}, \bibinfo
  {author} {\bibfnamefont {L.}~\bibnamefont {Asteria}}, \bibinfo {author}
  {\bibfnamefont {N.}~\bibnamefont {Fl{\"a}schner}}, \bibinfo {author}
  {\bibfnamefont {C.}~\bibnamefont {Becker}}, \bibinfo {author} {\bibfnamefont
  {K.}~\bibnamefont {Sengstock}},\ and\ \bibinfo {author} {\bibfnamefont
  {C.}~\bibnamefont {Weitenberg}},\ }\bibfield  {title} {\bibinfo {title}
  {Identifying quantum phase transitions using artificial neural networks on
  experimental data},\ }\href {https://doi.org/10.1038/s41567-019-0554-0}
  {\bibfield  {journal} {\bibinfo  {journal} {Nat. Phys.}\ }\textbf {\bibinfo
  {volume} {15}},\ \bibinfo {pages} {917} (\bibinfo {year} {2019})}\BibitemShut
  {NoStop}%
\bibitem [{\citenamefont {Bohrdt}\ \emph {et~al.}(2021)\citenamefont {Bohrdt},
  \citenamefont {Kim}, \citenamefont {Lukin}, \citenamefont {Rispoli},
  \citenamefont {Schittko}, \citenamefont {Knap}, \citenamefont {Greiner},\
  and\ \citenamefont {L\'eonard}}]{bohrdt:2021}%
  \BibitemOpen
  \bibfield  {author} {\bibinfo {author} {\bibfnamefont {A.}~\bibnamefont
  {Bohrdt}}, \bibinfo {author} {\bibfnamefont {S.}~\bibnamefont {Kim}},
  \bibinfo {author} {\bibfnamefont {A.}~\bibnamefont {Lukin}}, \bibinfo
  {author} {\bibfnamefont {M.}~\bibnamefont {Rispoli}}, \bibinfo {author}
  {\bibfnamefont {R.}~\bibnamefont {Schittko}}, \bibinfo {author}
  {\bibfnamefont {M.}~\bibnamefont {Knap}}, \bibinfo {author} {\bibfnamefont
  {M.}~\bibnamefont {Greiner}},\ and\ \bibinfo {author} {\bibfnamefont
  {J.}~\bibnamefont {L\'eonard}},\ }\bibfield  {title} {\bibinfo {title}
  {Analyzing {N}onequilibrium {Q}uantum {S}tates through {S}napshots with
  {A}rtificial {N}eural {N}etworks},\ }\href
  {https://doi.org/10.1103/PhysRevLett.127.150504} {\bibfield  {journal}
  {\bibinfo  {journal} {Phys. Rev. Lett.}\ }\textbf {\bibinfo {volume} {127}},\
  \bibinfo {pages} {150504} (\bibinfo {year} {2021})}\BibitemShut {NoStop}%
\bibitem [{\citenamefont {Miles}\ \emph {et~al.}(2023)\citenamefont {Miles},
  \citenamefont {Samajdar}, \citenamefont {Ebadi}, \citenamefont {Wang},
  \citenamefont {Pichler}, \citenamefont {Sachdev}, \citenamefont {Lukin},
  \citenamefont {Greiner}, \citenamefont {Weinberger},\ and\ \citenamefont
  {Kim}}]{miles:2023}%
  \BibitemOpen
  \bibfield  {author} {\bibinfo {author} {\bibfnamefont {C.}~\bibnamefont
  {Miles}}, \bibinfo {author} {\bibfnamefont {R.}~\bibnamefont {Samajdar}},
  \bibinfo {author} {\bibfnamefont {S.}~\bibnamefont {Ebadi}}, \bibinfo
  {author} {\bibfnamefont {T.~T.}\ \bibnamefont {Wang}}, \bibinfo {author}
  {\bibfnamefont {H.}~\bibnamefont {Pichler}}, \bibinfo {author} {\bibfnamefont
  {S.}~\bibnamefont {Sachdev}}, \bibinfo {author} {\bibfnamefont {M.~D.}\
  \bibnamefont {Lukin}}, \bibinfo {author} {\bibfnamefont {M.}~\bibnamefont
  {Greiner}}, \bibinfo {author} {\bibfnamefont {K.~Q.}\ \bibnamefont
  {Weinberger}},\ and\ \bibinfo {author} {\bibfnamefont {E.-A.}\ \bibnamefont
  {Kim}},\ }\bibfield  {title} {\bibinfo {title} {Machine learning discovery of
  new phases in programmable quantum simulator snapshots},\ }\href
  {https://doi.org/10.1103/PhysRevResearch.5.013026} {\bibfield  {journal}
  {\bibinfo  {journal} {Phys. Rev. Res.}\ }\textbf {\bibinfo {volume} {5}},\
  \bibinfo {pages} {013026} (\bibinfo {year} {2023})}\BibitemShut {NoStop}%
\bibitem [{sup()}]{suppl}%
  \BibitemOpen
  \href@noop {} {}\bibinfo {note} {See Supplemental Material, which includes
  Refs.~\cite{van:2017,liu:2018,beach:2018,suchsland:2018,lee:2019,kharkov:2020,guo:2020,greplova:2020,bohrdt:2021,richter:2022,gavreev:2022,zvyagintseva:2022,guo:2023,schlomer:2023,arnold:2022,schaefer:2019,greplova:2020,arnold:2021,paszke:2019,kass:1988,kingma:2014,baydin:2018,innes:2018,casella:2002,itensor},
  for (1) a discussion on the modifications of the phase-classification
  methods; (2) a comparison between the discriminative and generative approach;
  (3) a derivation of indicator signals for classical equilibrium systems; and
  (4) details on the data generation process for the anisotropic Ising model
  and cluster-Ising model.}\BibitemShut {Stop}%
\bibitem [{\citenamefont {Onsager}(1944)}]{onsager:1944}%
  \BibitemOpen
  \bibfield  {author} {\bibinfo {author} {\bibfnamefont {L.}~\bibnamefont
  {Onsager}},\ }\bibfield  {title} {\bibinfo {title} {Crystal {S}tatistics. i.
  {A} {T}wo-{D}imensional {M}odel with an {O}rder-{D}isorder {T}ransition},\
  }\href {https://doi.org/10.1103/PhysRev.65.117} {\bibfield  {journal}
  {\bibinfo  {journal} {Phys. Rev.}\ }\textbf {\bibinfo {volume} {65}},\
  \bibinfo {pages} {117} (\bibinfo {year} {1944})}\BibitemShut {NoStop}%
\bibitem [{\citenamefont {Arnold}\ and\ \citenamefont
  {Sch\"afer}(2022)}]{arnold:2022}%
  \BibitemOpen
  \bibfield  {author} {\bibinfo {author} {\bibfnamefont {J.}~\bibnamefont
  {Arnold}}\ and\ \bibinfo {author} {\bibfnamefont {F.}~\bibnamefont
  {Sch\"afer}},\ }\bibfield  {title} {\bibinfo {title} {Replacing neural
  networks by optimal analytical predictors for the detection of phase
  transitions},\ }\href {https://doi.org/10.1103/PhysRevX.12.031044} {\bibfield
   {journal} {\bibinfo  {journal} {Phys. Rev. X}\ }\textbf {\bibinfo {volume}
  {12}},\ \bibinfo {pages} {031044} (\bibinfo {year} {2022})}\BibitemShut
  {NoStop}%
\bibitem [{\citenamefont {Goodfellow}(2016)}]{goodfellow2:2016}%
  \BibitemOpen
  \bibfield  {author} {\bibinfo {author} {\bibfnamefont {I.}~\bibnamefont
  {Goodfellow}},\ }\bibfield  {title} {\bibinfo {title} {{NIPS 2016 Tutorial:
  Generative Adversarial Networks}},\ }\href {https://arxiv.org/abs/1701.00160}
  {\bibfield  {journal} {\bibinfo  {journal} {arXiv:1701.00160}\ } (\bibinfo
  {year} {2016})},\ \bibinfo {note} {pp. 9 -- 17}\BibitemShut {NoStop}%
\bibitem [{\citenamefont {Devroye}\ \emph {et~al.}(1996)\citenamefont
  {Devroye}, \citenamefont {Gy{\"o}rfi},\ and\ \citenamefont
  {Lugosi}}]{devroye2:1996}%
  \BibitemOpen
  \bibfield  {author} {\bibinfo {author} {\bibfnamefont {L.}~\bibnamefont
  {Devroye}}, \bibinfo {author} {\bibfnamefont {L.}~\bibnamefont
  {Gy{\"o}rfi}},\ and\ \bibinfo {author} {\bibfnamefont {G.}~\bibnamefont
  {Lugosi}},\ }\bibinfo {title} {{The Bayes Error}},\ in\ \href
  {https://doi.org/10.1007/978-1-4612-0711-5_2} {\emph {\bibinfo {booktitle}
  {{A Probabilistic Theory of Pattern Recognition}}}}\ (\bibinfo  {publisher}
  {Springer},\ \bibinfo {address} {New York, NY},\ \bibinfo {year} {1996})\
  pp.\ \bibinfo {pages} {9--20}\BibitemShut {NoStop}%
\bibitem [{\citenamefont {Goodfellow}\ \emph {et~al.}(2016)\citenamefont
  {Goodfellow}, \citenamefont {Bengio},\ and\ \citenamefont
  {Courville}}]{goodfellow:2016}%
  \BibitemOpen
  \bibfield  {author} {\bibinfo {author} {\bibfnamefont {I.}~\bibnamefont
  {Goodfellow}}, \bibinfo {author} {\bibfnamefont {Y.}~\bibnamefont {Bengio}},\
  and\ \bibinfo {author} {\bibfnamefont {A.}~\bibnamefont {Courville}},\ }\href
  {http://www.deeplearningbook.org} {\emph {\bibinfo {title} {Deep
  {L}earning}}}\ (\bibinfo  {publisher} {MIT Press},\ \bibinfo {year}
  {2016})\BibitemShut {NoStop}%
\bibitem [{non()}]{non_optimal_compression}%
  \BibitemOpen
  \href@noop {} {}\bibinfo {note} {While the sufficient statistic requires full
  knowledge of the underlying Hamiltonian, knowledge of the symmetries alone
  may also be used to perform a lossless compression of the state space,
  see~\cite{suppl} for an example.}\BibitemShut {Stop}%
\bibitem [{\citenamefont {Sharir}\ \emph {et~al.}(2020)\citenamefont {Sharir},
  \citenamefont {Levine}, \citenamefont {Wies}, \citenamefont {Carleo},\ and\
  \citenamefont {Shashua}}]{sharir:2020}%
  \BibitemOpen
  \bibfield  {author} {\bibinfo {author} {\bibfnamefont {O.}~\bibnamefont
  {Sharir}}, \bibinfo {author} {\bibfnamefont {Y.}~\bibnamefont {Levine}},
  \bibinfo {author} {\bibfnamefont {N.}~\bibnamefont {Wies}}, \bibinfo {author}
  {\bibfnamefont {G.}~\bibnamefont {Carleo}},\ and\ \bibinfo {author}
  {\bibfnamefont {A.}~\bibnamefont {Shashua}},\ }\bibfield  {title} {\bibinfo
  {title} {{Deep Autoregressive Models for the Efficient Variational Simulation
  of Many-Body Quantum Systems}},\ }\href
  {https://doi.org/10.1103/PhysRevLett.124.020503} {\bibfield  {journal}
  {\bibinfo  {journal} {Phys. Rev. Lett.}\ }\textbf {\bibinfo {volume} {124}},\
  \bibinfo {pages} {020503} (\bibinfo {year} {2020})}\BibitemShut {NoStop}%
\bibitem [{\citenamefont {Smacchia}\ \emph {et~al.}(2011)\citenamefont
  {Smacchia}, \citenamefont {Amico}, \citenamefont {Facchi}, \citenamefont
  {Fazio}, \citenamefont {Florio}, \citenamefont {Pascazio},\ and\
  \citenamefont {Vedral}}]{smacchia:2011}%
  \BibitemOpen
  \bibfield  {author} {\bibinfo {author} {\bibfnamefont {P.}~\bibnamefont
  {Smacchia}}, \bibinfo {author} {\bibfnamefont {L.}~\bibnamefont {Amico}},
  \bibinfo {author} {\bibfnamefont {P.}~\bibnamefont {Facchi}}, \bibinfo
  {author} {\bibfnamefont {R.}~\bibnamefont {Fazio}}, \bibinfo {author}
  {\bibfnamefont {G.}~\bibnamefont {Florio}}, \bibinfo {author} {\bibfnamefont
  {S.}~\bibnamefont {Pascazio}},\ and\ \bibinfo {author} {\bibfnamefont
  {V.}~\bibnamefont {Vedral}},\ }\bibfield  {title} {\bibinfo {title}
  {{Statistical mechanics of the cluster Ising model}},\ }\href
  {https://doi.org/10.1103/PhysRevA.84.022304} {\bibfield  {journal} {\bibinfo
  {journal} {Phys. Rev. A}\ }\textbf {\bibinfo {volume} {84}},\ \bibinfo
  {pages} {022304} (\bibinfo {year} {2011})}\BibitemShut {NoStop}%
\bibitem [{\citenamefont {Verresen}\ \emph {et~al.}(2017)\citenamefont
  {Verresen}, \citenamefont {Moessner},\ and\ \citenamefont
  {Pollmann}}]{verresen:2017}%
  \BibitemOpen
  \bibfield  {author} {\bibinfo {author} {\bibfnamefont {R.}~\bibnamefont
  {Verresen}}, \bibinfo {author} {\bibfnamefont {R.}~\bibnamefont {Moessner}},\
  and\ \bibinfo {author} {\bibfnamefont {F.}~\bibnamefont {Pollmann}},\
  }\bibfield  {title} {\bibinfo {title} {One-dimensional symmetry protected
  topological phases and their transitions},\ }\href
  {https://doi.org/10.1103/PhysRevB.96.165124} {\bibfield  {journal} {\bibinfo
  {journal} {Phys. Rev. B}\ }\textbf {\bibinfo {volume} {96}},\ \bibinfo
  {pages} {165124} (\bibinfo {year} {2017})}\BibitemShut {NoStop}%
\bibitem [{\citenamefont {Briegel}\ and\ \citenamefont
  {Raussendorf}(2001)}]{briegel:2001}%
  \BibitemOpen
  \bibfield  {author} {\bibinfo {author} {\bibfnamefont {H.~J.}\ \bibnamefont
  {Briegel}}\ and\ \bibinfo {author} {\bibfnamefont {R.}~\bibnamefont
  {Raussendorf}},\ }\bibfield  {title} {\bibinfo {title} {Persistent
  entanglement in arrays of interacting particles},\ }\href
  {https://doi.org/10.1103/PhysRevLett.86.910} {\bibfield  {journal} {\bibinfo
  {journal} {Phys. Rev. Lett.}\ }\textbf {\bibinfo {volume} {86}},\ \bibinfo
  {pages} {910} (\bibinfo {year} {2001})}\BibitemShut {NoStop}%
\bibitem [{\citenamefont {Azses}\ \emph {et~al.}(2020)\citenamefont {Azses},
  \citenamefont {Haenel}, \citenamefont {Naveh}, \citenamefont {Raussendorf},
  \citenamefont {Sela},\ and\ \citenamefont {Dalla~Torre}}]{azses:2020}%
  \BibitemOpen
  \bibfield  {author} {\bibinfo {author} {\bibfnamefont {D.}~\bibnamefont
  {Azses}}, \bibinfo {author} {\bibfnamefont {R.}~\bibnamefont {Haenel}},
  \bibinfo {author} {\bibfnamefont {Y.}~\bibnamefont {Naveh}}, \bibinfo
  {author} {\bibfnamefont {R.}~\bibnamefont {Raussendorf}}, \bibinfo {author}
  {\bibfnamefont {E.}~\bibnamefont {Sela}},\ and\ \bibinfo {author}
  {\bibfnamefont {E.~G.}\ \bibnamefont {Dalla~Torre}},\ }\bibfield  {title}
  {\bibinfo {title} {Identification of symmetry-protected topological states on
  noisy quantum computers},\ }\href
  {https://doi.org/10.1103/PhysRevLett.125.120502} {\bibfield  {journal}
  {\bibinfo  {journal} {Phys. Rev. Lett.}\ }\textbf {\bibinfo {volume} {125}},\
  \bibinfo {pages} {120502} (\bibinfo {year} {2020})}\BibitemShut {NoStop}%
\bibitem [{\citenamefont {Herrmann}\ \emph {et~al.}(2022)\citenamefont
  {Herrmann}, \citenamefont {Llima}, \citenamefont {Remm}, \citenamefont
  {Zapletal}, \citenamefont {McMahon}, \citenamefont {Scarato}, \citenamefont
  {Swiadek}, \citenamefont {Andersen}, \citenamefont {Hellings}, \citenamefont
  {Krinner} \emph {et~al.}}]{herrmann:2022}%
  \BibitemOpen
  \bibfield  {author} {\bibinfo {author} {\bibfnamefont {J.}~\bibnamefont
  {Herrmann}}, \bibinfo {author} {\bibfnamefont {S.~M.}\ \bibnamefont {Llima}},
  \bibinfo {author} {\bibfnamefont {A.}~\bibnamefont {Remm}}, \bibinfo {author}
  {\bibfnamefont {P.}~\bibnamefont {Zapletal}}, \bibinfo {author}
  {\bibfnamefont {N.~A.}\ \bibnamefont {McMahon}}, \bibinfo {author}
  {\bibfnamefont {C.}~\bibnamefont {Scarato}}, \bibinfo {author} {\bibfnamefont
  {F.}~\bibnamefont {Swiadek}}, \bibinfo {author} {\bibfnamefont {C.~K.}\
  \bibnamefont {Andersen}}, \bibinfo {author} {\bibfnamefont {C.}~\bibnamefont
  {Hellings}}, \bibinfo {author} {\bibfnamefont {S.}~\bibnamefont {Krinner}},
  \emph {et~al.},\ }\bibfield  {title} {\bibinfo {title} {Realizing quantum
  convolutional neural networks on a superconducting quantum processor to
  recognize quantum phases},\ }\href
  {https://doi.org/10.1038/s41467-022-31679-5} {\bibfield  {journal} {\bibinfo
  {journal} {Nat. Commun.}\ }\textbf {\bibinfo {volume} {13}},\ \bibinfo
  {pages} {1} (\bibinfo {year} {2022})}\BibitemShut {NoStop}%
\bibitem [{\citenamefont {Smith}\ \emph {et~al.}(2022)\citenamefont {Smith},
  \citenamefont {Jobst}, \citenamefont {Green},\ and\ \citenamefont
  {Pollmann}}]{smith:2022}%
  \BibitemOpen
  \bibfield  {author} {\bibinfo {author} {\bibfnamefont {A.}~\bibnamefont
  {Smith}}, \bibinfo {author} {\bibfnamefont {B.}~\bibnamefont {Jobst}},
  \bibinfo {author} {\bibfnamefont {A.~G.}\ \bibnamefont {Green}},\ and\
  \bibinfo {author} {\bibfnamefont {F.}~\bibnamefont {Pollmann}},\ }\bibfield
  {title} {\bibinfo {title} {Crossing a topological phase transition with a
  quantum computer},\ }\href
  {https://doi.org/10.1103/PhysRevResearch.4.L022020} {\bibfield  {journal}
  {\bibinfo  {journal} {Phys. Rev. Res.}\ }\textbf {\bibinfo {volume} {4}},\
  \bibinfo {pages} {L022020} (\bibinfo {year} {2022})}\BibitemShut {NoStop}%
\bibitem [{\citenamefont {Zapletal}\ \emph {et~al.}(2023)\citenamefont
  {Zapletal}, \citenamefont {McMahon},\ and\ \citenamefont
  {Hartmann}}]{zapletal:2023}%
  \BibitemOpen
  \bibfield  {author} {\bibinfo {author} {\bibfnamefont {P.}~\bibnamefont
  {Zapletal}}, \bibinfo {author} {\bibfnamefont {N.~A.}\ \bibnamefont
  {McMahon}},\ and\ \bibinfo {author} {\bibfnamefont {M.~J.}\ \bibnamefont
  {Hartmann}},\ }\bibfield  {title} {\bibinfo {title} {Error-tolerant quantum
  convolutional neural networks for symmetry-protected topological phases},\
  }\href {https://arxiv.org/abs/2307.03711} {\bibfield  {journal} {\bibinfo
  {journal} {arXiv:2307.03711}\ } (\bibinfo {year} {2023})}\BibitemShut
  {NoStop}%
\bibitem [{\citenamefont {Sadoune}\ \emph {et~al.}(2023)\citenamefont
  {Sadoune}, \citenamefont {Giudici}, \citenamefont {Liu},\ and\ \citenamefont
  {Pollet}}]{sadoune:2023}%
  \BibitemOpen
  \bibfield  {author} {\bibinfo {author} {\bibfnamefont {N.}~\bibnamefont
  {Sadoune}}, \bibinfo {author} {\bibfnamefont {G.}~\bibnamefont {Giudici}},
  \bibinfo {author} {\bibfnamefont {K.}~\bibnamefont {Liu}},\ and\ \bibinfo
  {author} {\bibfnamefont {L.}~\bibnamefont {Pollet}},\ }\bibfield  {title}
  {\bibinfo {title} {Unsupervised interpretable learning of phases from
  many-qubit systems},\ }\href
  {https://doi.org/10.1103/PhysRevResearch.5.013082} {\bibfield  {journal}
  {\bibinfo  {journal} {Phys. Rev. Res.}\ }\textbf {\bibinfo {volume} {5}},\
  \bibinfo {pages} {013082} (\bibinfo {year} {2023})}\BibitemShut {NoStop}%
\bibitem [{\citenamefont {Luo}\ \emph {et~al.}(2022)\citenamefont {Luo},
  \citenamefont {Chen}, \citenamefont {Carrasquilla},\ and\ \citenamefont
  {Clark}}]{luo:2022}%
  \BibitemOpen
  \bibfield  {author} {\bibinfo {author} {\bibfnamefont {D.}~\bibnamefont
  {Luo}}, \bibinfo {author} {\bibfnamefont {Z.}~\bibnamefont {Chen}}, \bibinfo
  {author} {\bibfnamefont {J.}~\bibnamefont {Carrasquilla}},\ and\ \bibinfo
  {author} {\bibfnamefont {B.~K.}\ \bibnamefont {Clark}},\ }\bibfield  {title}
  {\bibinfo {title} {{Autoregressive Neural Network for Simulating Open Quantum
  Systems via a Probabilistic Formulation}},\ }\href
  {https://doi.org/10.1103/PhysRevLett.128.090501} {\bibfield  {journal}
  {\bibinfo  {journal} {Phys. Rev. Lett.}\ }\textbf {\bibinfo {volume} {128}},\
  \bibinfo {pages} {090501} (\bibinfo {year} {2022})}\BibitemShut {NoStop}%
\bibitem [{\citenamefont {Reh}\ \emph {et~al.}(2021)\citenamefont {Reh},
  \citenamefont {Schmitt},\ and\ \citenamefont {G\"arttner}}]{reh:2021}%
  \BibitemOpen
  \bibfield  {author} {\bibinfo {author} {\bibfnamefont {M.}~\bibnamefont
  {Reh}}, \bibinfo {author} {\bibfnamefont {M.}~\bibnamefont {Schmitt}},\ and\
  \bibinfo {author} {\bibfnamefont {M.}~\bibnamefont {G\"arttner}},\ }\bibfield
   {title} {\bibinfo {title} {{Time-Dependent Variational Principle for Open
  Quantum Systems with Artificial Neural Networks}},\ }\href
  {https://doi.org/10.1103/PhysRevLett.127.230501} {\bibfield  {journal}
  {\bibinfo  {journal} {Phys. Rev. Lett.}\ }\textbf {\bibinfo {volume} {127}},\
  \bibinfo {pages} {230501} (\bibinfo {year} {2021})}\BibitemShut {NoStop}%
\bibitem [{\citenamefont {Carrasquilla}\ \emph {et~al.}(2019)\citenamefont
  {Carrasquilla}, \citenamefont {Torlai}, \citenamefont {Melko},\ and\
  \citenamefont {Aolita}}]{carrasquilla:2019}%
  \BibitemOpen
  \bibfield  {author} {\bibinfo {author} {\bibfnamefont {J.}~\bibnamefont
  {Carrasquilla}}, \bibinfo {author} {\bibfnamefont {G.}~\bibnamefont
  {Torlai}}, \bibinfo {author} {\bibfnamefont {R.~G.}\ \bibnamefont {Melko}},\
  and\ \bibinfo {author} {\bibfnamefont {L.}~\bibnamefont {Aolita}},\
  }\bibfield  {title} {\bibinfo {title} {Reconstructing quantum states with
  generative models},\ }\href {https://doi.org/10.1038/s42256-019-0028-1}
  {\bibfield  {journal} {\bibinfo  {journal} {Nat. Mach. Intell.}\ }\textbf
  {\bibinfo {volume} {1}},\ \bibinfo {pages} {155} (\bibinfo {year}
  {2019})}\BibitemShut {NoStop}%
\bibitem [{\citenamefont {Gomez}\ \emph {et~al.}(2022)\citenamefont {Gomez},
  \citenamefont {Yelin},\ and\ \citenamefont {Najafi}}]{gomez:2022}%
  \BibitemOpen
  \bibfield  {author} {\bibinfo {author} {\bibfnamefont {A.~M.}\ \bibnamefont
  {Gomez}}, \bibinfo {author} {\bibfnamefont {S.~F.}\ \bibnamefont {Yelin}},\
  and\ \bibinfo {author} {\bibfnamefont {K.}~\bibnamefont {Najafi}},\
  }\bibfield  {title} {\bibinfo {title} {{Reconstructing Quantum States Using
  Basis-Enhanced Born Machines}},\ }\href {https://arxiv.org/abs/2206.01273}
  {\bibfield  {journal} {\bibinfo  {journal} {arXiv:2206.01273}\ } (\bibinfo
  {year} {2022})}\BibitemShut {NoStop}%
\bibitem [{\citenamefont {Bohrdt}\ \emph {et~al.}(2019)\citenamefont {Bohrdt},
  \citenamefont {Chiu}, \citenamefont {Ji}, \citenamefont {Xu}, \citenamefont
  {Greif}, \citenamefont {Greiner}, \citenamefont {Demler}, \citenamefont
  {Grusdt},\ and\ \citenamefont {Knap}}]{bohrdt:2019}%
  \BibitemOpen
  \bibfield  {author} {\bibinfo {author} {\bibfnamefont {A.}~\bibnamefont
  {Bohrdt}}, \bibinfo {author} {\bibfnamefont {C.~S.}\ \bibnamefont {Chiu}},
  \bibinfo {author} {\bibfnamefont {G.}~\bibnamefont {Ji}}, \bibinfo {author}
  {\bibfnamefont {M.}~\bibnamefont {Xu}}, \bibinfo {author} {\bibfnamefont
  {D.}~\bibnamefont {Greif}}, \bibinfo {author} {\bibfnamefont
  {M.}~\bibnamefont {Greiner}}, \bibinfo {author} {\bibfnamefont
  {E.}~\bibnamefont {Demler}}, \bibinfo {author} {\bibfnamefont
  {F.}~\bibnamefont {Grusdt}},\ and\ \bibinfo {author} {\bibfnamefont
  {M.}~\bibnamefont {Knap}},\ }\bibfield  {title} {\bibinfo {title}
  {Classifying snapshots of the doped {H}ubbard model with machine learning},\
  }\href {https://doi.org/10.1038/s41567-019-0565-x} {\bibfield  {journal}
  {\bibinfo  {journal} {Nat. Phys.}\ }\textbf {\bibinfo {volume} {15}},\
  \bibinfo {pages} {921} (\bibinfo {year} {2019})}\BibitemShut {NoStop}%
\bibitem [{\citenamefont {Zhang}\ \emph {et~al.}(2019)\citenamefont {Zhang},
  \citenamefont {Mesaros}, \citenamefont {Fujita}, \citenamefont {Edkins},
  \citenamefont {Hamidian}, \citenamefont {Ch’ng}, \citenamefont {Eisaki},
  \citenamefont {Uchida}, \citenamefont {Davis}, \citenamefont {Khatami} \emph
  {et~al.}}]{zhang2:2019}%
  \BibitemOpen
  \bibfield  {author} {\bibinfo {author} {\bibfnamefont {Y.}~\bibnamefont
  {Zhang}}, \bibinfo {author} {\bibfnamefont {A.}~\bibnamefont {Mesaros}},
  \bibinfo {author} {\bibfnamefont {K.}~\bibnamefont {Fujita}}, \bibinfo
  {author} {\bibfnamefont {S.}~\bibnamefont {Edkins}}, \bibinfo {author}
  {\bibfnamefont {M.}~\bibnamefont {Hamidian}}, \bibinfo {author}
  {\bibfnamefont {K.}~\bibnamefont {Ch’ng}}, \bibinfo {author} {\bibfnamefont
  {H.}~\bibnamefont {Eisaki}}, \bibinfo {author} {\bibfnamefont
  {S.}~\bibnamefont {Uchida}}, \bibinfo {author} {\bibfnamefont {J.~S.}\
  \bibnamefont {Davis}}, \bibinfo {author} {\bibfnamefont {E.}~\bibnamefont
  {Khatami}}, \emph {et~al.},\ }\bibfield  {title} {\bibinfo {title} {Machine
  learning in electronic-quantum-matter imaging experiments},\ }\href
  {https://doi.org/10.1038/s41586-019-1319-8} {\bibfield  {journal} {\bibinfo
  {journal} {Nature}\ }\textbf {\bibinfo {volume} {570}},\ \bibinfo {pages}
  {484} (\bibinfo {year} {2019})}\BibitemShut {NoStop}%
\bibitem [{\citenamefont {Mu{\~n}oz-Gil}\ \emph {et~al.}(2021)\citenamefont
  {Mu{\~n}oz-Gil}, \citenamefont {Volpe}, \citenamefont {Garcia-March},
  \citenamefont {Aghion}, \citenamefont {Argun}, \citenamefont {Hong},
  \citenamefont {Bland}, \citenamefont {Bo}, \citenamefont {Conejero},
  \citenamefont {Firbas} \emph {et~al.}}]{munoz:2021}%
  \BibitemOpen
  \bibfield  {author} {\bibinfo {author} {\bibfnamefont {G.}~\bibnamefont
  {Mu{\~n}oz-Gil}}, \bibinfo {author} {\bibfnamefont {G.}~\bibnamefont
  {Volpe}}, \bibinfo {author} {\bibfnamefont {M.~A.}\ \bibnamefont
  {Garcia-March}}, \bibinfo {author} {\bibfnamefont {E.}~\bibnamefont
  {Aghion}}, \bibinfo {author} {\bibfnamefont {A.}~\bibnamefont {Argun}},
  \bibinfo {author} {\bibfnamefont {C.~B.}\ \bibnamefont {Hong}}, \bibinfo
  {author} {\bibfnamefont {T.}~\bibnamefont {Bland}}, \bibinfo {author}
  {\bibfnamefont {S.}~\bibnamefont {Bo}}, \bibinfo {author} {\bibfnamefont
  {J.~A.}\ \bibnamefont {Conejero}}, \bibinfo {author} {\bibfnamefont
  {N.}~\bibnamefont {Firbas}}, \emph {et~al.},\ }\bibfield  {title} {\bibinfo
  {title} {Objective comparison of methods to decode anomalous diffusion},\
  }\href {https://doi.org/10.1038/s41467-021-26320-w} {\bibfield  {journal}
  {\bibinfo  {journal} {Nat. Commun.}\ }\textbf {\bibinfo {volume} {12}},\
  \bibinfo {pages} {6253} (\bibinfo {year} {2021})}\BibitemShut {NoStop}%
\bibitem [{\citenamefont {Lu}\ \emph {et~al.}(2018)\citenamefont {Lu},
  \citenamefont {Huang}, \citenamefont {Li}, \citenamefont {Li}, \citenamefont
  {Chen}, \citenamefont {Lu}, \citenamefont {Ji}, \citenamefont {Shen},
  \citenamefont {Zhou},\ and\ \citenamefont {Zeng}}]{lu:2018}%
  \BibitemOpen
  \bibfield  {author} {\bibinfo {author} {\bibfnamefont {S.}~\bibnamefont
  {Lu}}, \bibinfo {author} {\bibfnamefont {S.}~\bibnamefont {Huang}}, \bibinfo
  {author} {\bibfnamefont {K.}~\bibnamefont {Li}}, \bibinfo {author}
  {\bibfnamefont {J.}~\bibnamefont {Li}}, \bibinfo {author} {\bibfnamefont
  {J.}~\bibnamefont {Chen}}, \bibinfo {author} {\bibfnamefont {D.}~\bibnamefont
  {Lu}}, \bibinfo {author} {\bibfnamefont {Z.}~\bibnamefont {Ji}}, \bibinfo
  {author} {\bibfnamefont {Y.}~\bibnamefont {Shen}}, \bibinfo {author}
  {\bibfnamefont {D.}~\bibnamefont {Zhou}},\ and\ \bibinfo {author}
  {\bibfnamefont {B.}~\bibnamefont {Zeng}},\ }\bibfield  {title} {\bibinfo
  {title} {Separability-entanglement classifier via machine learning},\ }\href
  {https://doi.org/10.1103/PhysRevA.98.012315} {\bibfield  {journal} {\bibinfo
  {journal} {Phys. Rev. A}\ }\textbf {\bibinfo {volume} {98}},\ \bibinfo
  {pages} {012315} (\bibinfo {year} {2018})}\BibitemShut {NoStop}%
\bibitem [{\citenamefont {Seif}\ \emph {et~al.}(2021)\citenamefont {Seif},
  \citenamefont {Hafezi},\ and\ \citenamefont {Jarzynski}}]{seif:2021}%
  \BibitemOpen
  \bibfield  {author} {\bibinfo {author} {\bibfnamefont {A.}~\bibnamefont
  {Seif}}, \bibinfo {author} {\bibfnamefont {M.}~\bibnamefont {Hafezi}},\ and\
  \bibinfo {author} {\bibfnamefont {C.}~\bibnamefont {Jarzynski}},\ }\bibfield
  {title} {\bibinfo {title} {Machine learning the thermodynamic arrow of
  time},\ }\href {https://doi.org/10.1038/s41567-020-1018-2} {\bibfield
  {journal} {\bibinfo  {journal} {Nat. Phys.}\ }\textbf {\bibinfo {volume}
  {17}},\ \bibinfo {pages} {105} (\bibinfo {year} {2021})}\BibitemShut
  {NoStop}%
\bibitem [{\citenamefont {Bezanson}\ \emph {et~al.}(2012)\citenamefont
  {Bezanson}, \citenamefont {Karpinski}, \citenamefont {Shah},\ and\
  \citenamefont {Edelman}}]{bezanson:2012}%
  \BibitemOpen
  \bibfield  {author} {\bibinfo {author} {\bibfnamefont {J.}~\bibnamefont
  {Bezanson}}, \bibinfo {author} {\bibfnamefont {S.}~\bibnamefont {Karpinski}},
  \bibinfo {author} {\bibfnamefont {V.~B.}\ \bibnamefont {Shah}},\ and\
  \bibinfo {author} {\bibfnamefont {A.}~\bibnamefont {Edelman}},\ }\bibfield
  {title} {\bibinfo {title} {{Julia: A fast dynamic language for technical
  computing}},\ }\href {https://arxiv.org/abs/1209.5145} {\bibfield  {journal}
  {\bibinfo  {journal} {arXiv:1209.5145}\ } (\bibinfo {year}
  {2012})}\BibitemShut {NoStop}%
\bibitem [{git()}]{github}%
  \BibitemOpen
  \href
  {https://github.com/arnoldjulian/Mapping-out-phase-diagrams-with-generative-classifiers}
  {\bibinfo {title}
  {{https://github.com/arnoldjulian/Mapping-out-phase-diagrams-with-generative-classifiers}}}\BibitemShut
  {NoStop}%
\bibitem [{\citenamefont {Richter-Laskowska}\ \emph {et~al.}(2023)\citenamefont
  {Richter-Laskowska}, \citenamefont {Kurpas},\ and\ \citenamefont
  {Ma\ifmmode~\acute{s}\else \'{s}\fi{}ka}}]{richter:2022}%
  \BibitemOpen
  \bibfield  {author} {\bibinfo {author} {\bibfnamefont {M.}~\bibnamefont
  {Richter-Laskowska}}, \bibinfo {author} {\bibfnamefont {M.}~\bibnamefont
  {Kurpas}},\ and\ \bibinfo {author} {\bibfnamefont {M.~M.}\ \bibnamefont
  {Ma\ifmmode~\acute{s}\else \'{s}\fi{}ka}},\ }\bibfield  {title} {\bibinfo
  {title} {Learning by confusion approach to identification of discontinuous
  phase transitions},\ }\href {https://doi.org/10.1103/PhysRevE.108.024113}
  {\bibfield  {journal} {\bibinfo  {journal} {Phys. Rev. E}\ }\textbf {\bibinfo
  {volume} {108}},\ \bibinfo {pages} {024113} (\bibinfo {year}
  {2023})}\BibitemShut {NoStop}%
\bibitem [{\citenamefont {Paszke}\ \emph {et~al.}(2019)\citenamefont {Paszke},
  \citenamefont {Gross}, \citenamefont {Massa}, \citenamefont {Lerer},
  \citenamefont {Bradbury}, \citenamefont {Chanan}, \citenamefont {Killeen},
  \citenamefont {Lin}, \citenamefont {Gimelshein}, \citenamefont {Antiga},
  \citenamefont {Desmaison}, \citenamefont {Kopf}, \citenamefont {Yang},
  \citenamefont {DeVito}, \citenamefont {Raison}, \citenamefont {Tejani},
  \citenamefont {Chilamkurthy}, \citenamefont {Steiner}, \citenamefont {Fang},
  \citenamefont {Bai},\ and\ \citenamefont {Chintala}}]{paszke:2019}%
  \BibitemOpen
  \bibfield  {author} {\bibinfo {author} {\bibfnamefont {A.}~\bibnamefont
  {Paszke}}, \bibinfo {author} {\bibfnamefont {S.}~\bibnamefont {Gross}},
  \bibinfo {author} {\bibfnamefont {F.}~\bibnamefont {Massa}}, \bibinfo
  {author} {\bibfnamefont {A.}~\bibnamefont {Lerer}}, \bibinfo {author}
  {\bibfnamefont {J.}~\bibnamefont {Bradbury}}, \bibinfo {author}
  {\bibfnamefont {G.}~\bibnamefont {Chanan}}, \bibinfo {author} {\bibfnamefont
  {T.}~\bibnamefont {Killeen}}, \bibinfo {author} {\bibfnamefont
  {Z.}~\bibnamefont {Lin}}, \bibinfo {author} {\bibfnamefont {N.}~\bibnamefont
  {Gimelshein}}, \bibinfo {author} {\bibfnamefont {L.}~\bibnamefont {Antiga}},
  \bibinfo {author} {\bibfnamefont {A.}~\bibnamefont {Desmaison}}, \bibinfo
  {author} {\bibfnamefont {A.}~\bibnamefont {Kopf}}, \bibinfo {author}
  {\bibfnamefont {E.}~\bibnamefont {Yang}}, \bibinfo {author} {\bibfnamefont
  {Z.}~\bibnamefont {DeVito}}, \bibinfo {author} {\bibfnamefont
  {M.}~\bibnamefont {Raison}}, \bibinfo {author} {\bibfnamefont
  {A.}~\bibnamefont {Tejani}}, \bibinfo {author} {\bibfnamefont
  {S.}~\bibnamefont {Chilamkurthy}}, \bibinfo {author} {\bibfnamefont
  {B.}~\bibnamefont {Steiner}}, \bibinfo {author} {\bibfnamefont
  {L.}~\bibnamefont {Fang}}, \bibinfo {author} {\bibfnamefont {J.}~\bibnamefont
  {Bai}},\ and\ \bibinfo {author} {\bibfnamefont {S.}~\bibnamefont
  {Chintala}},\ }\bibfield  {title} {\bibinfo {title} {{PyTorch: An Imperative
  Style, High-Performance Deep Learning Library}},\ }in\ \href
  {https://proceedings.neurips.cc/paper_files/paper/2019/hash/bdbca288fee7f92f2bfa9f7012727740-Abstract.html}
  {\emph {\bibinfo {booktitle} {Adv. Neural Inf. Process. Syst.}}},\
  Vol.~\bibinfo {volume} {32},\ \bibinfo {editor} {edited by\ \bibinfo {editor}
  {\bibfnamefont {H.}~\bibnamefont {Wallach}}, \bibinfo {editor} {\bibfnamefont
  {H.}~\bibnamefont {Larochelle}}, \bibinfo {editor} {\bibfnamefont
  {A.}~\bibnamefont {Beygelzimer}}, \bibinfo {editor} {\bibfnamefont
  {F.}~\bibnamefont {d\textquotesingle Alch\'{e}-Buc}}, \bibinfo {editor}
  {\bibfnamefont {E.}~\bibnamefont {Fox}},\ and\ \bibinfo {editor}
  {\bibfnamefont {R.}~\bibnamefont {Garnett}}}\ (\bibinfo  {publisher} {Curran
  Associates, Inc.},\ \bibinfo {year} {2019})\BibitemShut {NoStop}%
\bibitem [{\citenamefont {Kass}\ \emph {et~al.}(1988)\citenamefont {Kass},
  \citenamefont {Witkin},\ and\ \citenamefont {Terzopoulos}}]{kass:1988}%
  \BibitemOpen
  \bibfield  {author} {\bibinfo {author} {\bibfnamefont {M.}~\bibnamefont
  {Kass}}, \bibinfo {author} {\bibfnamefont {A.}~\bibnamefont {Witkin}},\ and\
  \bibinfo {author} {\bibfnamefont {D.}~\bibnamefont {Terzopoulos}},\
  }\bibfield  {title} {\bibinfo {title} {{Snakes: Active contour models}},\
  }\href {https://doi.org/10.1007/BF00133570} {\bibfield  {journal} {\bibinfo
  {journal} {Int. J. Comput. Vision}\ }\textbf {\bibinfo {volume} {1}},\
  \bibinfo {pages} {321} (\bibinfo {year} {1988})}\BibitemShut {NoStop}%
\bibitem [{\citenamefont {Kingma}\ and\ \citenamefont
  {Ba}(2014)}]{kingma:2014}%
  \BibitemOpen
  \bibfield  {author} {\bibinfo {author} {\bibfnamefont {D.~P.}\ \bibnamefont
  {Kingma}}\ and\ \bibinfo {author} {\bibfnamefont {J.}~\bibnamefont {Ba}},\
  }\bibfield  {title} {\bibinfo {title} {{Adam: A method for stochastic
  optimization}},\ }\href {https://arxiv.org/abs/1412.6980} {\bibfield
  {journal} {\bibinfo  {journal} {arXiv:1412.6980}\ } (\bibinfo {year}
  {2014})}\BibitemShut {NoStop}%
\bibitem [{\citenamefont {Baydin}\ \emph {et~al.}(2018)\citenamefont {Baydin},
  \citenamefont {Pearlmutter}, \citenamefont {Radul},\ and\ \citenamefont
  {Siskind}}]{baydin:2018}%
  \BibitemOpen
  \bibfield  {author} {\bibinfo {author} {\bibfnamefont {A.~G.}\ \bibnamefont
  {Baydin}}, \bibinfo {author} {\bibfnamefont {B.~A.}\ \bibnamefont
  {Pearlmutter}}, \bibinfo {author} {\bibfnamefont {A.~A.}\ \bibnamefont
  {Radul}},\ and\ \bibinfo {author} {\bibfnamefont {J.~M.}\ \bibnamefont
  {Siskind}},\ }\bibfield  {title} {\bibinfo {title} {{Automatic
  Differentiation in Machine Learning: a Survey}},\ }\href
  {http://jmlr.org/papers/v18/17-468.html} {\bibfield  {journal} {\bibinfo
  {journal} {J. Mach. Learn. Res.}\ }\textbf {\bibinfo {volume} {18}},\
  \bibinfo {pages} {1} (\bibinfo {year} {2018})}\BibitemShut {NoStop}%
\bibitem [{\citenamefont {Innes}(2018)}]{innes:2018}%
  \BibitemOpen
  \bibfield  {author} {\bibinfo {author} {\bibfnamefont {M.}~\bibnamefont
  {Innes}},\ }\bibfield  {title} {\bibinfo {title} {{Flux: Elegant machine
  learning with Julia}},\ }\href {https://doi.org/10.21105/joss.00602}
  {\bibfield  {journal} {\bibinfo  {journal} {J. Open Source Softw.}\ }\textbf
  {\bibinfo {volume} {3}},\ \bibinfo {pages} {602} (\bibinfo {year}
  {2018})}\BibitemShut {NoStop}%
\bibitem [{\citenamefont {Casella}\ and\ \citenamefont
  {Berger}(2002)}]{casella:2002}%
  \BibitemOpen
  \bibfield  {author} {\bibinfo {author} {\bibfnamefont {G.}~\bibnamefont
  {Casella}}\ and\ \bibinfo {author} {\bibfnamefont {R.~L.}\ \bibnamefont
  {Berger}},\ }\href {https://worldcat.org/title/67327073} {\emph {\bibinfo
  {title} {Statistical inference}}}\ (\bibinfo  {publisher} {Duxbury},\
  \bibinfo {year} {2002})\BibitemShut {NoStop}%
\bibitem [{\citenamefont {Fishman}\ \emph {et~al.}(2022)\citenamefont
  {Fishman}, \citenamefont {White},\ and\ \citenamefont
  {Stoudenmire}}]{itensor}%
  \BibitemOpen
  \bibfield  {author} {\bibinfo {author} {\bibfnamefont {M.}~\bibnamefont
  {Fishman}}, \bibinfo {author} {\bibfnamefont {S.~R.}\ \bibnamefont {White}},\
  and\ \bibinfo {author} {\bibfnamefont {E.~M.}\ \bibnamefont {Stoudenmire}},\
  }\bibfield  {title} {\bibinfo {title} {{The ITensor Software Library for
  Tensor Network Calculations}},\ }\href
  {https://doi.org/10.21468/SciPostPhysCodeb.4} {\bibfield  {journal} {\bibinfo
   {journal} {SciPost Phys. Codebases}\ ,\ \bibinfo {pages} {4}} (\bibinfo
  {year} {2022})}\BibitemShut {NoStop}%
\end{thebibliography}%


\begin{thebibliography}{24}%
\makeatletter
\providecommand \@ifxundefined [1]{%
 \@ifx{#1\undefined}
}%
\providecommand \@ifnum [1]{%
 \ifnum #1\expandafter \@firstoftwo
 \else \expandafter \@secondoftwo
 \fi
}%
\providecommand \@ifx [1]{%
 \ifx #1\expandafter \@firstoftwo
 \else \expandafter \@secondoftwo
 \fi
}%
\providecommand \natexlab [1]{#1}%
\providecommand \enquote  [1]{``#1''}%
\providecommand \bibnamefont  [1]{#1}%
\providecommand \bibfnamefont [1]{#1}%
\providecommand \citenamefont [1]{#1}%
\providecommand \href@noop [0]{\@secondoftwo}%
\providecommand \href [0]{\begingroup \@sanitize@url \@href}%
\providecommand \@href[1]{\@@startlink{#1}\@@href}%
\providecommand \@@href[1]{\endgroup#1\@@endlink}%
\providecommand \@sanitize@url [0]{\catcode `\\12\catcode `\$12\catcode
  `\&12\catcode `\#12\catcode `\^12\catcode `\_12\catcode `\%12\relax}%
\providecommand \@@startlink[1]{}%
\providecommand \@@endlink[0]{}%
\providecommand \url  [0]{\begingroup\@sanitize@url \@url }%
\providecommand \@url [1]{\endgroup\@href {#1}{\urlprefix }}%
\providecommand \urlprefix  [0]{URL }%
\providecommand \Eprint [0]{\href }%
\providecommand \doibase [0]{https://doi.org/}%
\providecommand \selectlanguage [0]{\@gobble}%
\providecommand \bibinfo  [0]{\@secondoftwo}%
\providecommand \bibfield  [0]{\@secondoftwo}%
\providecommand \translation [1]{[#1]}%
\providecommand \BibitemOpen [0]{}%
\providecommand \bibitemStop [0]{}%
\providecommand \bibitemNoStop [0]{.\EOS\space}%
\providecommand \EOS [0]{\spacefactor3000\relax}%
\providecommand \BibitemShut  [1]{\csname bibitem#1\endcsname}%
\let\auto@bib@innerbib\@empty
\bibitem [{\citenamefont {van Nieuwenburg}\ \emph {et~al.}(2017)\citenamefont
  {van Nieuwenburg}, \citenamefont {Liu},\ and\ \citenamefont
  {Huber}}]{van:2017}%
  \BibitemOpen
  \bibfield  {author} {\bibinfo {author} {\bibfnamefont {E.~P.~L.}\
  \bibnamefont {van Nieuwenburg}}, \bibinfo {author} {\bibfnamefont {Y.-H.}\
  \bibnamefont {Liu}},\ and\ \bibinfo {author} {\bibfnamefont {S.~D.}\
  \bibnamefont {Huber}},\ }\bibfield  {title} {\bibinfo {title} {Learning phase
  transitions by confusion},\ }\href {https://doi.org/10.1038/nphys4037}
  {\bibfield  {journal} {\bibinfo  {journal} {Nat. Phys.}\ }\textbf {\bibinfo
  {volume} {13}},\ \bibinfo {pages} {435} (\bibinfo {year} {2017})}\BibitemShut
  {NoStop}%
\bibitem [{\citenamefont {Liu}\ and\ \citenamefont {van
  Nieuwenburg}(2018)}]{liu:2018}%
  \BibitemOpen
  \bibfield  {author} {\bibinfo {author} {\bibfnamefont {Y.-H.}\ \bibnamefont
  {Liu}}\ and\ \bibinfo {author} {\bibfnamefont {E.~P.~L.}\ \bibnamefont {van
  Nieuwenburg}},\ }\bibfield  {title} {\bibinfo {title} {{D}iscriminative
  {C}ooperative {N}etworks for {D}etecting {P}hase {T}ransitions},\ }\href
  {https://doi.org/10.1103/PhysRevLett.120.176401} {\bibfield  {journal}
  {\bibinfo  {journal} {Phys. Rev. Lett.}\ }\textbf {\bibinfo {volume} {120}},\
  \bibinfo {pages} {176401} (\bibinfo {year} {2018})}\BibitemShut {NoStop}%
\bibitem [{\citenamefont {Beach}\ \emph {et~al.}(2018)\citenamefont {Beach},
  \citenamefont {Golubeva},\ and\ \citenamefont {Melko}}]{beach:2018}%
  \BibitemOpen
  \bibfield  {author} {\bibinfo {author} {\bibfnamefont {M.~J.~S.}\
  \bibnamefont {Beach}}, \bibinfo {author} {\bibfnamefont {A.}~\bibnamefont
  {Golubeva}},\ and\ \bibinfo {author} {\bibfnamefont {R.~G.}\ \bibnamefont
  {Melko}},\ }\bibfield  {title} {\bibinfo {title} {Machine learning vortices
  at the {K}osterlitz-{T}houless transition},\ }\href
  {https://doi.org/10.1103/PhysRevB.97.045207} {\bibfield  {journal} {\bibinfo
  {journal} {Phys. Rev. B}\ }\textbf {\bibinfo {volume} {97}},\ \bibinfo
  {pages} {045207} (\bibinfo {year} {2018})}\BibitemShut {NoStop}%
\bibitem [{\citenamefont {Suchsland}\ and\ \citenamefont
  {Wessel}(2018)}]{suchsland:2018}%
  \BibitemOpen
  \bibfield  {author} {\bibinfo {author} {\bibfnamefont {P.}~\bibnamefont
  {Suchsland}}\ and\ \bibinfo {author} {\bibfnamefont {S.}~\bibnamefont
  {Wessel}},\ }\bibfield  {title} {\bibinfo {title} {Parameter diagnostics of
  phases and phase transition learning by neural networks},\ }\href
  {https://doi.org/10.1103/PhysRevB.97.174435} {\bibfield  {journal} {\bibinfo
  {journal} {Phys. Rev. B}\ }\textbf {\bibinfo {volume} {97}},\ \bibinfo
  {pages} {174435} (\bibinfo {year} {2018})}\BibitemShut {NoStop}%
\bibitem [{\citenamefont {Lee}\ and\ \citenamefont {Kim}(2019)}]{lee:2019}%
  \BibitemOpen
  \bibfield  {author} {\bibinfo {author} {\bibfnamefont {S.~S.}\ \bibnamefont
  {Lee}}\ and\ \bibinfo {author} {\bibfnamefont {B.~J.}\ \bibnamefont {Kim}},\
  }\bibfield  {title} {\bibinfo {title} {Confusion scheme in machine learning
  detects double phase transitions and quasi-long-range order},\ }\href
  {https://doi.org/10.1103/PhysRevE.99.043308} {\bibfield  {journal} {\bibinfo
  {journal} {Phys. Rev. E}\ }\textbf {\bibinfo {volume} {99}},\ \bibinfo
  {pages} {043308} (\bibinfo {year} {2019})}\BibitemShut {NoStop}%
\bibitem [{\citenamefont {Kharkov}\ \emph {et~al.}(2020)\citenamefont
  {Kharkov}, \citenamefont {Sotskov}, \citenamefont {Karazeev}, \citenamefont
  {Kiktenko},\ and\ \citenamefont {Fedorov}}]{kharkov:2020}%
  \BibitemOpen
  \bibfield  {author} {\bibinfo {author} {\bibfnamefont {Y.~A.}\ \bibnamefont
  {Kharkov}}, \bibinfo {author} {\bibfnamefont {V.~E.}\ \bibnamefont
  {Sotskov}}, \bibinfo {author} {\bibfnamefont {A.~A.}\ \bibnamefont
  {Karazeev}}, \bibinfo {author} {\bibfnamefont {E.~O.}\ \bibnamefont
  {Kiktenko}},\ and\ \bibinfo {author} {\bibfnamefont {A.~K.}\ \bibnamefont
  {Fedorov}},\ }\bibfield  {title} {\bibinfo {title} {Revealing quantum chaos
  with machine learning},\ }\href {https://doi.org/10.1103/PhysRevB.101.064406}
  {\bibfield  {journal} {\bibinfo  {journal} {Phys. Rev. B}\ }\textbf {\bibinfo
  {volume} {101}},\ \bibinfo {pages} {064406} (\bibinfo {year}
  {2020})}\BibitemShut {NoStop}%
\bibitem [{\citenamefont {Guo}\ \emph {et~al.}(2020)\citenamefont {Guo},
  \citenamefont {Ai},\ and\ \citenamefont {He}}]{guo:2020}%
  \BibitemOpen
  \bibfield  {author} {\bibinfo {author} {\bibfnamefont {W.}~\bibnamefont
  {Guo}}, \bibinfo {author} {\bibfnamefont {B.}~\bibnamefont {Ai}},\ and\
  \bibinfo {author} {\bibfnamefont {L.}~\bibnamefont {He}},\ }\bibfield
  {title} {\bibinfo {title} {Reveal flocking of birds flying in fog by machine
  learning},\ }\href {https://arxiv.org/abs/2005.10505} {\bibfield  {journal}
  {\bibinfo  {journal} {arXiv:2005.10505}\ } (\bibinfo {year}
  {2020})}\BibitemShut {NoStop}%
\bibitem [{\citenamefont {Greplova}\ \emph {et~al.}(2020)\citenamefont
  {Greplova}, \citenamefont {Valenti}, \citenamefont {Boschung}, \citenamefont
  {Schäfer}, \citenamefont {Lörch},\ and\ \citenamefont
  {Huber}}]{greplova:2020}%
  \BibitemOpen
  \bibfield  {author} {\bibinfo {author} {\bibfnamefont {E.}~\bibnamefont
  {Greplova}}, \bibinfo {author} {\bibfnamefont {A.}~\bibnamefont {Valenti}},
  \bibinfo {author} {\bibfnamefont {G.}~\bibnamefont {Boschung}}, \bibinfo
  {author} {\bibfnamefont {F.}~\bibnamefont {Schäfer}}, \bibinfo {author}
  {\bibfnamefont {N.}~\bibnamefont {Lörch}},\ and\ \bibinfo {author}
  {\bibfnamefont {S.~D.}\ \bibnamefont {Huber}},\ }\bibfield  {title} {\bibinfo
  {title} {Unsupervised identification of topological phase transitions using
  predictive models},\ }\href {https://doi.org/10.1088/1367-2630/ab7771}
  {\bibfield  {journal} {\bibinfo  {journal} {New J. Phys.}\ }\textbf {\bibinfo
  {volume} {22}},\ \bibinfo {pages} {045003} (\bibinfo {year}
  {2020})}\BibitemShut {NoStop}%
\bibitem [{\citenamefont {Bohrdt}\ \emph {et~al.}(2021)\citenamefont {Bohrdt},
  \citenamefont {Kim}, \citenamefont {Lukin}, \citenamefont {Rispoli},
  \citenamefont {Schittko}, \citenamefont {Knap}, \citenamefont {Greiner},\
  and\ \citenamefont {L\'eonard}}]{bohrdt:2021}%
  \BibitemOpen
  \bibfield  {author} {\bibinfo {author} {\bibfnamefont {A.}~\bibnamefont
  {Bohrdt}}, \bibinfo {author} {\bibfnamefont {S.}~\bibnamefont {Kim}},
  \bibinfo {author} {\bibfnamefont {A.}~\bibnamefont {Lukin}}, \bibinfo
  {author} {\bibfnamefont {M.}~\bibnamefont {Rispoli}}, \bibinfo {author}
  {\bibfnamefont {R.}~\bibnamefont {Schittko}}, \bibinfo {author}
  {\bibfnamefont {M.}~\bibnamefont {Knap}}, \bibinfo {author} {\bibfnamefont
  {M.}~\bibnamefont {Greiner}},\ and\ \bibinfo {author} {\bibfnamefont
  {J.}~\bibnamefont {L\'eonard}},\ }\bibfield  {title} {\bibinfo {title}
  {Analyzing {N}onequilibrium {Q}uantum {S}tates through {S}napshots with
  {A}rtificial {N}eural {N}etworks},\ }\href
  {https://doi.org/10.1103/PhysRevLett.127.150504} {\bibfield  {journal}
  {\bibinfo  {journal} {Phys. Rev. Lett.}\ }\textbf {\bibinfo {volume} {127}},\
  \bibinfo {pages} {150504} (\bibinfo {year} {2021})}\BibitemShut {NoStop}%
\bibitem [{\citenamefont {Richter-Laskowska}\ \emph {et~al.}(2023)\citenamefont
  {Richter-Laskowska}, \citenamefont {Kurpas},\ and\ \citenamefont
  {Ma\ifmmode~\acute{s}\else \'{s}\fi{}ka}}]{richter:2022}%
  \BibitemOpen
  \bibfield  {author} {\bibinfo {author} {\bibfnamefont {M.}~\bibnamefont
  {Richter-Laskowska}}, \bibinfo {author} {\bibfnamefont {M.}~\bibnamefont
  {Kurpas}},\ and\ \bibinfo {author} {\bibfnamefont {M.~M.}\ \bibnamefont
  {Ma\ifmmode~\acute{s}\else \'{s}\fi{}ka}},\ }\bibfield  {title} {\bibinfo
  {title} {Learning by confusion approach to identification of discontinuous
  phase transitions},\ }\href {https://doi.org/10.1103/PhysRevE.108.024113}
  {\bibfield  {journal} {\bibinfo  {journal} {Phys. Rev. E}\ }\textbf {\bibinfo
  {volume} {108}},\ \bibinfo {pages} {024113} (\bibinfo {year}
  {2023})}\BibitemShut {NoStop}%
\bibitem [{\citenamefont {Gavreev}\ \emph {et~al.}(2022)\citenamefont
  {Gavreev}, \citenamefont {Mastiukova}, \citenamefont {Kiktenko},\ and\
  \citenamefont {Fedorov}}]{gavreev:2022}%
  \BibitemOpen
  \bibfield  {author} {\bibinfo {author} {\bibfnamefont {M.~A.}\ \bibnamefont
  {Gavreev}}, \bibinfo {author} {\bibfnamefont {A.~S.}\ \bibnamefont
  {Mastiukova}}, \bibinfo {author} {\bibfnamefont {E.~O.}\ \bibnamefont
  {Kiktenko}},\ and\ \bibinfo {author} {\bibfnamefont {A.~K.}\ \bibnamefont
  {Fedorov}},\ }\bibfield  {title} {\bibinfo {title} {Learning entanglement
  breakdown as a phase transition by confusion},\ }\href
  {https://doi.org/10.1088/1367-2630/ac7fb2} {\bibfield  {journal} {\bibinfo
  {journal} {New J. Phys.}\ }\textbf {\bibinfo {volume} {24}},\ \bibinfo
  {pages} {073045} (\bibinfo {year} {2022})}\BibitemShut {NoStop}%
\bibitem [{\citenamefont {Zvyagintseva}\ \emph {et~al.}(2022)\citenamefont
  {Zvyagintseva}, \citenamefont {Sigurdsson}, \citenamefont {Kozin},
  \citenamefont {Iorsh}, \citenamefont {Shelykh}, \citenamefont {Ulyantsev},\
  and\ \citenamefont {Kyriienko}}]{zvyagintseva:2022}%
  \BibitemOpen
  \bibfield  {author} {\bibinfo {author} {\bibfnamefont {D.}~\bibnamefont
  {Zvyagintseva}}, \bibinfo {author} {\bibfnamefont {H.}~\bibnamefont
  {Sigurdsson}}, \bibinfo {author} {\bibfnamefont {V.~K.}\ \bibnamefont
  {Kozin}}, \bibinfo {author} {\bibfnamefont {I.}~\bibnamefont {Iorsh}},
  \bibinfo {author} {\bibfnamefont {I.~A.}\ \bibnamefont {Shelykh}}, \bibinfo
  {author} {\bibfnamefont {V.}~\bibnamefont {Ulyantsev}},\ and\ \bibinfo
  {author} {\bibfnamefont {O.}~\bibnamefont {Kyriienko}},\ }\bibfield  {title}
  {\bibinfo {title} {Machine learning of phase transitions in nonlinear
  polariton lattices},\ }\href {https://doi.org/10.1038/s42005-021-00755-5}
  {\bibfield  {journal} {\bibinfo  {journal} {Commun. Phys.}\ }\textbf
  {\bibinfo {volume} {5}},\ \bibinfo {pages} {8} (\bibinfo {year}
  {2022})}\BibitemShut {NoStop}%
\bibitem [{\citenamefont {Guo}\ and\ \citenamefont {He}(2023)}]{guo:2023}%
  \BibitemOpen
  \bibfield  {author} {\bibinfo {author} {\bibfnamefont {W.}~\bibnamefont
  {Guo}}\ and\ \bibinfo {author} {\bibfnamefont {L.}~\bibnamefont {He}},\
  }\bibfield  {title} {\bibinfo {title} {Learning phase transitions from
  regression uncertainty: a new regression-based machine learning approach for
  automated detection of phases of matter},\ }\href
  {https://doi.org/10.1088/1367-2630/acef4e} {\bibfield  {journal} {\bibinfo
  {journal} {New J. Phys.}\ }\textbf {\bibinfo {volume} {25}},\ \bibinfo
  {pages} {083037} (\bibinfo {year} {2023})}\BibitemShut {NoStop}%
\bibitem [{\citenamefont {Schlömer}\ and\ \citenamefont
  {Bohrdt}(2023)}]{schlomer:2023}%
  \BibitemOpen
  \bibfield  {author} {\bibinfo {author} {\bibfnamefont {H.}~\bibnamefont
  {Schlömer}}\ and\ \bibinfo {author} {\bibfnamefont {A.}~\bibnamefont
  {Bohrdt}},\ }\bibfield  {title} {\bibinfo {title} {{Fluctuation based
  interpretable analysis scheme for quantum many-body snapshots}},\ }\href
  {https://doi.org/10.21468/SciPostPhys.15.3.099} {\bibfield  {journal}
  {\bibinfo  {journal} {SciPost Phys.}\ }\textbf {\bibinfo {volume} {15}},\
  \bibinfo {pages} {099} (\bibinfo {year} {2023})}\BibitemShut {NoStop}%
\bibitem [{\citenamefont {Arnold}\ and\ \citenamefont
  {Sch\"afer}(2022)}]{arnold:2022}%
  \BibitemOpen
  \bibfield  {author} {\bibinfo {author} {\bibfnamefont {J.}~\bibnamefont
  {Arnold}}\ and\ \bibinfo {author} {\bibfnamefont {F.}~\bibnamefont
  {Sch\"afer}},\ }\bibfield  {title} {\bibinfo {title} {Replacing neural
  networks by optimal analytical predictors for the detection of phase
  transitions},\ }\href {https://doi.org/10.1103/PhysRevX.12.031044} {\bibfield
   {journal} {\bibinfo  {journal} {Phys. Rev. X}\ }\textbf {\bibinfo {volume}
  {12}},\ \bibinfo {pages} {031044} (\bibinfo {year} {2022})}\BibitemShut
  {NoStop}%
\bibitem [{\citenamefont {Sch\"afer}\ and\ \citenamefont
  {L\"orch}(2019)}]{schaefer:2019}%
  \BibitemOpen
  \bibfield  {author} {\bibinfo {author} {\bibfnamefont {F.}~\bibnamefont
  {Sch\"afer}}\ and\ \bibinfo {author} {\bibfnamefont {N.}~\bibnamefont
  {L\"orch}},\ }\bibfield  {title} {\bibinfo {title} {Vector field divergence
  of predictive model output as indication of phase transitions},\ }\href
  {https://doi.org/10.1103/PhysRevE.99.062107} {\bibfield  {journal} {\bibinfo
  {journal} {Phys. Rev. E}\ }\textbf {\bibinfo {volume} {99}},\ \bibinfo
  {pages} {062107} (\bibinfo {year} {2019})}\BibitemShut {NoStop}%
\bibitem [{\citenamefont {Arnold}\ \emph {et~al.}(2021)\citenamefont {Arnold},
  \citenamefont {Sch\"afer}, \citenamefont {\ifmmode~\check{Z}\else
  \v{Z}\fi{}onda},\ and\ \citenamefont {Lode}}]{arnold:2021}%
  \BibitemOpen
  \bibfield  {author} {\bibinfo {author} {\bibfnamefont {J.}~\bibnamefont
  {Arnold}}, \bibinfo {author} {\bibfnamefont {F.}~\bibnamefont {Sch\"afer}},
  \bibinfo {author} {\bibfnamefont {M.}~\bibnamefont {\ifmmode~\check{Z}\else
  \v{Z}\fi{}onda}},\ and\ \bibinfo {author} {\bibfnamefont {A.~U.~J.}\
  \bibnamefont {Lode}},\ }\bibfield  {title} {\bibinfo {title} {Interpretable
  and unsupervised phase classification},\ }\href
  {https://doi.org/10.1103/PhysRevResearch.3.033052} {\bibfield  {journal}
  {\bibinfo  {journal} {Phys. Rev. Res.}\ }\textbf {\bibinfo {volume} {3}},\
  \bibinfo {pages} {033052} (\bibinfo {year} {2021})}\BibitemShut {NoStop}%
\bibitem [{\citenamefont {Kass}\ \emph {et~al.}(1988)\citenamefont {Kass},
  \citenamefont {Witkin},\ and\ \citenamefont {Terzopoulos}}]{kass:1988}%
  \BibitemOpen
  \bibfield  {author} {\bibinfo {author} {\bibfnamefont {M.}~\bibnamefont
  {Kass}}, \bibinfo {author} {\bibfnamefont {A.}~\bibnamefont {Witkin}},\ and\
  \bibinfo {author} {\bibfnamefont {D.}~\bibnamefont {Terzopoulos}},\
  }\bibfield  {title} {\bibinfo {title} {{Snakes: Active contour models}},\
  }\href {https://doi.org/10.1007/BF00133570} {\bibfield  {journal} {\bibinfo
  {journal} {Int. J. Comput. Vision}\ }\textbf {\bibinfo {volume} {1}},\
  \bibinfo {pages} {321} (\bibinfo {year} {1988})}\BibitemShut {NoStop}%
\bibitem [{\citenamefont {Kingma}\ and\ \citenamefont
  {Ba}(2014)}]{kingma:2014}%
  \BibitemOpen
  \bibfield  {author} {\bibinfo {author} {\bibfnamefont {D.~P.}\ \bibnamefont
  {Kingma}}\ and\ \bibinfo {author} {\bibfnamefont {J.}~\bibnamefont {Ba}},\
  }\bibfield  {title} {\bibinfo {title} {{Adam: A method for stochastic
  optimization}},\ }\href {https://arxiv.org/abs/1412.6980} {\bibfield
  {journal} {\bibinfo  {journal} {arXiv:1412.6980}\ } (\bibinfo {year}
  {2014})}\BibitemShut {NoStop}%
\bibitem [{\citenamefont {Paszke}\ \emph {et~al.}(2019)\citenamefont {Paszke},
  \citenamefont {Gross}, \citenamefont {Massa}, \citenamefont {Lerer},
  \citenamefont {Bradbury}, \citenamefont {Chanan}, \citenamefont {Killeen},
  \citenamefont {Lin}, \citenamefont {Gimelshein}, \citenamefont {Antiga},
  \citenamefont {Desmaison}, \citenamefont {Kopf}, \citenamefont {Yang},
  \citenamefont {DeVito}, \citenamefont {Raison}, \citenamefont {Tejani},
  \citenamefont {Chilamkurthy}, \citenamefont {Steiner}, \citenamefont {Fang},
  \citenamefont {Bai},\ and\ \citenamefont {Chintala}}]{paszke:2019}%
  \BibitemOpen
  \bibfield  {author} {\bibinfo {author} {\bibfnamefont {A.}~\bibnamefont
  {Paszke}}, \bibinfo {author} {\bibfnamefont {S.}~\bibnamefont {Gross}},
  \bibinfo {author} {\bibfnamefont {F.}~\bibnamefont {Massa}}, \bibinfo
  {author} {\bibfnamefont {A.}~\bibnamefont {Lerer}}, \bibinfo {author}
  {\bibfnamefont {J.}~\bibnamefont {Bradbury}}, \bibinfo {author}
  {\bibfnamefont {G.}~\bibnamefont {Chanan}}, \bibinfo {author} {\bibfnamefont
  {T.}~\bibnamefont {Killeen}}, \bibinfo {author} {\bibfnamefont
  {Z.}~\bibnamefont {Lin}}, \bibinfo {author} {\bibfnamefont {N.}~\bibnamefont
  {Gimelshein}}, \bibinfo {author} {\bibfnamefont {L.}~\bibnamefont {Antiga}},
  \bibinfo {author} {\bibfnamefont {A.}~\bibnamefont {Desmaison}}, \bibinfo
  {author} {\bibfnamefont {A.}~\bibnamefont {Kopf}}, \bibinfo {author}
  {\bibfnamefont {E.}~\bibnamefont {Yang}}, \bibinfo {author} {\bibfnamefont
  {Z.}~\bibnamefont {DeVito}}, \bibinfo {author} {\bibfnamefont
  {M.}~\bibnamefont {Raison}}, \bibinfo {author} {\bibfnamefont
  {A.}~\bibnamefont {Tejani}}, \bibinfo {author} {\bibfnamefont
  {S.}~\bibnamefont {Chilamkurthy}}, \bibinfo {author} {\bibfnamefont
  {B.}~\bibnamefont {Steiner}}, \bibinfo {author} {\bibfnamefont
  {L.}~\bibnamefont {Fang}}, \bibinfo {author} {\bibfnamefont {J.}~\bibnamefont
  {Bai}},\ and\ \bibinfo {author} {\bibfnamefont {S.}~\bibnamefont
  {Chintala}},\ }\bibfield  {title} {\bibinfo {title} {{PyTorch: An Imperative
  Style, High-Performance Deep Learning Library}},\ }in\ \href
  {https://proceedings.neurips.cc/paper_files/paper/2019/hash/bdbca288fee7f92f2bfa9f7012727740-Abstract.html}
  {\emph {\bibinfo {booktitle} {Adv. Neural Inf. Process. Syst.}}},\
  Vol.~\bibinfo {volume} {32},\ \bibinfo {editor} {edited by\ \bibinfo {editor}
  {\bibfnamefont {H.}~\bibnamefont {Wallach}}, \bibinfo {editor} {\bibfnamefont
  {H.}~\bibnamefont {Larochelle}}, \bibinfo {editor} {\bibfnamefont
  {A.}~\bibnamefont {Beygelzimer}}, \bibinfo {editor} {\bibfnamefont
  {F.}~\bibnamefont {d\textquotesingle Alch\'{e}-Buc}}, \bibinfo {editor}
  {\bibfnamefont {E.}~\bibnamefont {Fox}},\ and\ \bibinfo {editor}
  {\bibfnamefont {R.}~\bibnamefont {Garnett}}}\ (\bibinfo  {publisher} {Curran
  Associates, Inc.},\ \bibinfo {year} {2019})\BibitemShut {NoStop}%
\bibitem [{\citenamefont {Innes}(2018)}]{innes:2018}%
  \BibitemOpen
  \bibfield  {author} {\bibinfo {author} {\bibfnamefont {M.}~\bibnamefont
  {Innes}},\ }\bibfield  {title} {\bibinfo {title} {{Flux: Elegant machine
  learning with Julia}},\ }\href {https://doi.org/10.21105/joss.00602}
  {\bibfield  {journal} {\bibinfo  {journal} {J. Open Source Softw.}\ }\textbf
  {\bibinfo {volume} {3}},\ \bibinfo {pages} {602} (\bibinfo {year}
  {2018})}\BibitemShut {NoStop}%
\bibitem [{\citenamefont {Baydin}\ \emph {et~al.}(2018)\citenamefont {Baydin},
  \citenamefont {Pearlmutter}, \citenamefont {Radul},\ and\ \citenamefont
  {Siskind}}]{baydin:2018}%
  \BibitemOpen
  \bibfield  {author} {\bibinfo {author} {\bibfnamefont {A.~G.}\ \bibnamefont
  {Baydin}}, \bibinfo {author} {\bibfnamefont {B.~A.}\ \bibnamefont
  {Pearlmutter}}, \bibinfo {author} {\bibfnamefont {A.~A.}\ \bibnamefont
  {Radul}},\ and\ \bibinfo {author} {\bibfnamefont {J.~M.}\ \bibnamefont
  {Siskind}},\ }\bibfield  {title} {\bibinfo {title} {{Automatic
  Differentiation in Machine Learning: a Survey}},\ }\href
  {http://jmlr.org/papers/v18/17-468.html} {\bibfield  {journal} {\bibinfo
  {journal} {J. Mach. Learn. Res.}\ }\textbf {\bibinfo {volume} {18}},\
  \bibinfo {pages} {1} (\bibinfo {year} {2018})}\BibitemShut {NoStop}%
\bibitem [{\citenamefont {Casella}\ and\ \citenamefont
  {Berger}(2002)}]{casella:2002}%
  \BibitemOpen
  \bibfield  {author} {\bibinfo {author} {\bibfnamefont {G.}~\bibnamefont
  {Casella}}\ and\ \bibinfo {author} {\bibfnamefont {R.~L.}\ \bibnamefont
  {Berger}},\ }\href {https://worldcat.org/title/67327073} {\emph {\bibinfo
  {title} {Statistical inference}}}\ (\bibinfo  {publisher} {Duxbury},\
  \bibinfo {year} {2002})\BibitemShut {NoStop}%
\bibitem [{\citenamefont {Fishman}\ \emph {et~al.}(2022)\citenamefont
  {Fishman}, \citenamefont {White},\ and\ \citenamefont
  {Stoudenmire}}]{itensor}%
  \BibitemOpen
  \bibfield  {author} {\bibinfo {author} {\bibfnamefont {M.}~\bibnamefont
  {Fishman}}, \bibinfo {author} {\bibfnamefont {S.~R.}\ \bibnamefont {White}},\
  and\ \bibinfo {author} {\bibfnamefont {E.~M.}\ \bibnamefont {Stoudenmire}},\
  }\bibfield  {title} {\bibinfo {title} {{The ITensor Software Library for
  Tensor Network Calculations}},\ }\href
  {https://doi.org/10.21468/SciPostPhysCodeb.4} {\bibfield  {journal} {\bibinfo
   {journal} {SciPost Phys. Codebases}\ ,\ \bibinfo {pages} {4}} (\bibinfo
  {year} {2022})}\BibitemShut {NoStop}%
\end{thebibliography}%
\end{document}


\setcounter{equation}{0}
	\setcounter{figure}{0}
	\setcounter{table}{0}
	\setcounter{page}{1}
	\makeatletter
	\renewcommand{\thesection}{S\arabic{section}}
	\renewcommand{\theequation}{S\arabic{equation}}
	\renewcommand{\thefigure}{S\arabic{figure}}
	
	\title{Supplemental Material: Mapping out phase diagrams with generative classifiers}
	\author{Julian Arnold}
	\affiliation{Department of Physics, University of Basel, Klingelbergstrasse 82, 4056 Basel, Switzerland}
	\author{Frank Sch\"afer}
	\affiliation{CSAIL, Massachusetts Institute of Technology, Cambridge, MA
		02139, USA}
	\author{Alan Edelman}
	\affiliation{CSAIL, Massachusetts Institute of Technology, Cambridge, MA
		02139, USA}
	\affiliation{Department of Mathematics, Massachusetts Institute of Technology, Cambridge, MA
		02139, USA}
	\author{Christoph Bruder}
	\affiliation{Department of Physics, University of Basel, Klingelbergstrasse 82, 4056 Basel, Switzerland}
	\date{\today}
	
	\maketitle
	
	\section{Additional details on phase-classification methods}
	\subsection{Assumptions when setting up phase-classification tasks}
	When setting up phase-classification tasks, we choose a uniform distribution over the set of parameters associated with each class, i.e., $P(\bm{\gamma}|y) = 1 / |\Gamma_{y}|$ for $\bm{\gamma}\in \Gamma_{y}$ and zero otherwise. This choice has been implicitly made via the definition of a loss function in previous works tackling the problem of mapping out phase diagrams with discriminative models (see Sec.~\ref{sec:discr_indicators}). Due to the application of Bayes' theorem within the generative approach, such probabilistic assumptions are brought to light. In principle, our framework allows for a free choice of $P(\bm{\gamma}|y)$. It represents our freedom to choose to what degree different points in parameter space are considered representative of the label $y$. Because the labeling procedure is different for the three schemes for mapping out phase diagrams discussed in the main text [Eqs. (3), (5), and (7) in the main text], its choice affects them differently. 
	
	In scheme 3 [Eq. (7) in the main text], each sampled value of the tuning parameter is considered its own class $\Gamma_{y} = \{ \bm{\gamma}_{y}\}$. As such, setting $P(\bm{\gamma}|y) = 1$ if $\bm{\gamma}  = \bm{\gamma}_{y}$ and zero otherwise is the only sensible choice. This corresponds to a special case of the uniform distribution.
	
	In scheme 2 [Eq. (5) in the main text], the sets $\Gamma_{1}^{(i)}(\bm{\gamma})$ and $\Gamma_{2}^{(i)}(\bm{\gamma})$ are each comprised of the $l$ points closest to $\bm{\gamma}$ in part 1 and 2 of the split parameter space. In the main text, we consider $l=1$ and each set is composed of a single point leaving only a single sensible choice as mentioned above for scheme 1. For $l > 1$, choosing $P(\bm{\gamma}|y)$ to be distinct from a uniform distribution would put different weights on different points within a given set. However, without additional system information, there is no compelling reason to do so. Moreover, note that for small $l$ and dense sampling of the parameter space, reweighting is expected to have a marginal effect since, in this case, $P(\bm{x}|\bm{\gamma})$ is similar for points $\bm{\gamma}$ within each of the two sets.
	
	In scheme 1 [Eq. (3) in the main text], the sets $\{ \Gamma_{y}\}_{y \in \mathcal{Y}}$ are chosen to be representative points within the $K$ distinct phases of the system. As such, scheme 1 already allows for flexibility in choosing points that are representative of a given label. While additional flexibility in the weighting of these points can be introduced, without additional system information there is no compelling reason to do so.
	
	\begin{figure}[bth!]
		\centering
		\includegraphics[width=0.9\linewidth]{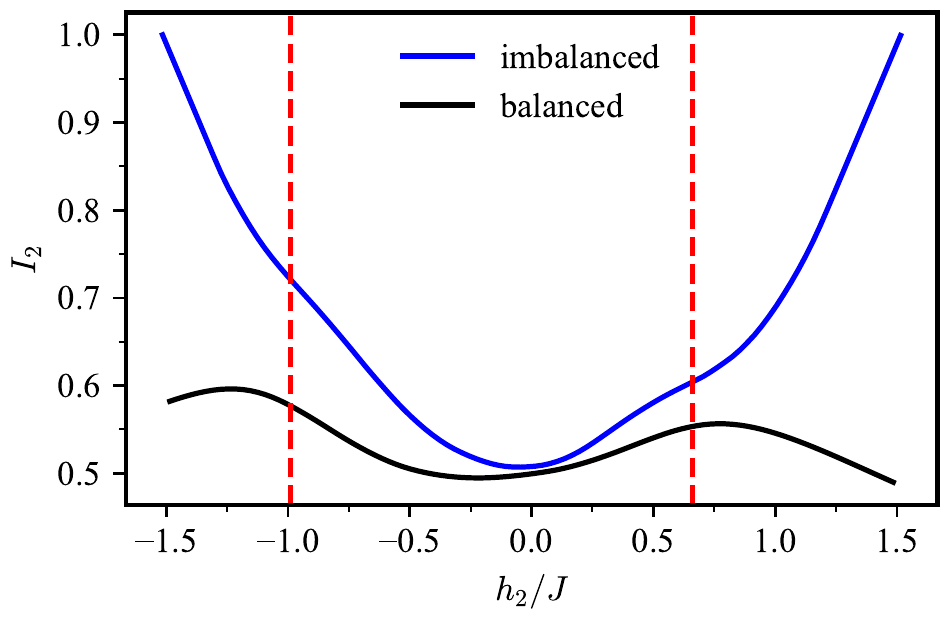}
		\caption{Results for the cluster-Ising model [Eq.~(9) in the main text, $L=7$] at $h_{1}/J = 0.2$ measured using a Pauli-6 POVM. The set $\Gamma$ is composed of a uniform grid with 101 points. Indicator $I_{2}(\gamma)$ with $l = 101$ (i.e., considering global bipartitions of the one-dimensional parameter space) and in the presence (blue) or absence (black) of class imbalance. Estimated critical points (red dashed lines) are determined from maxima in $|\partial^2 \langle H \rangle_{\bm{\gamma}} /\partial \gamma_{2}^2|$. Here, the expected value involved in $I_{2}(\gamma)$ is computed exactly and the ground state is obtained via exact diagonalization.}
		\label{fig:LBC_imbalance}
	\end{figure}
	
	\subsection{Class imbalance}\label{sec_class_imb}
	In traditional discriminative machine learning (ML) tasks, class imbalance occurs when the number of samples representing each class within the data set differs. In phase-classification tasks, this can, for example, occur when one does not have access to an equal amount of data at each point in parameter space. Moreover, in scheme 1 (with indicator $I_{1}$) and 2 (with indicator $I_{2}$) discussed in the main text, an imbalance can also occur when sampling an \emph{equal} amount of data at each point in parameter space. In both cases, the imbalance arises because the sets $\{ \Gamma_{y}\}_{y \in \mathcal{Y}}$ are not of equal size. In scheme 1, this can be circumvented by choosing an equal number of points in parameter space to represent each phase. In scheme 2, data sets of unequal size occur as the bipartitions artificially divide the uniformly sampled one-dimensional parameter space. In previous works~\cite{van:2017,liu:2018,beach:2018,suchsland:2018,lee:2019,kharkov:2020,guo:2020,greplova:2020,bohrdt:2021,richter:2022,gavreev:2022,zvyagintseva:2022,guo:2023,schlomer:2023}, this class imbalance was not accounted for. As a result, the corresponding indicator $I_{2}$ will exhibit trivial local maxima ($I_{2}=1$) at the edges of the sampled region of the parameter space where the entire data is given the same label. If no phase transition is present, the indicator will thus exhibit a characteristic V-shape. In the presence of a phase transition, a W-shape is expected, where the location of the intermediate maximum corresponds to the predicted critical point. Such behavior is undesirable for an indicator of phase transitions, and can even lead to a failure to detect certain transitions (or misdetection). In Ref.~\cite{bohrdt:2021}, an attempt has been made to correct this pathological behavior. However, \citet{arnold:2022} have found that this procedure biases the transition point towards the center of the parameter range under consideration.
	
	In this work, we addressed class imbalances arising from sets $\{ \Gamma_{y}\}_{y \in \mathcal{Y}}$ of unequal size by choosing a uniform prior distribution $P(y)$. In particular, this removes the trivial local maxima at the edges of the sampled region of the parameter space previously encountered for the indicator $I_{2}$ and we instead expect a single peak in the presence of a phase transition. Figure~\ref{fig:LBC_imbalance} shows an example of transitions that are not detected using scheme 2 in the presence of class imbalance highlighting the importance of our proposed modification.
	
	\begin{figure}[bth!]
		\centering
		\includegraphics[width=0.99\linewidth]{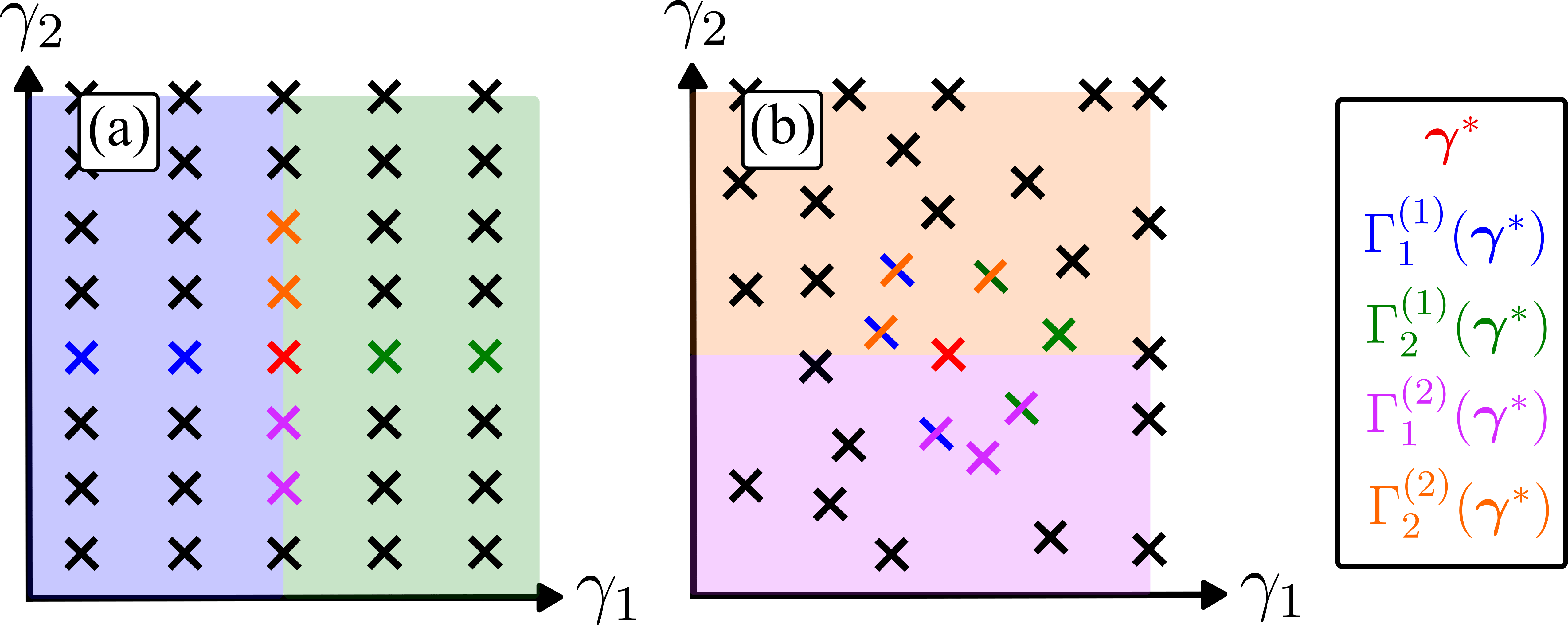}
		\caption{Illustration for constructing local bipartitions for scheme 2 [Eq. (5) in the main text] of a two-dimensional parameter space ($d=2$) that is sampled (a) uniformly on a grid ($l=2$) or (b) non-uniformly ($l=3$). Crosses denote sampled points in parameter space $\bm{\gamma} \in \Gamma$. Sampled points belonging to one of the four sets $\Gamma_{1}^{(1)}(\bm{\gamma}^*)$,  $\Gamma_{2}^{(1)}(\bm{\gamma}^*)$, $\Gamma_{1}^{(2)}(\bm{\gamma}^*)$,  $\Gamma_{2}^{(2)}(\bm{\gamma}^*)$ are colored blue, green, pink, and orange, respectively. The corresponding splits of the parameter space are denoted by shaded colored regions.}
		\label{fig:LBC_bipartitions}
	\end{figure}

	\subsection{Scheme 2: How to choose local sets of points}
	In scheme 2 [Eq. (5) in the main text], at each sampled point $\bm{\gamma}^* = \left( \gamma_{1}^*,\gamma_{2}^*,\dots, \gamma_{d}^* \right) \in \Gamma$, the parameter space is split along each direction. For a given direction $1 \leq i \leq d$, this yields two sets, $\Gamma_{1}^{(i)}(\bm{\gamma}^*)$ and $\Gamma_{2}^{(i)}(\bm{\gamma}^*)$, each comprised of the $l$ points closest to $\bm{\gamma}^*$ in part 1 and 2 of the split parameter space, respectively. Using different norms for measuring this distance results in different local sets. Figure~\ref{fig:LBC_bipartitions} illustrates two distinct strategies. In this work, we choose the $l$ points $\bm{\gamma}$ closest to $\bm{\gamma}^*$ in each part of the split parameter space with respect to $|\gamma_{i} - \gamma_{i}^*|$, see Fig.~\ref{fig:LBC_bipartitions}(a). This is suitable for parameter spaces sampled on a uniform grid. In the case of non-uniformly sampled parameter spaces, one may instead consider a Euclidean distance $\norm{\bm{\gamma} - \bm{\gamma}^*}_{2}$, see Fig.~\ref{fig:LBC_bipartitions}(b). Note that in both cases, sets with a numbers of points lower than $l$ may arise at the edges of the parameter space. The resulting class imbalance is corrected for, see Sec.~\ref{sec_class_imb}.
	
	\subsection{Scheme 3: Dividing by the standard deviation}
\begin{figure}[tbh!]
		\centering
		\includegraphics[width=0.99\linewidth]{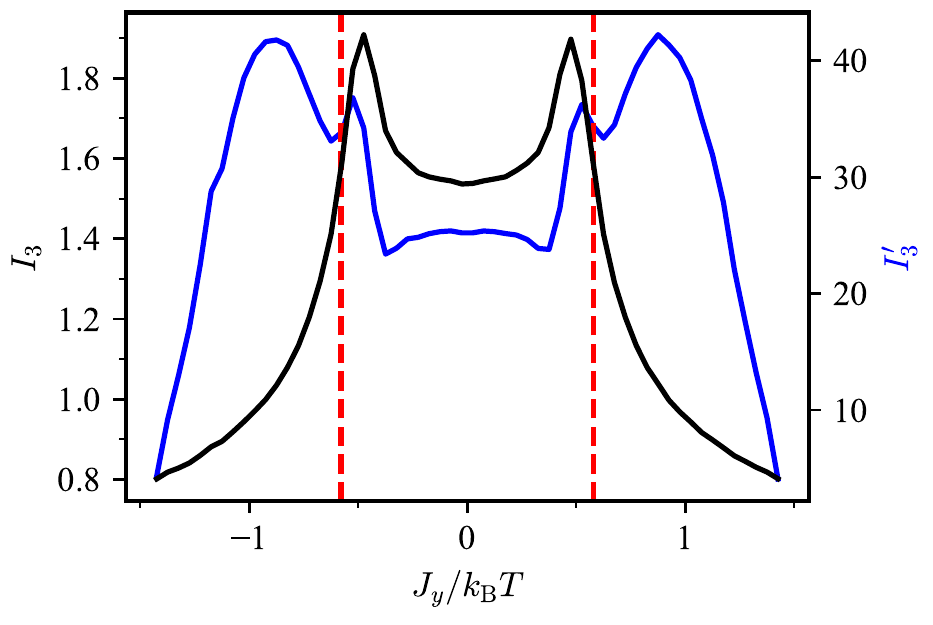}
		\caption{Results for the anisotropic Ising model on a square lattice ($L=20$) at $J_{x}/k_{\rm B}T = -0.325$. The indicator $I_{3}(\gamma)$ (black, left $y$-axis) and $I_{3}'(\gamma)$ (blue, right $y$-axis) are computed based on predictions $\hat{\bm{\gamma}}$ that are estimated element-wise from line scans across the entire two-dimensional phase diagram. The set $\Gamma$ is composed of a uniform grid with 60 points for each axis and $|\mathcal{D}_{\bm{\gamma}}| = 10^5 \; \forall \bm{\gamma} \in \Gamma$. Onsager's analytical solution for the phase boundaries is shown as a red dashed line.}
		\label{fig:fig_PBM_mod}
	\end{figure}

	For the third phase-classification scheme discussed in the main text (with indicator $I_{3}$), we have proposed to divide the corresponding signal by its standard deviation. We find this modified indicator to be superior compared to its unmodified version
	\begin{align*}\label{eq:IPBM}
		I_{3}'(\bm{\gamma}) &= \sqrt{\sum_{i=1}^{d} \left(\frac{\partial \hat{\gamma}_{i}(\bm{\gamma})}{\partial \gamma_{i}}\right)^2},\\
		&=  \norm{ \mathbb{E}_{\bm{x} \sim P(\bm{x}|\bm{\gamma})}\left[\hat{\bm{\gamma}}(\bm{x}) \nabla_{\bm{\gamma}} \ln P(\bm{x}|\bm{\gamma}) \right]}_{2} \numberthis,
	\end{align*}
	which (up to a constant offset) has been considered in previous studies~\cite{schaefer:2019,greplova:2020,arnold:2021,arnold:2022}. As an example, Fig.~\ref{fig:fig_PBM_mod} compares the two indicators for the anisotropic Ising model. The indicator $I_{3}'$ shows four distinct peaks with the dominant two occurring within the two ordered phases (i.e., at larger $|J_{y}/k_{\rm B}T|$). Such spurious signals have also been observed in previous studies using indicators where the standard deviation is not taken into account~\cite{schaefer:2019,arnold:2022}. In contrast, the modified indicator $I_{3}$ we propose in this work shows two distinct peaks that qualitatively agree with the two critical points.
	
	\subsection{Discriminative cooperative networks}
	\indent In Ref.~\cite{liu:2018}, a scheme for mapping out phase diagrams in two-dimensional parameter spaces was proposed. Instead of a brute-force search of the entire parameter space for all possible phase boundaries, the predicted phase boundary is modeled as a parametrized curve using an active contour model called \emph{snake}~\cite{kass:1988}. This snake partitions the parameter space locally and is driven via internal forces that, e.g., prevent bending and stretching, as well as external forces aiming to minimize the overall classification error. In Ref.~\cite{liu:2018}, the external force is generated in an interplay between a guesser network and learner network, together called \emph{discriminative cooperative networks}, that are optimized via a joint cost function. The guesser provides labels for the data which the learner should reproduce. In each step, the guesser tries to provide a better set of labels based on the predictions of the learner to cooperatively minimize the classification error (i.e., the corresponding loss).
	
	\indent Let us first consider a one-dimensional parameter space. Following Ref.~\cite{liu:2018}, we use the following sigmoidal guesser
	\begin{equation}
		\tilde{P}_{\bm{\theta}_{G}}(1|\gamma) = \frac{1}{1+e^{(\gamma_{G} - \gamma)/\sigma_{G}}},
	\end{equation}
	where $\tilde{P}_{\bm{\theta}_{G}}(2|\gamma) = 1 - \tilde{P}_{\bm{\theta}_{G}}(1|\gamma)$. The guesser is characterized by two parameters $\bm{\theta}_{G}=(\gamma_{G},\sigma_{G})$, where $\gamma_{G}$ corresponds to the guessed transition point and $\sigma_{G}$ determines the sharpness of the transition. The learner is given by $\tilde{P}_{\bm{\theta}_{L}}(y|\bm{x})$. The guesser and learner are optimized jointly using a cross-entropy loss function
	\begin{widetext}
		\begin{equation}\label{eq:DCN}
			\mathcal{L}(\bm{\theta}) = -  \frac{1}{|\mathcal{Y}|}\sum_{y \in \mathcal{Y}} \frac{1}{|\Gamma_{y}|}\sum_{\gamma \in \Gamma_{y}}\frac{1}{|\mathcal{D}_{\gamma}|}\sum_{\bm{x} \in \mathcal{D}_{\gamma}} \tilde{P}_{\bm{\theta}_{G}}(y|\gamma) \log\left(\tilde{P}_{\bm{\theta}_{L}}(y|\bm{x})\right),
		\end{equation}
	\end{widetext}
	where $\bm{\theta} = \left( {\bm{\theta}_{G}}, {\bm{\theta}_{L}}\right)$. The parameters of the two networks can now be optimized using gradient descent on the loss function in Eq.~\eqref{eq:DCN}. This corresponds to a discriminative approach and has been used in Ref.~\cite{liu:2018}.
	
\begin{figure*}[htb!]
		\begin{center}
			\includegraphics[width=0.9\linewidth]{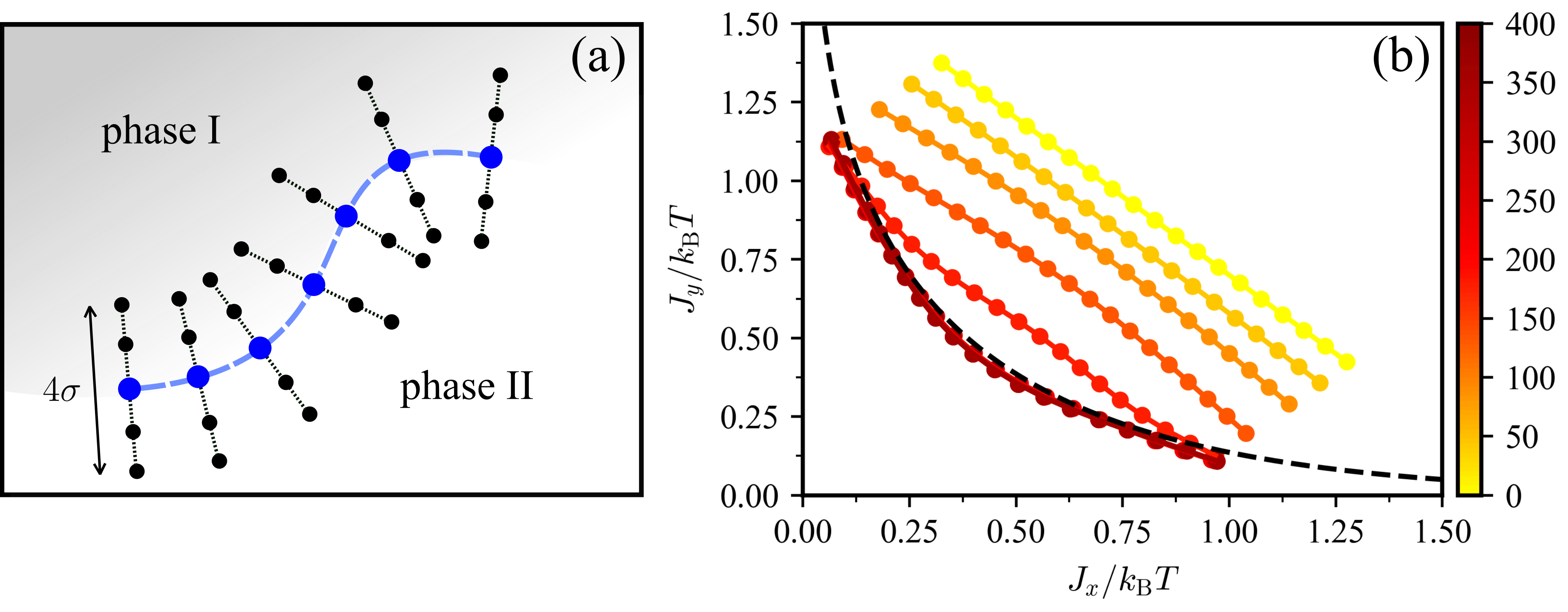}
			\caption{(a) Schematic illustration of the snake model. The blue circles represent the snake nodes (here 7). The normal (sensing) direction at each node is shown as a black-dashed line and the relevant $2l$ sampled points within a distance $2l \sigma$ are shown as black circles (here $l=2$). (b) Snake boundary for the anisotropic Ising model on a square lattice ($L=20$; $J_{x}/k_{\rm B}T,J_{y}/k_{\rm B}T\geq 0$) obtained using a generative classifier where the color bar denotes the training epoch. The snake consists of 20 nodes with $l=4$ and hyperparameters $\alpha = 0.002$, $\beta = 0.4$, $\kappa = 0.9$, and learning rates of $10^{-4}$ and $5 \times 10^{-4}$ associated with $E_{\rm int}$ and $E_{\rm ext}$, respectively. The set $\Gamma$ is composed of a uniform grid with 30 points for each axis and $|\mathcal{D}_{\bm{\gamma}}| = 10^5 \; \forall \bm{\gamma} \in \Gamma$. Onsager's analytical solution for the phase boundary is shown as a black-dashed line.
			}
			\label{fig:DCNs_LBC}
		\end{center}
	\end{figure*}

	\indent Moving to two-dimensional parameter spaces, one can utilize the snake as a parametrization for the guesser which is a discretized curve of linked nodes, $\bm{r}(s)=\left(\gamma_{1}(s),\gamma_{2}(s)\right)$, parametrized by $s \in [0,1]$ (assuming an open snake), see Fig.~\ref{fig:DCNs_LBC}(a). The snake moves in order to minimize its total energy $E_{\rm tot}=E_{\rm int} + E_{\rm ext}$. The internal energy
	\begin{equation}
		E_{\rm int} = \int_0^1\left(\alpha \norm{\frac{\partial \bm{r}}{\partial s}}_{2}^2+\beta\norm{\frac{\partial^2 \bm{r}}{\partial s^2}}_{2}^2\right) ds
	\end{equation}
	is introduced to make the snake smoother, where the hyperparameter $\alpha$ penalizes the stretching of the snake and $\beta$ penalizes its bending. The snake can sense its surroundings at each node through $2l$ sampled points perpendicular to the snake within a distance $l \sigma$ [see Fig.~\ref{fig:DCNs_LBC}(b)]. The overall guesser function is comprised of local guesser functions evaluated at each node that sense the direction perpendicular to the snake. The external energy $E_{\rm ext}$ of the snake corresponds to the overall loss obtained by summing the cross-entropy losses of all individual one-dimensional guessers [cf. Eq.~\eqref{eq:DCN}] and gives rise to an external force $-\delta E_{\rm ext}/\delta \mathbf{r}$ pointing perpendicular to the snake at each node.
	
	\indent In this work, we replace the learner network at each node by a corresponding generative classifier which makes updating the learner in each step obsolete. To construct the generative classifier, we derive the (empirically) optimal predictor based on the loss in Eq.~\eqref{eq:DCN}. Following the procedure outline in Sec.~\ref{sec:discr_indicators}, we obtain 
	\begin{equation}
		P_{\rm emp}(y|\bm{x}) = \frac{\sum_{\gamma \in \Gamma_{y}} \tilde{P}_{\bm{\theta}_{G}}(y|\gamma) \frac{M_{\gamma}(\bm{x})}{|\mathcal{D}_{\gamma}|} \frac{1}{|\mathcal{Y}|}\frac{1}{|\Gamma_{y}|}}{\sum_{y' \in \mathcal{Y}} \sum_{\gamma' \in \Gamma_{y'}} \tilde{P}_{\bm{\theta}_{G}}(y'|\gamma') \frac{M_{\gamma'}(\bm{x})}{|\mathcal{D}_{\bm{\gamma}'}|} \frac{1}{|\mathcal{Y}|}\frac{1}{|\Gamma_{y'}|}}.
	\end{equation}
	In the infinite-data limit, this converges to 
	\begin{equation}\label{eq:DCN_gen}
		P(y|\bm{x}) = \frac{\sum_{\gamma \in \Gamma_{y}} \tilde{P}_{\bm{\theta}_{G}}(y|\gamma)P(\bm{x}|\gamma)P(y)P(\gamma|y)}{\sum_{y'\in \mathcal{Y}} \sum_{\gamma' \in \Gamma_{y'}}\tilde{P}_{\bm{\theta}_{G}}(y'|\gamma')P(\bm{x}|\gamma')P(y)P(\gamma'|y')},
	\end{equation}
	with a uniform prior over the classes $P(y) = 1/|\mathcal{Y}|\; \forall y \in \mathcal{Y}$ and $P(\gamma|y) = 1/|\Gamma_{y}|$ if $\gamma \in \Gamma_{y}$ and zero otherwise. Thus, we can obtain a generative classifier by modeling $P(\bm{x}|\gamma)$ in Eq.~\eqref{eq:DCN_gen}. Note that the expression in Eq.~\eqref{eq:SI_optt} for binary labels is recovered as a special case.\\
	
	In our implementation, we fix the width parameter $\sigma_{G}$ of the sigmoid guesser to be $\sigma_{G} = \sigma/10$, where $\sigma$ is the width parameter of the snake. We reduce $\sigma$ exponentially during the optimization according to 
	\begin{equation}
		\sigma_{k} = \sigma_{\rm end}+(\sigma_{\rm start}-\sigma_{\rm end}) \kappa^{k},
	\end{equation}
	where $k$ denotes the current epoch, $\sigma_{\rm start/end}$ are the start and end values, and $\kappa$ is the decay rate. We set $\sigma_{\rm end}=\Delta \gamma$, where $\Delta \gamma$ is the spacing between neighboring sampled points in parameter space and $\sigma_{\rm start} = 5 \sigma_{\rm end}$. This choice yields a strong gradient signal early on that becomes weaker but more accurate at later stages, which facilitates the convergence of the snake to the true underlying phase boundary. We construct models for $P(\bm{x}|\bm{\gamma})$ at points $\bm{\gamma}$ not contained within the initial sampled set $\Gamma$ from $\{  \tilde{P}(\bm{x}|\bm{\gamma})\}_{\bm{\gamma} \in \Gamma}$ using bilinear interpolation. Gradients of the energy with respect to the node positions are calculated using finite differencing. We minimize the snake's total energy using gradient-based optimization with Adam~\cite{kingma:2014}.\\ 
	
	The results obtained using this scheme with a generative learner for the anisotropic Ising model are shown in Fig.~\ref{fig:DCNs_LBC}(b). While the snake eventually finds the underlying phase boundary, there are several downsides to this scheme. First, the number of unique classification tasks that need to be solved over the course of the training is $N_{\rm epochs} \times N_{\rm nodes} = 400 \times 20 = 8000$. This is larger than the number of unique classification tasks that need to be solved within scheme 2 proposed in the main text which is given by $|\Gamma| \times 2 = 900 \times 2 = 1800$. Therefore, there is no gain in computation time compared to a brute-force search of the parameter space (as performed within scheme 2). The snake scheme also has a multitude of hyperparameters: $\alpha$, $\beta$, $\sigma_{\rm start}$, $\kappa$, $l$, learning rates, as well as the initial snake positioning and snake topology (e.g., whether the snake is open or closed and whether certain nodes should remain fixed throughout training). Thus, to run the snake scheme one has to perform hyperparameter tuning. This involves additional computational effort and/or prior knowledge of the underlying phase diagram. In particular, we find the results to depend heavily on $\alpha$, $\beta$, as well as the learning rates and initial position of the snake. Moreover, the scheme is not reliable in the presence of more than two phases as the snake will typically converge to one of the phase boundaries, missing the remaining ones [see Fig.~\ref{fig:DCNs_LBC}(b)]. To get around this, the scheme must be run multiple times with different snake initializations (possibly guided by prior knowledge of the phase diagram).\\  
	
	\section{Discriminative approach to mapping out phase diagrams}\label{sec:discr_indicators}
	The computation of all three indicators of phase transitions [Eqs. (3), (5), and (7) in the main text] boils down to solving classification tasks. In the main text, we discuss how such tasks can be solved in a generative manner. Here, we discuss how they can be solved in a discriminative manner. In this case, we look for the parameters $\bm{\theta}$ of a parametric model $\tilde{P}_{\bm{\theta}}(y|\bm{x})$ that minimize the following cross-entropy loss function
	\begin{equation}\label{eq:SI_x1}
		\mathcal{L}(\bm{\theta}) = - \frac{1}{|\mathcal{Y}|} \sum_{y \in \mathcal{Y}} \frac{1}{|\mathcal{D}_{y}|} \sum_{\bm{x} \in \mathcal{D}_{y}}  {\rm ln}\left(\tilde{P}_{\bm{\theta}}(y|\bm{x})\right),
	\end{equation}
	where $\mathcal{D}_{y} = \{ \bm{x} \in \mathcal{D}_{\bm{\gamma}} |\bm{\gamma} \in \Gamma_{y}\}$ is the relevant data set drawn from $P(\bm{x}|y)=1/|\Gamma_{y}| \sum_{\bm{\gamma} \in \Gamma_{y}} P(\bm{x}|\bm{\gamma})$ and $\mathcal{D}_{\bm{\gamma}}$ corresponding to a set of samples drawn from $P(\bm{x}|\bm{\gamma})$. Here, $|\mathcal{D}_{\bm{\gamma}}|$ is the same for all $\bm{\gamma} \in \Gamma_{y}$ (corresponding to the choice $P(\bm{\gamma}|y) = 1 / |\Gamma_{y}|$ for $\bm{\gamma}\in \Gamma_{y}$ and zero otherwise). Class imbalance is addressed by the rescaling factors $1/|\mathcal{D}_{y}|$ as is common in neural-network training~\cite{paszke:2019}. 
	
	Based on the trained parametric model, in scheme 1 we can estimate the central quantity of interest $P(y|\bm{\gamma})$ as 
	\begin{equation}\label{eq:SL_1}
		P(y|\bm{\gamma}) \approx \frac{1}{|\mathcal{D}_{\bm{\gamma}}|}\sum_{\bm{x} \in \mathcal{D}_{\bm{\gamma}}} \tilde{P}_{\bm{\theta}}(y|\bm{x})
	\end{equation}
	In scheme 2, the relevant quantity $p_{\rm err}$ can, for example, be estimated as
	\begin{equation}\label{eq:LBC_err_SI}
		p_{\rm err} \approx \frac{1}{2}\sum_{y \in \{1,2 \}} \frac{1}{|\mathcal{D}_{y}|}\sum_{\bm{x} \in \mathcal{D}_{y}} \left|y- \argmax_{y'} \tilde{P}_{\bm{\theta}}(y'|\bm{x})\right|.
	\end{equation}
	In scheme 3, the key quantity is $\hat{\bm{\gamma}}(\bm{x})$ which can be estimated as
	\begin{equation}\label{eq:PBM_pred_SI}
		\hat{\bm{\gamma}}(\bm{x}) \approx \sum_{y \in \mathcal{Y}} \tilde{P}_{\bm{\theta}}(y|\bm{x})\bm{\gamma}_{y}
	\end{equation}
	Alternatively, it may also be approximated directly using a parametric predictive model $\hat{\bm{\gamma}}_{\bm{\theta}}(\bm{x})$ that is trained to solve a regression task instead of a classification task (as has been done in previous works~\cite{schaefer:2019,greplova:2020,arnold:2021,arnold:2022}). In this case, the relevant loss function is
	\begin{equation}\label{eq:SI_x2}
		\mathcal{L}(\bm{\theta}) = \frac{1}{|\mathcal{Y}|} \sum_{y \in \mathcal{Y}} \frac{1}{|\mathcal{D}_{y}|} \sum_{\bm{x} \in \mathcal{D}_{y}}  \norm{\hat{\bm{\gamma}}_{\bm{\theta}}(\bm{x}) - \bm{\gamma}_{y}}_{2}^2.
	\end{equation}
	
	Above we have introduced the loss functions that are used to optimize parametric predictive models in the discriminative approach. In the following, we prove that in the infinite-data limit, an optimal predictive model, i.e., a model that minimizes the corresponding loss functions, yields Bayes optimal predictions. Analyzing its predictions yields the indicator signals discussed in the main text. Moreover, we show that an optimal discriminative classifier yields the same predictions as a generative classifier with $\tilde{P}(\bm{x}|\bm{\gamma})$ given by the empirical distribution obtained from the dataset $\mathcal{D}_{\bm{\gamma}}$.\\
	
	\indent For a given sample $\bm{x} \in \{x \in \mathcal{D}_{y}|y \in \mathcal{Y} \}$, we can determine the corresponding empirically optimal model prediction $P_{\rm emp}(y|\bm{x})$ by minimizing the loss function in Eq.~\eqref{eq:SI_x1} with respect to $\tilde{P}_{\bm{\theta}}(y|\bm{x})$ subjected to the equality constraint $\sum_{y\in\mathcal{Y}} \tilde{P}_{\bm{\theta}}(y|\bm{x}) - 1 = 0$. Using the method of Lagrange multipliers, the stationary points satisfy the following conditions
	\begin{equation}
		\lambda = \frac{1}{|\mathcal{Y}|} \frac{M_{y}(\bm{x})}{|\mathcal{D}_{y}|} \frac{1}{P_{\rm emp}(y|\bm{x})}\; \forall y \in \mathcal{Y},
	\end{equation}
	where $M_{y}(\bm{x})$ denotes the number of times $\bm{x}$ appears in $\mathcal{D}_{y}$ and $\lambda$ is the corresponding Lagrange multiplier. Together with the equality constraint, this yields
	\begin{equation}
		P_{\rm emp}(y|\bm{x}) = \frac{\frac{M_{y}(\bm{x})}{|\mathcal{D}_{y}|} \frac{1}{|\mathcal{Y}|}}{\sum_{y' \in \mathcal{Y}} \frac{M_{y'}(\bm{x})}{|\mathcal{D}_{y'}|} \frac{1}{|\mathcal{Y}|}}.
	\end{equation}
	Identifying the empirical distribution $P_{\rm emp}(\bm{x}|y) = M_{y}(\bm{x})/|\mathcal{D}_{y}|$ as well as the uniform prior over the classes $P(y) = 1/|\mathcal{Y}|\; \forall y \in \mathcal{Y}$, we have
	\begin{equation}\label{eq:SI_emp}
		P_{\rm emp}(y|\bm{x}) = \frac{P_{\rm emp}(\bm{x}|y)P(y)}{\sum_{y' \in \mathcal{Y}} P_{\rm emp}(\bm{x}|y')P(y')}.
	\end{equation}
	In the infinite-data limit $|\mathcal{D}_{y}| \rightarrow \infty$, we have $P_{\rm emp}(\bm{x}|y) \rightarrow P(\bm{x}|y)$. Thus, the predictions of the empirically optimal model converge to the predictions of a Bayes optimal model
	\begin{equation}\label{eq:SI_optt}
		P_{\rm emp}(y|\bm{x}) \rightarrow P(y|\bm{x}) = \frac{P(\bm{x}|y)P(y)}{\sum_{y' \in \mathcal{Y}} P(\bm{x}|y')P(y')},
	\end{equation}
	and we have recovered Eq.~(1) of the main text. Moreover, based on these optimal predictions, we obtain the optimal indicator of scheme 1 [Eq.~(3) of the main text]. To recover the optimal indicator of scheme 2, it remains to be shown that the estimated error rate in Eq.~\eqref{eq:LBC_err_SI} converges to the error rate as defined in Eq.~(4) of the main text in the case of an optimal discriminative model and infinite data. In this limit, Eq.~\eqref{eq:LBC_err_SI} transforms to 
	\begin{align*}\label{eq:perr_deriv}
		p_{\rm err} &= \frac{1}{2} \sum_{y \in \{1,2\}} \sum_{\bm{x}\in \mathcal{X}} P(\bm{x}|y)\left|y- \argmax_{y'} P(y'|\bm{x})\right|\\
		&= \frac{1}{2} \sum_{\substack{\bm{x}\in \mathcal{X}\\
				P(\bm{x}|1) \geq P(\bm{x}|2)}} P(\bm{x}|2) + \frac{1}{2} \sum_{\substack{\bm{x}\in \mathcal{X}\\
				P(\bm{x}|2) > P(\bm{x}|1)}} P(\bm{x}|1)\\
		&= \frac{1}{2} \left( 1- \frac{1}{2}\sum_{\bm{x}\in \mathcal{X}} \left|P(\bm{x}|1) - P(\bm{x}|2)\right|\right)\numberthis.
	\end{align*}
	For the last step, we use the fact that for any two probability distributions $p$ and $q$ over a discrete variable $x$, we have
	\begin{equation}
		\frac{1}{2}\sum_{x} \left|p(x)-q(x)\right| = 1-\sum_{\substack{x\\ q(x) > p(x)}} p(x) -\sum_{\substack{x\\ p(x) \geq q(x)}} q(x)\numberthis.
	\end{equation}
	Continuing with Eq.~\eqref{eq:perr_deriv}, we have
	\begin{align*}\label{eq:perr_deriv_2}
		p_{\rm err} &= \frac{1}{2} \left( \sum_{\bm{x}\in \mathcal{X}} P(\bm{x}) - \sum_{\bm{x}\in \mathcal{X}}P(\bm{x}) \left|P(1|\bm{x}) - P(2|\bm{x})\right|\right)\\
		&= \sum_{\bm{x}\in \mathcal{X}} P(\bm{x})\left( \frac{1}{2}\left( 1-  \left|P(1|\bm{x}) - P(2|\bm{x})\right| \right)\right)\\
		&= \sum_{\bm{x}\in \mathcal{X}}P(\bm{x}) \min\{P(1|\bm{x}), P(2|\bm{x})\},\numberthis
	\end{align*}
	where in step 2 we used Bayes' theorem $P(\bm{x}|y)/P(\bm{x})=P(y|\bm{x})/P(y)$ with $P(y)=1/2$. Plugging in the definition of
	\begin{align*}
		P(\bm{x}) &= \sum_{y \in \{ 1,2\}}P(\bm{x}|y)P(y) = \frac{1}{2}\sum_{y \in \{ 1,2\}} \sum_{\bm{\gamma} \in \Gamma }P(\bm{x}|\bm{\gamma})P(\bm{\gamma}|y)\\
		&= \frac{1}{2} \sum_{y \in \{ 1,2\}}\frac{1}{|\Gamma_{y}|} \sum_{\bm{\gamma} \in \Gamma_{y} }P(\bm{x}|\bm{\gamma})\numberthis
	\end{align*}
	in Eq.~\eqref{eq:perr_deriv_2} and defining $p_{\rm err}(\bm{x}) = \min\{P(1|\bm{x}), P(2|\bm{x})\}$ we recover the error rate in Eq.~(4) of the main text.\\
	
	\indent Following a similar minimization procedure for the loss function in Eq.~\eqref{eq:SI_x2}, we obtain
	\begin{align*}
		\hat{\bm{\gamma}}_{\rm emp}(\bm{x}) &= \sum_{y \in \mathcal{Y}} \frac{ \frac{M_{y}(\bm{x})}{|\mathcal{D}_{y}|} \frac{1}{|\mathcal{Y}|}}{\sum_{y' \in \mathcal{Y}} \frac{M_{y'}(\bm{x})}{|\mathcal{D}_{y'}|} \frac{1}{|\mathcal{Y}|}} \bm{\gamma}_{y}\\
		&= \sum_{y \in \mathcal{Y}} \frac{ P_{\rm emp}(\bm{x}|y) P(y)}{\sum_{y' \in \mathcal{Y}} P_{\rm emp}(\bm{x}|y') P(y')} \bm{\gamma}_{y},\numberthis
	\end{align*}
	which in the infinite-data limit converges to 
	\begin{equation}
		\hat{\bm{\gamma}}(\bm{x}) = \sum_{y\in \mathcal{Y}} \frac{P(\bm{x}'|y)P(y)}{\sum_{y'\in \mathcal{Y}} P(\bm{x}'|y')P(y')} \bm{\gamma}_{y} = \sum_{y\in \mathcal{Y}} P(y|\bm{x}) \bm{\gamma}_{y},
	\end{equation}
	corresponding to Eq.~(6) in the main text. Based on these predictions, we directly recover the indicator of scheme 3 [Eq.~(7) of the main text].\\
	
	\begin{figure*}[tbh!]
		\begin{center}
			\includegraphics[width=0.8\textwidth]{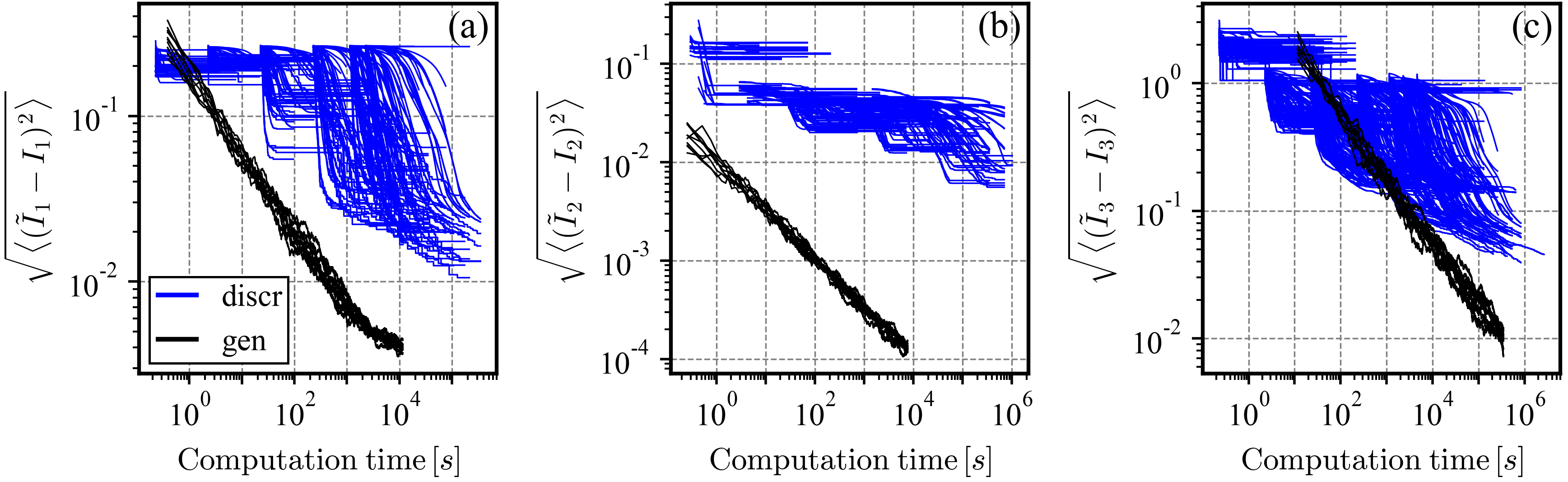}
			\caption{Root-mean-square error (RMSE) of estimated indicator of phase transitions $\tilde{I}$ as a function of the computation time associated with the discriminative (blue) and generative approach (black) to (a) scheme 1, (b) scheme 2, and (c) scheme 3 for the cluster-Ising model [Eq.~(9) in the main text, $L=7$] at $h_{1}/J = 0.2$. The total computation time is comprised of the time associated with data generation, training, and computation of the indicator based on the predictions, i.e., $\tilde{P}(y|\bm{x})$. For the data generation and construction of generative classifiers, we consider generative models based on matrix product states optimized via the density matrix renormalization group algorithm. The indicators $\tilde{I}$ are estimated from datasets $\{ \mathcal{D}_{\bm{\gamma}} \}_{\bm{\gamma} \in \Gamma}$ of various size. The reference indicators $I$ are computed exactly based on ground states obtained via exact diagonalization. The set $\Gamma$ is composed of a uniform grid with 101 points. In scheme 1, $\Gamma_{1} = \{ -1.5 \; h_{2}/J \}$, $\Gamma_{2} = \{ -0.03 \; h_{2}/J \}$, and $\Gamma_{3} = \{ 1.47 \; h_{2}/J \}$. In scheme 2, we consider $l=2$. The discriminative approach to scheme 3 is performed according to Eq.~\eqref{eq:SI_x2}. For the generative approach, 10 independent runs with distinct datasets are shown. For the discriminative approach, we considered feedforward NNs of various sizes (ranging from a single hidden layer with 64 nodes to five hidden layers with 64 nodes each; containing rectified linear units as activation functions), different dataset sizes $|\mathcal{D}_{\bm{\gamma}}|$ (ranging from 10 to $2 \times 10^{5}$), learning rates (ranging from $5 \times 10^{-5}$ to $5 \times 10^{-1}$), and number of training epochs (ranging from 1 to $5 \times 10^{4}$). Each blue curve corresponds to a training run with a fixed choice of hyperparameters and we keep track of the best RMSE throughout training. The NNs are implemented using Flux~\cite{innes:2018} in \texttt{Julia}, where the weights and biases are optimized via gradient descent with Adam~\cite{kingma:2014}. Gradients are calculated using backpropagation~\cite{baydin:2018}. The NN inputs are composed of six integers per qubit encoding the POVM element as well as the corresponding measurement outcome and are standardized before training.}
			\label{fig:CI_timing}
		\end{center}
	\end{figure*}
	
	\indent Note that using
	\begin{align*}
		P_{\rm emp}(\bm{x}|y) &= \frac{1}{|\Gamma_{y}|} \sum_{\bm{\gamma} \in \Gamma_{y}} \frac{M_{\bm{\gamma}}(\bm{x})}{|\mathcal{D}_{\bm{\gamma}}|}\\
		&= \frac{1}{|\Gamma_{y}|} \sum_{\bm{\gamma} \in \Gamma_{y}} P_{\rm emp}(\bm{x}|\bm{\gamma})\numberthis,
	\end{align*}
	Eq.~\eqref{eq:SI_emp} coincides with Eq. (2) in the main text with $P(\bm{x}|\bm{\gamma})$ replaced by $P_{\rm emp}(\bm{x}|\bm{\gamma})$. That is, the predictions of an optimal discriminative model are equivalent to the ones of a generative model with $\tilde{P}(\bm{x}|\bm{\gamma}) = P_{\rm emp}(\bm{x}|\bm{\gamma})$.
	
	\section{Comparison between discriminative and generative approach}
	In this section, we compare the computational cost associated with the discriminative and generative approaches to mapping out phase diagrams. Let $t^{\rm gen}_{\rm eval}$ and $t^{\rm discr}_{\rm eval}$ be the times corresponding to evaluating the generative model $\tilde{P}(\bm{x}|\bm{\gamma})$ or discriminative model $\tilde{P}(y|\bm{x})$ for a given $\bm{x}$, respectively. These times depend heavily on the choice of model, implementation, and hardware. Here, the reported computation times were assessed in \texttt{Julia} (version 1.8.2) on a single 3.70 GHz Intel Xeon W-2135 CPU. For the non-parametric generative model of the anisotropic Ising model ($L=20$), the evaluation time is negligible $t^{\rm gen}_{\rm eval} \approx 40$ ns. In the case of the matrix product state-based generative models for the ground states of the cluster-Ising Hamiltonian, the evaluation time $t^{\rm gen}_{\rm eval}$ ranges from $\approx 96$~${\rm \mu s}$ for $L=7$ (on average across the sampled parameter space) to $\approx 3$~ms for $L=71$. For comparison, the evaluation time associated with a feedforward NN is $t^{\rm discr}_{\rm eval} \approx 7$~${\rm \mu s}$ in the case of a single hidden layer containing a single node and, for example, $t^{\rm discr}_{\rm eval} \approx 2$~ms in the case of five hidden layers containing 1024 nodes each (assuming only a single input and output node for simplicity).
	
	\indent The computation time of a single prediction via the generative approach, i.e., computation of $\tilde{P}_{\rm gen}(y|\bm{x}) \; \forall y \in \mathcal{Y}$ for a given sample $\bm{x}$ via Eq. (2) in the main text, scales as $t^{\rm gen}_{\rm pred} =  t^{\rm gen}_{\rm eval} | \Gamma_{\mathcal{Y}}| $, where $\Gamma_{\mathcal{Y}}= \{ \bm{\gamma} \in \Gamma_{y}| y \in \mathcal{Y}\}$. In the main text, in scheme 1 we have $| \Gamma_{\mathcal{Y}}| = K = 5$ and $3$, in scheme 2, $| \Gamma_{\mathcal{Y}}| = l = 1$, and in scheme 3, $| \Gamma_{\mathcal{Y}}| = 60$ and $101$, in the case of the anisotropic Ising and cluster-Ising model, respectively. The time associated with computing $\tilde{P}_{\rm discr}(y|\bm{x})$ via the discriminative approach is $t^{\rm discr}_{\rm pred} = t^{\rm discr}_{\rm eval}$. Given a model $\tilde{P}(y|\bm{x})$ (either discriminative or generative), the time associated with computing an indicator $I(\bm{\gamma}) \; \forall \bm{\gamma} \in \Gamma$ scales as $t_{\rm pred} |\Gamma| |\mathcal{D}_{\bm{\gamma}}|$ in case of scheme 1 and 3, and $t_{\rm pred} |\Gamma| |\mathcal{D}_{\bm{\gamma}}||\Gamma_{y}|$ in case of scheme 2. Based on this, for the anisotropic Ising model, the generative approach yields a speedup in the indicator computation of \textit{at least} (assuming a feedforward NN of minimal size as the discriminative model) a factor of $ t_{\rm pred}^{\rm discr}/t_{\rm pred}^{\rm gen} \approx 3$, $35$, and $87$ in the case of scheme 3, 1, and 2, respectively.
	
	\indent In scenarios where the generative model acts as a data source, such as in the two cases considered in the main text, the discriminative approach is guaranteed to have an overhead with respect to the generative approach due to training. We can estimate this overhead, i.e., the computation time required until the first evaluation of $\tilde{P}_{\rm discr}(y|\bm{x})$, to scale as $t^{\rm discr}_{\rm eval} | \Gamma_{\mathcal{Y}}| |\mathcal{D}_{\bm{\gamma}}|N_{\rm epochs}$. Note that to predict the indicator in scheme 2, $d |\Gamma|$ such discriminative models need to be trained (one for each bipartition at each point in parameter space), making the overhead scale accordingly. The number of training epochs is denoted by $N_{\rm epochs}$. In practice, $N_{\rm epochs}$ may range from $\mathcal{O}(10^2) - \mathcal{O}(10^4)$ and above. In cases where $\tilde{P}(\bm{x}|\bm{\gamma}) \approx P(\bm{x}|\bm{\gamma})$, for the discriminative approach to yield a predictive model of comparable quality, one expects it to be highly expressive (resulting in a high $t^{\rm discr}_{\rm eval}$), trained for a long time, and with a large dataset, resulting in a large overhead overall. The requirement of a larger dataset comes with an additional cost (with respect to the generative approach) associated with sample generation. 
	
	Because of this overhead, the generative approach can be more computationally efficient than the discriminative approach \textit{even if} $t_{\rm pred}^{\rm gen} > t_{\rm pred}^{\rm discr}$. This is illustrated in Fig.~\ref{fig:CI_timing} for the case of the cluster-Ising model. For scheme 2 [Fig.~\ref{fig:CI_timing}(b)], the overhead of the discriminative approach is so large that the generative approach is more efficient irrespective of the error tolerance. In particular, in scheme 2, for small $l$ it is expected to be difficult to achieve accurate results using the discriminative approach with a small dataset, because of the small differences in the underlying distributions. For schemes 1 and 3 [Fig.~\ref{fig:CI_timing}(a) and (c)], there exists a crossover point in terms of the error of the estimated indicator below which the generative approach is computationally more efficient. Note that hyperparameter tuning is required for the discriminative approach to yield accurate estimates of the indicator, which is not accounted for in the overall computation time. In contrast, using the generative approach, the estimated indicator can be systematically improved by sampling more data.
	
	\section{Mapping out phase diagrams of classical equilibrium systems}
	\subsection{Optimal lossless compression of state space}
	In this section, we prove that to map out phase diagrams of a large class of classical equilibrium systems by computing one of the three indicators discussed in the main text, it suffices to model $\{ P(\bm{X}|\bm{\gamma})\}_{\bm{\gamma} \in \Gamma}$, where $\bm{X}$ is the corresponding sufficient statistic.
	
	Consider probability distributions of the exponential family
	\begin{equation}\label{eq:SM_exp_fam}
		P_{\rm exp}(\bm{x}|\bm{\gamma}) = h(\bm{x})\exp(\sum_{i=1}^{d} \eta_{i}(\bm{\gamma}) X_{i}(\bm{x})-A(\bm{\gamma})),
	\end{equation}
	where $h(\bm{x})\geq 0$ is the carrier measure, $\bm{X}(\bm{x})$ is a sufficient statistic, $\bm{\eta}(\bm{\gamma})$ are the natural parameters, and $A(\bm{\gamma})$ is the log-partition function. The statistic $\bm{X}(\bm{x})$ is a \emph{minimal} sufficient statistic for $\bm{\gamma}$ if the set of allowed natural parameters $\bm{\eta}(\bm{\gamma})$ spans a $d$-dimensional space~\cite{casella:2002}. The distribution over the sufficient statistic is given by
	\begin{align*}
		P_{\rm exp}(\bm{X}'|\bm{\gamma}) &= \sum_{\bm{x} \in \mathcal{X}} P_{\rm exp}(\bm{x}|\bm{\gamma}) \delta(\bm{X}(\bm{x}) - \bm{X}')\\
		&= g(\bm{X}')\exp(\bm{\eta}(\bm{\gamma}) \cdot \bm{X}' - A(\bm{\gamma}))\numberthis,
	\end{align*}
	where $g(\bm{X}') = \sum_{\bm{x} \in \mathcal{X}} h(\bm{x})\delta(\bm{X}(\bm{x}) - \bm{X}')$. We have 
	\begin{equation}\label{eq:SM_der}
		P_{\rm exp}(\bm{x}|\bm{\gamma})/P_{\rm exp}(\bm{X}(\bm{x})|\bm{\gamma})=h(\bm{x})/g(\bm{X}(\bm{x})).
	\end{equation}
	\begin{widetext}
		Crucially, this ratio is independent of $\bm{\gamma}$. Using Eq.~\eqref{eq:SM_der}, the conditional probability of a label in phase-classification tasks [Eq.~(2) in the main text] can be written as
		\begin{equation}\label{eq:Bayes_2_SM}
			P(y|\bm{x}) =  \frac{\frac{1}{|\Gamma_{y}|}\sum_{\bm{\gamma} \in \Gamma_{y}} P_{\rm exp}(\bm{x}|\bm{\gamma})}{\sum_{y' \in \mathcal{Y}}\frac{1}{|\Gamma_{y'}|}\sum_{\bm{\gamma}' \in \Gamma_{y'}} P_{\rm exp}(\bm{x}|\bm{\gamma}')}
			=  \frac{\frac{1}{|\Gamma_{y}|}\sum_{\bm{\gamma} \in \Gamma_{y}} P_{\rm exp}(\bm{X}(\bm{x})|\bm{\gamma})}{\sum_{y' \in \mathcal{Y}}\frac{1}{|\Gamma_{y'}|}\sum_{\bm{\gamma}' \in \Gamma_{y'}} P_{\rm exp}(\bm{X}(\bm{x})|\bm{\gamma}')}
			= P(y|\bm{X}(\bm{x})).
		\end{equation}
		Thus,
		\begin{align*}\label{eq:SM_der_1}
			P(y|\bm{\gamma}) &= \mathbb{E}_{\bm{x} \sim P_{\rm exp}(\bm{x}|\bm{\gamma})}\left[ P(y|\bm{x}) \right] =  \sum_{\bm{x} \in \mathcal{X}} P_{\rm exp}(\bm{x}|\bm{\gamma})  P(y|\bm{x}) = \sum_{\bm{x} \in \mathcal{X}}  P_{\rm exp}(\bm{X}(\bm{x})|\bm{\gamma})  P(y|\bm{X}(\bm{x})) h(\bm{x})/g(\bm{X}(\bm{x}))\\
			&= \sum_{\bm{X}' \in \mathcal{X}_{\rm suff}}\sum_{\bm{x} \in \mathcal{X}} P_{\rm exp}(\bm{X}'|\bm{\gamma})  P(y|\bm{X}') \delta(\bm{X}(\bm{x})-\bm{X}') h(\bm{x})/g(\bm{X}')= \sum_{\bm{X} \in \mathcal{X}_{\rm suff}} P_{\rm exp}(\bm{X}|\bm{\gamma})  P(y|\bm{X}) = \mathbb{E}_{\bm{X} \sim P_{\rm exp}(\bm{X}|\bm{\gamma})}\left[  P(y|\bm{X})\right] \numberthis, \\
		\end{align*}
	\end{widetext}
	where $\mathcal{X}_{\rm suff}$ is the state space (without duplicates) associated with the sufficient statistic $\mathcal{X}_{\rm suff} = \{\bm{X}(\bm{x}) | \bm{x} \in \mathcal{X} \}$. Similarly, using Eq.~\eqref{eq:Bayes_2_SM}, we have
	\begin{align*}
		\hat{\bm{\gamma}}(\bm{x}) &=  \sum_{y \in \mathcal{Y}} P(y|\bm{x})\bm{\gamma}_{y} = \sum_{y \in \mathcal{Y}} P(y|\bm{x})\bm{\gamma}_{y} P(y|\bm{X}(\bm{x}))\bm{\gamma}_{y}\\
		&= \hat{\bm{\gamma}}\left(\bm{X}(\bm{x})\right) \numberthis,
	\end{align*}
	and
	\begin{align*}\label{eq:SM_der_2}
		\hat{\bm{\gamma}}(\bm{\gamma}) &= \mathbb{E}_{\bm{x} \sim P_{\rm exp}(\bm{x}|\bm{\gamma})}\left[ \hat{\bm{\gamma}}(\bm{x})\right] = \sum_{\bm{x} \in \mathcal{X}} P_{\rm exp}(\bm{x}|\bm{\gamma})\hat{\bm{\gamma}}(\bm{x})\\
		&=  \sum_{\bm{X} \in \mathcal{X}_{\rm suff}} P_{\rm exp}(\bm{X}|\bm{\gamma}) \hat{\bm{\gamma}}(\bm{X}) = \mathbb{E}_{\bm{X} \sim P_{\rm exp}(\bm{X}|\bm{\gamma})}\left[  \hat{\bm{\gamma}}(\bm{X})\right]\numberthis.
	\end{align*}
	The relevant indicators [Eqs. (6), (8), and (10)] of the three phase-classification methods presented in the main text can be straightforwardly computed based on the quantities in Eqs.~\eqref{eq:SM_der_1}, and~\eqref{eq:SM_der_2}, which are expressed solely in terms of the distribution over the sufficient statistic $P(\bm{X}|\bm{\gamma})$ (instead of the full distribution over $\bm{x}$).
	
	In the main text, we considered classical systems with dimensionless Hamiltonians of the form
	\begin{equation}\label{eq:exp}
		\mathcal{H}(\bm{x},\bm{\gamma}) = \sum_{i=1}^{d} \gamma_{i} X_{i}(\bm{x}),
	\end{equation}
	at equilibrium with a large thermal reservoir. In this case, the probability of finding the system in state $\bm{x} \in \mathcal{X}$ is given by
	\begin{equation}\label{eq:Boltzmann_SM}
		P(\bm{x}| \bm{\gamma}) = e^{-\mathcal{H}(\bm{x},\bm{\gamma})}/Z(\bm{\gamma}).
	\end{equation}
	Such Boltzmann distributions belong to the exponential family [Eq.~\eqref{eq:SM_exp_fam}] with $h(\bm{x}) = 1$, $A(\bm{\gamma}) = {\rm ln}\,Z(\bm{\gamma})$, and $\eta_{i}(\bm{\gamma}) = \gamma_{i}$. Thus, as a special case of the above, it follows that phase diagrams of such systems can be mapped out given distributions over the sufficient statistic $\{P(\bm{X}|\bm{\gamma})\}_{\bm{\gamma} \in \Gamma}$ (instead of the full distributions over $\bm{x}$).
	
	\subsection{Utilizing knowledge of symmetries}
	The sufficient statistic allows for an optimal lossless compression of the state space $\mathcal{X}$. However, it requires knowledge of the underlying Hamiltonian. In the following, we demonstrate how one can achieve a lossless compression of the state space without explicitly knowing the underlying Hamiltonian, but only utilizing knowledge of its symmetries. Let us consider a symmetry operation $S: \mathcal{X} \rightarrow \mathcal{X}$ that leaves the energy of the system invariant $\mathcal{H}\left(S(\bm{x})\right) = \mathcal{H}(\bm{x}) \;\forall \bm{x} \in \mathcal{X}$. Considering dimensionless Hamiltonians of the form given in Eq.~\eqref{eq:exp}, we have
	\begin{equation}
		\sum_{i=1}^{d} \gamma_{i} \left[X_{i}(\bm{x}) - X_{i}\left(S(\bm{x})\right)\right] = 0
	\end{equation}
	for all allowed parameters $\bm{\gamma}$. Assuming that this set spans a $d$-dimensional space, we have $\bm{X}(\bm{x}) = \bm{X}\left(S(\bm{x})\right)\; \forall \bm{x} \in \mathcal{X}$. That is, the sufficient statistic is also invariant under any symmetry operation $S$. Thus, we can perform a lossless compression by adopting a representation that is unique for all samples related by symmetry operations. Let us denote the associated state space by $\mathcal{X}_{\rm symm}$. Note that while this compression is lossless, it is not necessarily optimal, i.e., $|\mathcal{X}_{\rm symm}| \geq |\mathcal{X}_{\rm suff}| $.
	
	\section{Data generation}
	Here, we provide further details on the data-generation process for the (classical) anisotropic Ising model and the (quantum) cluster-Ising model.
	\subsection{Anisotropic Ising model}
	Given a set of parameters $\bm{\gamma}=(J_{x}/k_{\rm B}T,J_{y}/k_{\rm B}T)$, we use the Metropolis-Hastings algorithm to sample spin configurations from the corresponding Boltzmann distribution. The lattice is updated by drawing a random spin, which is flipped with probability ${\rm min}(1, e^{-\Delta E/k_{\rm B}T})$, where $\Delta E$ is the energy difference resulting from the considered flip. After a thermalization period of $10^5$ lattice sweeps, we collect $10^{5}$ samples. We treat each quadrant of the phase diagram separately. In each quadrant, we initialize the system in one of the two corresponding ground states and increase $\gamma_{1}$ at constant $\gamma_{2}$. When increasing $\gamma_{1}$, we use the spin configuration from the preceding value as an initial condition for the Markov chain. When increasing $\gamma_{2}$, we reset the system to the corresponding ground state.
	
	\subsection{Cluster-Ising model}
	Given a set of parameters $\bm{\gamma}=(h_{1}/J,h_{2}/J)$, we compute ground states of the corresponding cluster-Ising Hamiltonian [Eq.~(9) in main text] with $L = 71$ spins using the finite-size density matrix renormalization group (DMRG) with a maximum bond dimension of 150. We utilize the ITensor package~\cite{itensor} in \texttt{Julia}.
	
	\bibliography{refs.bib}